\documentclass[aps,prd,twocolumn,superscriptaddress,showpacs,amsmath,amssymb]{revtex4-2}

\usepackage{upgreek}
\usepackage{standalone}
\usepackage{booktabs}
\usepackage{physics}
\usepackage{siunitx}
\usepackage{tabularx}
\usepackage{comment}
\usepackage{multirow}
\usepackage{soul}
\usepackage{CJKutf8}
\usepackage[caption=false]{subfig}
\usepackage{diagbox}
\usepackage{graphicx}

\usepackage{graphicx} % Include figure files
\usepackage{epstopdf} % Allow to include eps files
\usepackage{dcolumn} % Align table columns on decimal point
\usepackage{xcolor} % Color support
\usepackage{amsmath,amssymb} % Extended symbol collection
\usepackage{hyperref} % Generate pdfs with hyperlinks
\usepackage[utf8]{inputenc} % Allow UTF8 characters
\usepackage[english]{babel} % All publications are in English
\usepackage[displaymath, mathlines]{lineno} % Line numbers for drafts
\usepackage{blindtext} % For testing
\usepackage{ifthen} % for conditional statements
\usepackage{orcidlink} % for ORCIDs in author list
\usepackage{cancel}
\usepackage{siunitx}
\usepackage{physics}
\AtBeginDocument{\RenewCommandCopy\qty\SI}
\usepackage{xspace}
\graphicspath{{figures/}} % Use figures directory for figures

\newboolean{articletitles}
\setboolean{articletitles}{true} % False removes titles in references

\addto\captionsenglish{}
\addto\captionsenglish{}

\input{belle2-symbols}

\newcommand{\Btag}{\ensuremath{B_\mathrm{tag}}\xspace}

\usepackage[capitalise]{cleveref}
\crefname{section}{Sec.}{Secs.}
\crefname{table}{Tab.}{Tabs.}

\def\Btaunu {\ensuremath{B^+\to\tau^+\nu_\tau}\xspace}   
\def\brbtaunu {\ensuremath{\mathcal{B}(B^+\to\tau^+\nu_\tau)}\xspace}
\def\btag {\ensuremath{B_\text{tag}}\xspace}
\def\bsig {\ensuremath{B_\text{sig}}\xspace}
\def\tagprob {\ensuremath{\mathcal{O}_{\rm FEI}}\xspace}
\def\eextra{\ensuremath{E_\text{ECL}^\text{extra}}\xspace}
\def\missM{\ensuremath{M_{\mathrm{miss}}^2}\xspace}
\def\costbto{\ensuremath{\cos \theta_\text{T}}\xspace}
\def\R{\ensuremath{R_2}\xspace}
\def\ng{\ensuremath{n_{\gamma\mathrm{extra}}}\xspace}
\def\nBB{\ensuremath{n_{\FourS}}\xspace}

\begin{document}
\title {Measurement of \texorpdfstring{\Btaunu}{Btaunu} branching fraction with a hadronic tagging method at \belletwo}
%%% Paper:    B+ to tau+ nu_tau
%%% Journal:  Physical Review D
%%% Contacts: M. Aversano, G. De Nardo, G. Gaudino, T. Iijima, M. Merola
%%% ====================================================================
%%% Use \input{pub075-orcid} to insert this material into your latex file.
  \author{I.~Adachi\,\orcidlink{0000-0003-2287-0173}} % 2590
  \author{K.~Adamczyk\,\orcidlink{0000-0001-6208-0876}} % 2239
% \author{L.~Aggarwal\,\orcidlink{0000-0002-0909-7537}} % 10083
% \author{P.~Ahlburg\,\orcidlink{0000-0002-9832-7604}} % 2367
  \author{H.~Ahmed\,\orcidlink{0000-0003-3976-7498}} % 11323
% \author{J.~K.~Ahn\,\orcidlink{0000-0002-5795-2243}} % 7423
  \author{Y.~Ahn\,\orcidlink{0000-0001-6820-0576}} % 14363
  \author{H.~Aihara\,\orcidlink{0000-0002-1907-5964}} % 2223
  \author{N.~Akopov\,\orcidlink{0000-0002-4425-2096}} % 9443
  \author{M.~Alhakami\,\orcidlink{0000-0002-2234-8628}} % 28103
  \author{A.~Aloisio\,\orcidlink{0000-0002-3883-6693}} % 2194
  \author{N.~Althubiti\,\orcidlink{0000-0003-1513-0409}} % 21524
% \author{L.~Andricek\,\orcidlink{0000-0003-1755-4475}} % 2098
  \author{M.~Angelsmark\,\orcidlink{0000-0003-4745-1020}} % 13963
% ...
% 13403
% \author{S.~Zou\,\orcidlink{0000-0003-3377-7222}} % 19363
\collaboration{The Belle II Collaboration}

\begin{abstract}
\vspace{-2.5ex}
We present a measurement of the branching fraction of \Btaunu decays using $\SI{387\pm6e6}{}$ \FourS collected between 2019 and 2022 %at the \FourS resonance 
with the Belle II detector at SuperKEKB \epem collider. We reconstruct the accompanying \Bm meson using the hadronic tagging method, while \Btaunu candidates are identified in the recoil. We find evidence for \Btaunu decays at 3.0 standard deviations, including systematic uncertainties. The measured branching fraction is $\brbtaunu = [1.24 \pm 0.41 \stat \pm 0.19 \syst] \times 10^{-4}$.
\end{abstract}

\maketitle

\section{Introduction}
The leptonic decay \Btaunu \footnote{Throughout this paper, the inclusion of the charge-conjugate decay mode is implied unless otherwise stated.} is a process with a clean theoretical prediction in the Standard Model (SM) and is potentially sensitive to contributions from beyond-the-standard-model (BSM) physics.
In the SM, the branching fraction is given by: 
\begin{equation}\label{eq:bf_taunu}
    \mathcal{B}(\Btaunu) = \frac{G_F^2 m_B m_\tau^2}{8\pi} \Biggl[1-\frac{m_\tau^2}{m_B^2}\Biggr]^2 f_B^2 \abs{V_{ub}}^2 \tau_B,
\end{equation}
where $G_F$ is the Fermi coupling constant, $m_B$ and $m_\tau$ are the masses of the charged \Bp meson and the $\tau$ lepton, respectively, $f_B$ is the \Bp meson decay constant, $V_{ub}$ is the Cabibbo-Kobayashi-Maskawa matrix element related to $u$ and $b$ quarks, and $\tau_B$ the lifetime of the \Bp meson. All of the quantities in Eq.~\ref{eq:bf_taunu} are measured experimentally \cite{2024pdg} except for $f_B$, which is determined from Lattice Quantum Chromodynamics (LQCD) simulations~\cite{aoki2024flagreview2024}.

Assuming the SM and the precise calculation of $f_B$ from LQCD, the \Btaunu decay mode provides a direct measurement of the CKM matrix element \Vub that is independent of exclusive and inclusive semileptonic $B \to X_u \ell \neul$ decays, which are typically studied for this purpose~\cite{banerjee2024averagesbhadronchadrontaulepton}. Moreover, in leptonic decays, the theoretical uncertainty will not be a limiting factor soon; the FLAG working group estimates an uncertainty below 1\%~\cite{aoki2024flagreview2024}.
The \Btaunu decay is sensitive to BSM contributions, such as those predicted by models with a charged Higgs boson, the two Higgs Doublet Model (2HDM) \cite{higgscarico,Crivellin:2012ye,Haller:2018nnx}, or various supersymmetric extensions of the SM~\cite{Bryman:2019ssi,Guerrera:2022ykl}. In these models the branching fraction of the \Btaunu decay can be enhanced (or suppressed) by a factor up to 4~\cite{Jung_2010}, taking into account experimental constraints from previous measurements. Therefore, a precise measurement of the branching fraction can also be used to constrain the parameter space of these models.

\belle and \babar measured \brbtaunu reconstructing the accompanying \Bm meson in hadronic decays~\cite{babar_had,belle_had} or semileptonic decays~\cite{belle_semi,babar_semi}. 
Table \ref{tab:pastmeasurements} shows past measurements and their current world average.

\begin{table}[htbp]
    \renewcommand{\arraystretch}{1.2}
    \centering
    \caption{Published results for \brbtaunu by \belle, \babar and the PDG average.}
    \begin{tabular*}{\linewidth}{@{\extracolsep{\fill}}ccc}
        \toprule
        \toprule
    Experiment & Tag & $\mathcal{B}(10^{-4})$\\
        \midrule
        \belle & Hadronic & $0.72^{+0.27}_{-0.25} \pm 0.11$ \\
        \babar & Hadronic & $1.83 ^{+0.53}_{-0.49} \pm 0.24$ \\
        \belle & Semileptonic & $1.25 \pm 0.28 \pm 0.27$ \\
        \babar & Semileptonic & $1.8 \pm 0.8 \pm 0.2$ \\
        \midrule 
        PDG & & $1.09 \pm 0.24$ \\
        \bottomrule
        \bottomrule
    \end{tabular*}
    \label{tab:pastmeasurements}
\end{table}

The measurement described in this paper is based on data collected by the \belletwo experiment at the SuperKEKB electron-positron collider between 2019 and 2022 and has an integrated luminosity of $\SI{365.4\pm1.7}{\invfb}$~\cite{Belle-II:2024vuc}, corresponding to a number of produced \FourS estimated to be $\nBB = (387 \pm 6) \times 10^6$. 
In addition, we use $42.3~\invfb$ of data collected at the slightly lower center-of-mass energy of 10.52 GeV (off-resonance) to calibrate the background from continuum \epem\to\qqbar (where $\Pq =\Pu, \Pd, \Ps, \Pc$) and \epem\to\tautau events in a data-driven way.
A \Bm meson is fully reconstructed in an exclusive hadronic decay (\Btag) and the remaining charged particle trajectories (tracks) and neutral energy deposit in the calorimeter (clusters) are examined for the signature of a \Btaunu decay (\bsig). We consider four \taup decays with a single charged particle in the final state: \taup\to\ep\neue\neutb, \taup\to\mup\neum\neutb, \taup\to\pip\neutb, and \taup\to\rhop\neutb channels, where $\rho$ is the $\rho(770)$. These modes account for approximately 72\% of all $\tau$ decays \cite{2024pdg}. Each mode is treated as a distinct signal category.
We define a set of selection requirements to suppress the backgrounds for which either the \Btag or the \bsig are misidentified. We optimize the signal selection on simulation, which is corrected and validated on several control samples. We extract the branching fraction using a simultaneous two-dimensional maximum likelihood fit to two discriminating variables, the residual energy in the electromagnetic calorimeter not associated with the reconstructed \BpBm pair, and the missing mass squared of the event.

\section{\belletwo detector and simulation}

The Belle~II experiment~\cite{Abe:2010gxa} is located at the SuperKEKB accelerator~\cite{Akai:2018mbz}, which collides $\SI{7}{\gev}$ electrons and $\SI{4}{\gev}$ positrons at and near the \FourS resonance. The \belletwo detector  \cite{Abe:2010gxa} has a cylindrical geometry arranged around the interaction point (IP), which is enclosed by a beryllium beam pipe with an inner radius of 1 cm, and includes a two-layer silicon-pixel detector~(PXD) surrounded by a four-layer double-sided silicon-strip detector~(SVD)  \cite{Belle-IISVD:2022upf} and a 56-layer central drift chamber~(CDC). These detectors reconstruct tracks of charged particles.  In this work, we analyze the data for the period, when only one-sixth of the second layer of the PXD was installed.  Surrounding the CDC, which also provides ionization-energy-loss measurements, is a time-of-propagation counter~(TOP)  \cite{Kotchetkov:2018qzw} in the central region and an aerogel-based ring-imaging Cherenkov counter~(ARICH) in the forward endcap region.  These detectors provide charged-particle identification (PID).  Surrounding the TOP and ARICH is an electromagnetic calorimeter~(ECL) based on CsI(Tl) crystals that primarily provide energy and timing measurements for photons and electrons. Outside of the ECL is a superconducting solenoid magnet, which provides an axial magnetic field of $1.5\,\text{T}$.
A \KL and muon identification system is located outside of the magnet and consists of flux-return iron plates interspersed with resistive plate chambers and plastic scintillators.
The central axis of the solenoid defines the $z$ axis of the laboratory frame, pointing approximately in the direction of the electron beam.

The analysis strategy is tested and optimized on Monte Carlo (MC) simulated event samples before being applied to the data.
Quark-antiquark pairs (\qqbar) from \epem collisions are generated using \textsc{KKMC}~\cite{Jadach:1999vf} with \textsc{Pythia8} ~\cite{Sjostrand:2014zea}, while hadron decays are simulated with \textsc{EvtGen} ~\cite{Lange:2001uf}. The detector response is simulated using \textsc{Geant4}~ \cite{Agostinelli:2002hh}. For all simulated events, electromagnetic final-state radiation is taken into account using the \textsc{PHOTOS} package  \cite{davidson2015photos, Barberio:1993qi}. 
For the simulated signal, we produce $\SI{4e7}{}$ events, in which one $B$ meson decays exclusively to the \Btaunu final state, and the other $B$ meson decays generically.
For the \BBbar, \qqbar, and \tautau backgrounds we use simulated samples with equivalent integrated luminosities of $\SI{2.8}{ab^{-1}}$, $\SI{1}{ab^{-1}}$ and $\SI{0.6}{ab^{-1}}$, respectively. Both experimental and simulated data are processed using the \belletwo analysis software framework~\cite{Kuhr:2018lps}. 

\section{Reconstruction and Event Selection}
\label{sec:rec}

To consider an event for further analysis we require the reconstruction of three or more tracks, each with an impact parameter with respect to the IP, less than $\SI{2}{cm}$ along the $z$ axis, and $\SI{0.5}{cm}$ in the transverse direction; three or more ECL clusters with $E > \SI{100}{\mev}$ and within the CDC acceptance; a minimum transverse momentum of $\SI{100}{\mev}/c$ for each of the charged particles; the total energy of reconstructed tracks and ECL clusters greater than $\SI{4}{GeV}$.
 
Simulation samples and data are then passed through the Full Event Interpretation (FEI)~\cite{FEI}, a hierarchical multivariate algorithm that fully reconstructs the \Btag in thousands of possible decay chains. The output of the FEI algorithm for each event is a set of reconstructed \btag candidates with an associated score (\tagprob); the higher the \tagprob, the higher the expected purity of the candidate. We require \tagprob$>0.001$ and the resulting fraction of \FourS events with a correctly reconstructed charged \btag candidate is estimated from simulations to be approximately 0.30\% with a purity of 29\%~\cite{belleiicollaboration2020calibration}.
The average \Btag candidate multiplicity in signal events is 1.02. We retain one \Btag candidate per event, choosing the candidate with the highest value of \tagprob.

We use two kinematic variables to discriminate between events with a correctly reconstructed \btag candidate and misreconstructed events: the beam-energy-constrained mass  $M_\text{bc} = \sqrt{(E^*_\text{beam})^2/c^4-|\vec{p}^{\ *}_\text{tag}|^2/c^2}$ and the energy difference $\Delta E =E^*_\text{tag}-E_\text{beam}^*$, where $\vec{p}^{\ *}_\text{tag}$ ($E^*_\text{tag}$) is the reconstructed momentum (energy) of the \btag and $E_\text{beam}^*$ is the beam energy, all evaluated in the center of mass (c.m.) frame. We require $M_\text{bc}>\SI{5.27}{GeV}/c^2$ and $-0.15 < \Delta E < \SI{0.1}{GeV}$.  
 
The \bsig is reconstructed from all the remaining tracks and neutral objects (ECL clusters with no tracks associated) not used to reconstruct the \btag. 
We consider four different signal categories identifying the following final states: \taup\to\ep\neue\neutb, \taup\to\mup\neum\neutb, \taup\to\pip\neutb, and \taup\to\rhop\neutb.
For this purpose, in addition to the tracks associated with the \btag, we require only another track in the event, with a charge opposite to the \btag charge and a momentum $p > \SI{0.5}{GeV}/c$. 
The PID criteria to distinguish between electron, muon, and pion hypotheses are based on multivariate classifiers that utilize information from all sub-detectors.
This information is combined into a boosted decision tree for electrons or a likelihood function for muons and pions. The likelihood for pion or muon hypothesis combines PID information from all sub-detectors except SVD and PXD.
From simulation, we estimate that PID selection criteria discriminate electrons, muons, and pions with efficiencies of 99\%, 82\%, and 97\% at misidentification rates of 1\%, 5\%, and 3\%, respectively.
A bremsstrahlung correction in the \taup\to\ep\neue\neutb decay mode is applied: the four-momentum of the electron candidates is corrected by adding the four-momenta of photons with an energy below 1.0 GeV within a cone of 0.05 rad around the electron-momentum vector.
From pairs of ECL clusters not used to reconstruct \btag, we reconstruct $\piz\to\g\g$ candidates. We require these candidates to have an invariant mass in the range $120 < m_{\g\g} < 145 \mev/c^2$, which corresponds to approximately $\pm 1~\sigma$ of the mass resolution. The \piz candidates are combined with a \pip candidate to form a \rhop candidate.
If the \rhop candidate has a reconstructed invariant mass within $625 < m_{\pip\piz} < 925\mev/c^2$, the \bsig is assigned to the \taup\to\rhop\neutb category. Otherwise, the event is assigned to the \taup\to\pip\neutb category. If multiple \rhop candidates are reconstructed, we choose the one closest to the \rhop mass~\cite{2024pdg}. 

Any cluster in the ECL not associated with the \btag nor with the \bsig is subjected to a procedure that rejects the clusters from beam-induced backgrounds, the interaction of hadronic particles with detector material (hadronic split-off showers), and neutral hadrons. The procedure uses two different multivariate classifiers, trained on a \Bz\to\Dstarm\ellp\neul data control sample as described in~\cite{refId0}. All the objects refined by this clean-up procedure represent the rest of the event (ROE).

The two most discriminating observables based on the ROE are the total residual energy from neutral clusters in the ECL (\eextra), and the square of missing four-momentum (\missM) calculated using the known beam energies and all the reconstructed objects:
\begin{equation}
p^*_\text{miss} = (2E_\text{beam}^*, 0, 0, 0) - p^*_\text{tag} - p^*_\text{sig} - p^*_\text{ROE},
\end{equation}
where $p^*_\text{tag}$, $p^*_\text{sig}$, $p^*_\text{ROE}$ are respectively the \btag, the signal candidate, and the ROE objects four-momenta in the c.m.~frame.

We suppress continuum \epem\to \qqbar and \epem\to\tautau backgrounds by applying a loose selection on two event-shape variables. 
The first variable, \costbto, is the cosine of the angle between the thrust axis of the \btag and the thrust axis of its recoil. The thrust axis is defined as the unit vector $\hat{t}$ that maximizes the thrust value $ \sum |\hat{t} \cdot \vec{p}^{\ *}_i|/\sum |\vec{p}^{\ *}_i|$ where $\vec{p}^{\ *}_i$ is the momentum of the $i$-th final-state particle in the $e^+e^-$ c.m. frame~\cite{Brandt:1964sa, Farhi:1977sg}.
This variable discriminates between signal and background because \BBbar events have an approximately uniform distribution, while continuum events consist of two back-to-back jets in the c.m. frame.
The second variable is the ratio of Fox-Wolfram polynomials \R~\cite{R2_citazione}, which measures the degree of sphericity of the event with lower values for \BBbar events and higher values for continuum events. 
We require $|\costbto| < 0.9$ and $\R < 0.6$, selecting 73\% of the signal and rejecting 99\% of the \tautau background events. 

\begin{figure*}[tbp]
    \centering
    \begin{minipage}{0.4\textwidth}
        \centering
        \includegraphics[width=\textwidth, keepaspectratio]{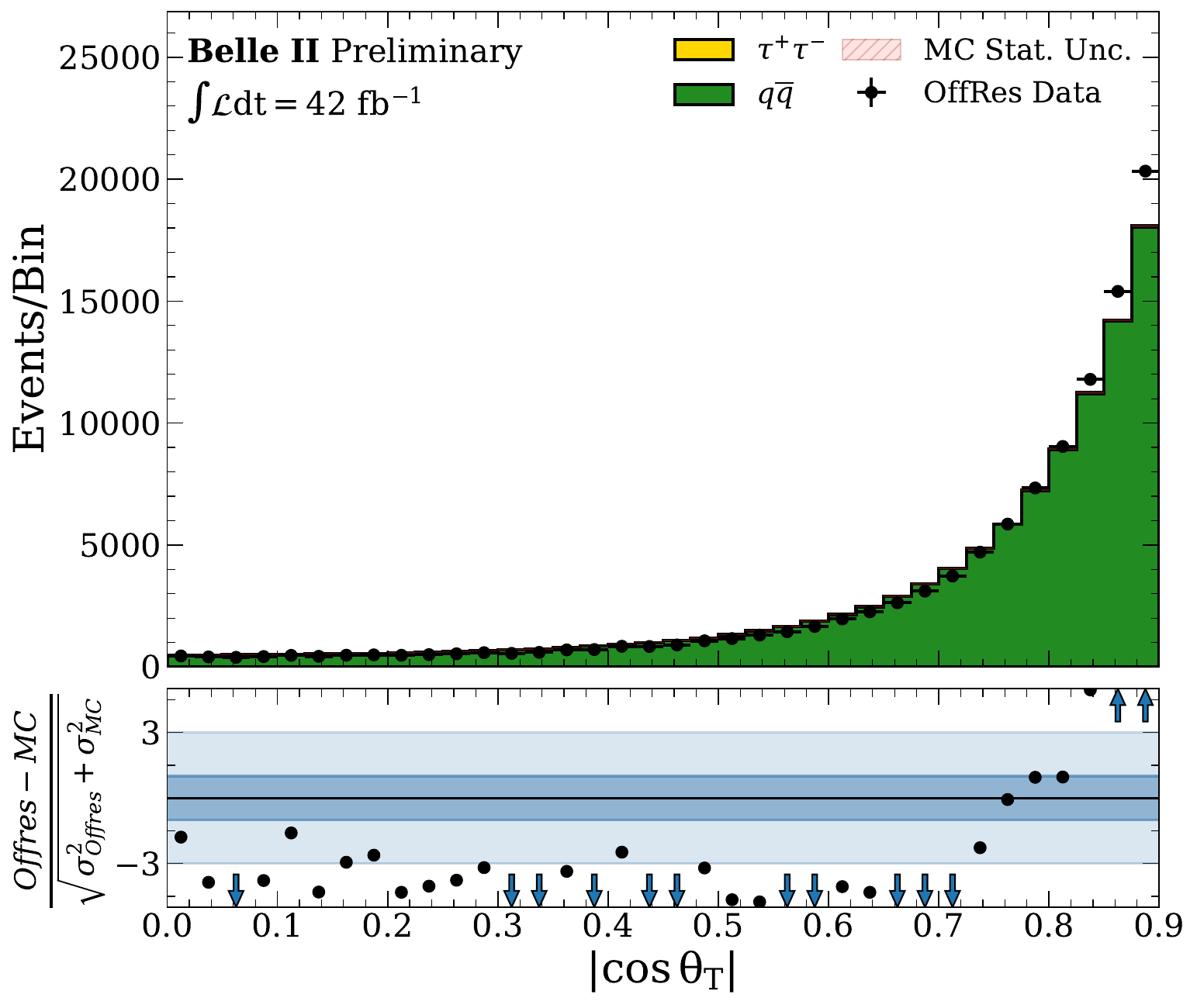}
    \end{minipage}
    \begin{minipage}{0.4\textwidth}
        \centering
        \includegraphics[width=\textwidth, keepaspectratio]{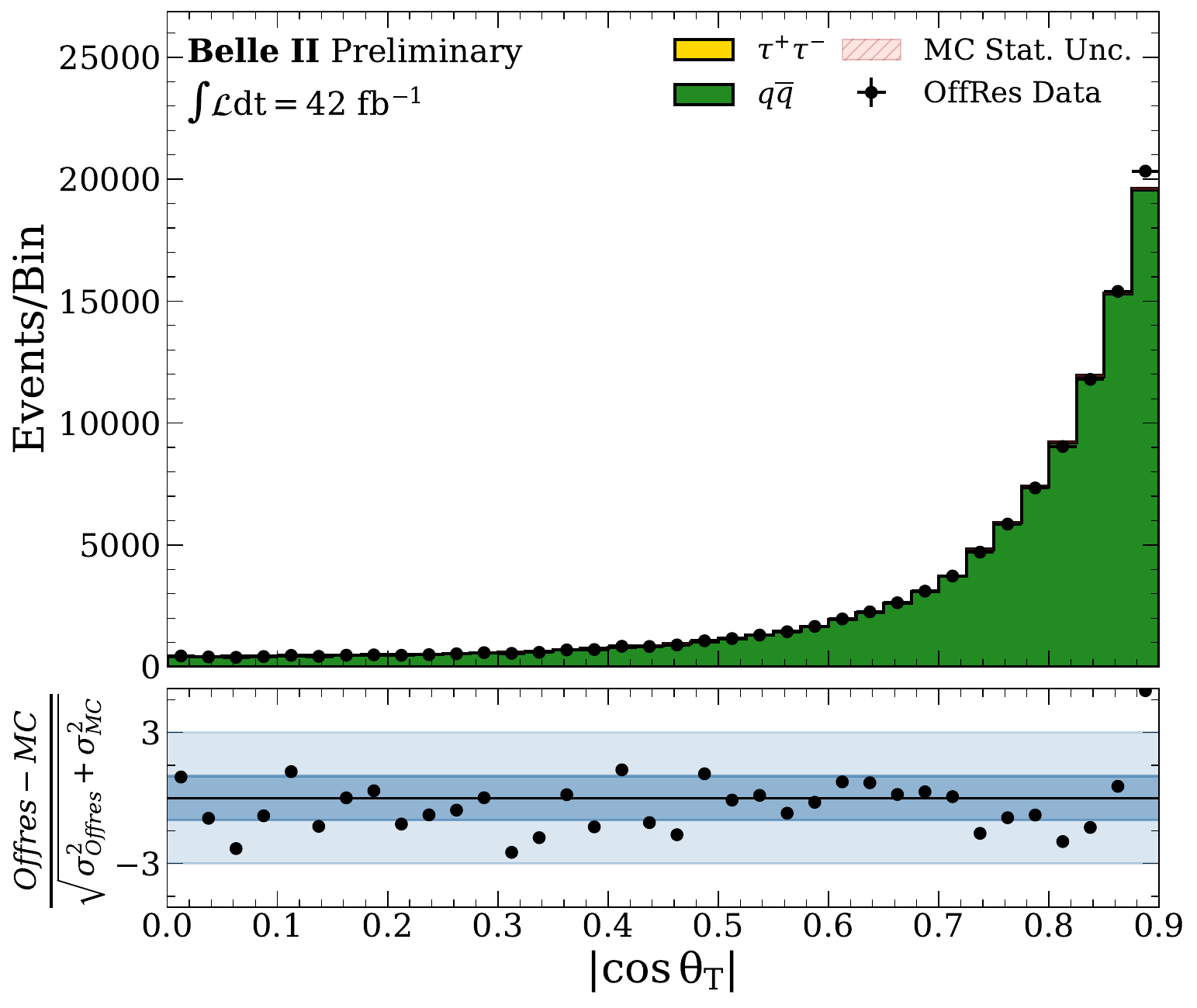}
    \end{minipage}\\
    \begin{minipage}{0.4\textwidth}
        \centering
        \includegraphics[width=\textwidth, keepaspectratio]{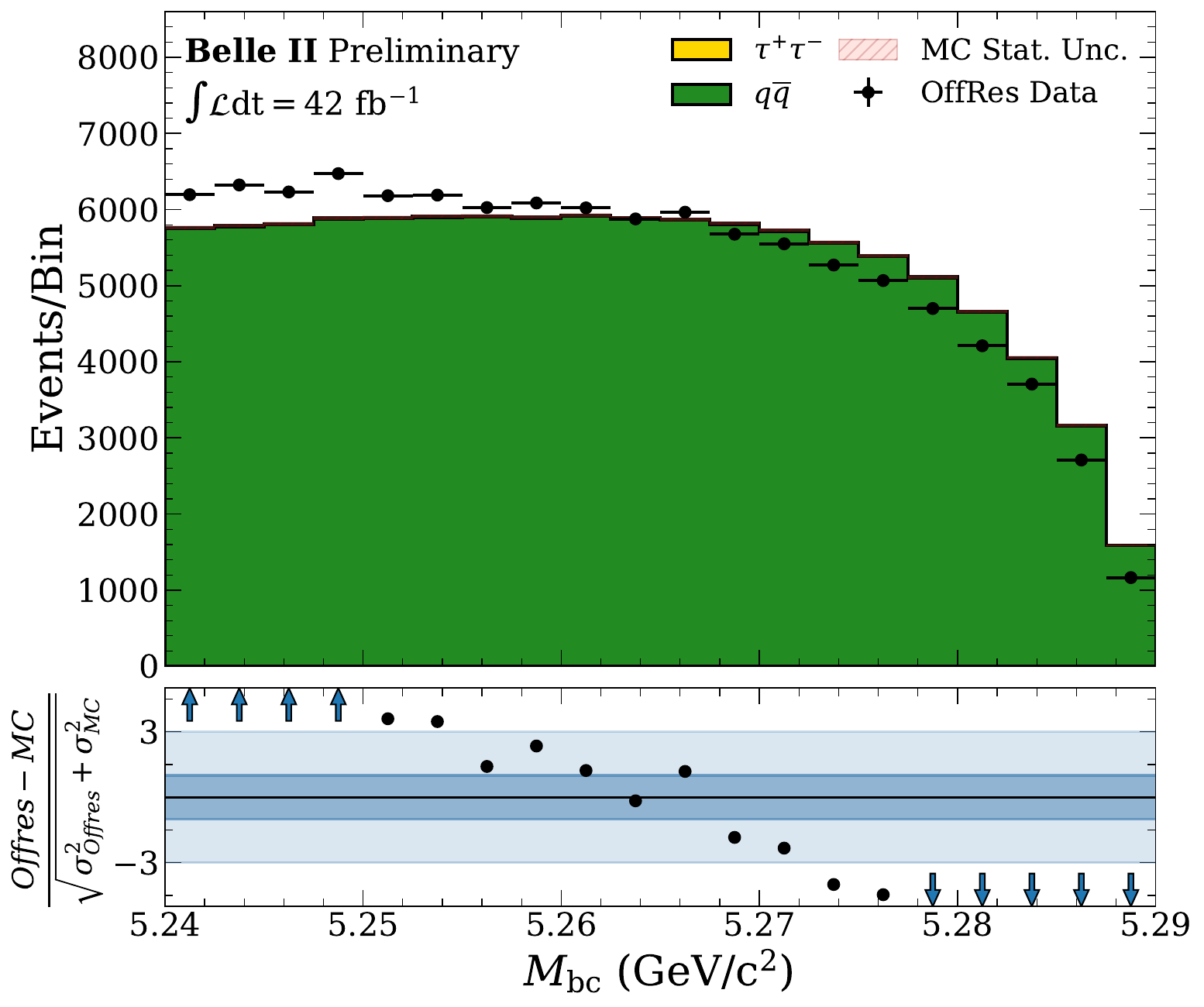}
    \end{minipage}
    \begin{minipage}{0.4\textwidth}
        \centering
        \includegraphics[width=\textwidth, keepaspectratio]{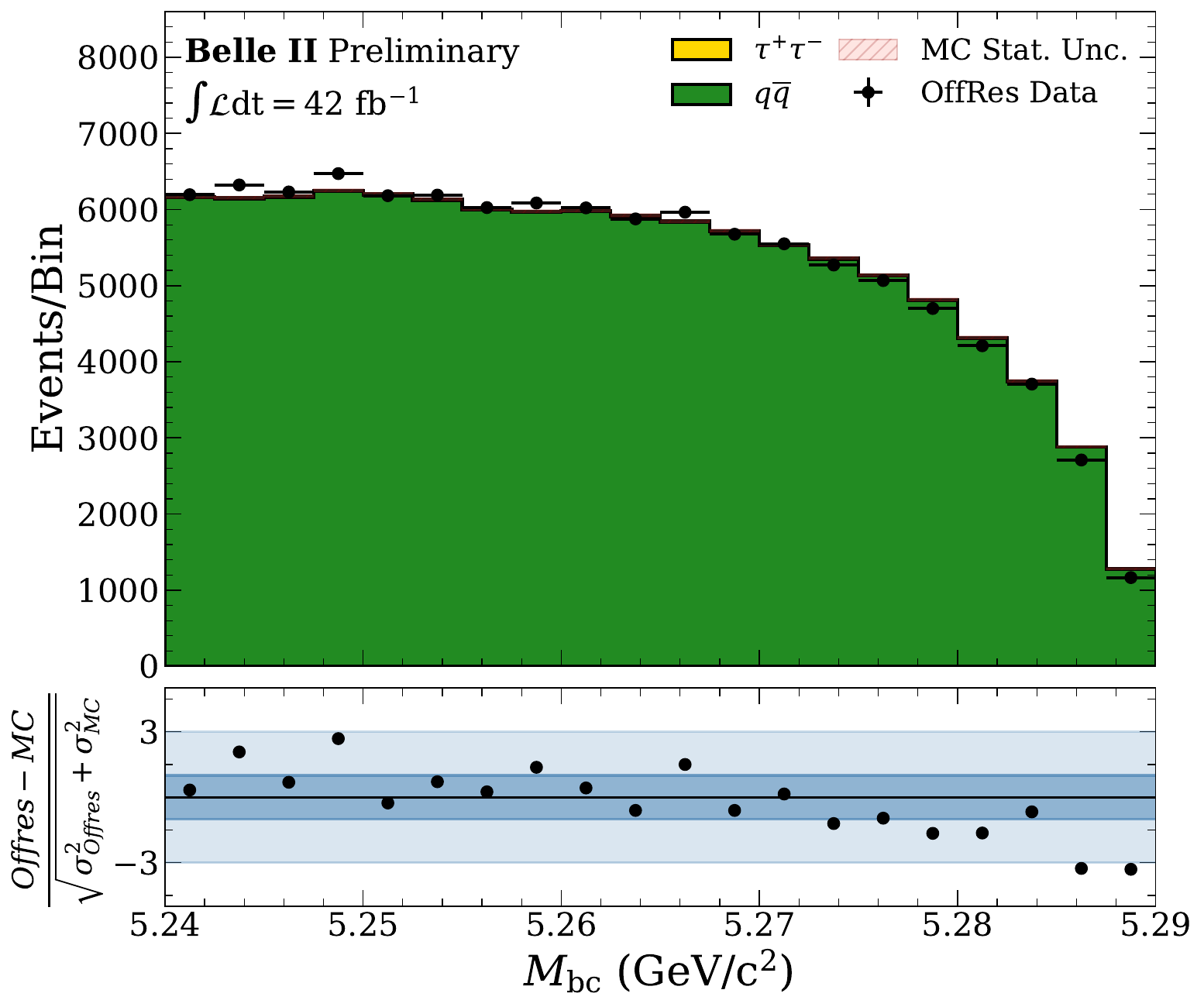}
    \end{minipage}\\
    \caption{Distributions of $|\costbto|$ (top) and $M_\text{bc}$ (bottom) in off-resonance data and continuum simulation before (left) and after (right) the continuum MC reweighting. The \tautau component is negligible after the requirement on \R.}
    \label{fig:cont_reweight}
\end{figure*}

After these loose requirements, the remaining continuum events are still a significant fraction of the total background, especially for the hadronic modes \taup\to\pip\neutb and \taup\to\rhop\neutb, and need further suppression. Since there are no \BBbar events in the off-resonance sample, we use it to estimate the expected yield of the continuum background. 
Due to the lower statistics of the off-resonance sample after the signal selection procedure, we use simulation instead to describe the shapes of all variables for continuum events. Nevertheless, comparing simulation and off-resonance data we observe that continuum simulation does not adequately reproduce the shapes of many variables, as it can be seen, for example, in Figure~\ref{fig:cont_reweight} (left plots).
We correct this mis-modeling of the shapes by reweighting the simulation using the off-resonance data.
We employ a fast boosted decision tree (FBDT) classifier~\cite{Keck:2017gsv} using the event shape variables described in~\cite{Bevan2014}.
We use a simulated sample of continuum events and off-resonance data, which both correspond to $5\times10^4$ events, for training and validation. 
The weight $w_\texttt{CR}^i$ for each event in simulation is defined as:
\begin{equation}
    w_\texttt{CR}^i= \frac{\OCR^i}{1-\OCR^i}
\end{equation}
where $\OCR^i$ is the output of the classifier for the event $i$~\cite{Martschei_2012,feld_lena_2020_22016}.
Figure \ref{fig:cont_reweight} shows the effect of the continuum reweighting on the simulated events for the  $|\costbto|$ and $M_\text{bc}$ distributions.

To further suppress continuum, simulated \BBbar and reweighted continuum events are used to train another FBDT, based on event-shape variables described in~\cite{Bevan2014}. The most discriminating variables are the Super-Fox-Wolfram moments and the harmonic moments \mbox{$B_l = \sum_i (p_i/\sqrt{s}) P_l (\cos\alpha_i)$}, where $P_l$ is the Legendre polynomial of order $l$, $p_i$ is the momentum of the particle $i$ and $\alpha_i$ is the angle calculated with respect to the thrust axis of the \btag recoil, as defined in~\cite{R2_citazione}.
As the leptonic and hadronic \taup channel inputs differ slightly, we train two independent FBDTs. To avoid bias, we use the classifier on simulated samples of continuum and simulated samples of \BBbar of the same size ($10^5$ events) for training and validation.

Finally, we exploit the following variables to further suppress backgrounds: \TagProb, the output of the continuum suppression FBDT \OCS, and the momentum of the reconstructed daughter of the \taup (\ep, \mup, \pip, and \rhop) in the laboratory frame $p_\text{cand}$.
More stringent requirements on \TagProb improve the purity of the \btag reconstruction; \OCS is used for discrimination against continuum, especially necessary for hadronic \taup decay modes; a tight requirement on $p_\text{cand}$ is effective to suppress \BBbar background for hadronic modes produced in \taup two-body decays.
We optimize the selection requirements independently for each channel to minimize the statistical uncertainty of signal yields. The signal yields are fitted (as explained in Sec. \ref{sec:sig_sel_opt}) on pseudo-datasets generated from signal and background simulations, assuming a signal branching fraction $\brbtaunu_\text{PDG}$ and integrated luminosity as in data.
Table \ref{tab:best_cuts} shows the optimized selection.
\begin{table}[htbp]
    \renewcommand{\arraystretch}{1.2}
    \centering
    \caption{Optimized signal selection. $\epsilon$ is the efficiency of a $\Btaunu$ event to be reconstructed in each signal category.}
    \begin{tabular*}{\linewidth}{@{\extracolsep{\fill}}c|ccc|c}
        \toprule
        \toprule
        Sig. category & \tagprob  & \OCS & $p_\text{cand}$ (GeV/c) & $\epsilon$ ($10^{-4}$)\\
        \hline
        \ep    &  \multirow{4}{*}{$>0.01$} & $<0.8$ & \multirow{2}{*}{$>0.5$}      & 7.3 \\
        \mup  &                           & $<0.6$ &                               & 7.6 \\
        \pip  &                           & $<0.6$ & $>1.4$                        & 3.4 \\
        \rhop &                           & $<0.7$ & $>1.65$                       & 3.1 \\
        \bottomrule
        \bottomrule
    \end{tabular*}
    \label{tab:best_cuts}
\end{table}

\section{Calibration and Model Validation}
\label{sec:calib}
In this section, we discuss the efficiency correction and \eextra calibration and validation using several control samples and signal sidebands.

\subsection{Efficiency correction}
We determine the hadronic FEI reconstruction efficiency from simulation and then we correct it with a data-driven procedure using two data control samples. 
For the first one, we examine the recoil of the \btag to select a sample of inclusive semileptonic decays \Bp\to$X$\ellp\neul, requiring the reconstruction of a high energy electron or muon. 
The details of the procedure are discussed elsewhere~\cite{belleiicollaboration2020calibration}.
A second sample is selected requiring the presence of a hadronic decay $\Bp\to D^{(*)}\xspace\pip$, searching for the $D^{(*)}$ resonance in the \pip recoil. The corrections to \btag reconstruction efficiency, extracted from the hadronic sample are consistent with the first ones, and, therefore, we use a combination of the correction factors from the two samples. The correction factors, obtained by comparing the yields in data and simulation, depend on the \btag decay mode and vary between 0.6 and 1.1.

Additional corrections are applied to account for mis-modeling in simulation of PID efficiencies and mis-identification probabilities. 
They are evaluated by comparing data and simulation on pure samples of electrons and muons from inclusive decays \jpsi\to \ellell, and low multiplicity processes \epem\to \ellell (\g) and \epem\to \epem \ellell. To calibrate the charged pion identification we use a sample of pions from inclusive decays of \KS\to\pipi, \Dstarp\to\Dz\pip, and \Lz\to\Pp\pim. The correction factors depend on the momentum and polar angle of the tracks.\\
\indent Finally, a photon efficiency correction is applied to the ECL clusters in the ROE using a random removal algorithm, since photon reconstruction efficiency in data is always slightly smaller than photon reconstruction efficiency in simulation. A photon in MC is excluded from the reconstruction with a probability $1-\omega$, where $\omega$ is the ratio of photon reconstruction efficiency in data and in simulation. To do so, we generate a repeatable random sequence of values between zero and one, and a given photon is removed if the corresponding random value is greater than $\omega$. The $\omega$ ratios, which vary between 0.8 and 1.0, are extracted from data and simulation samples of \epem\to\mumu\g events as a function of missing momentum and its angular direction ($\theta$,$\phi$) as shown in~\cite{Svidras:473367}.

\begin{figure*}[htbp]
    \centering
    \captionsetup{justification=centering}
    \subfloat{(a)\label{fig:extra_mul_a}}{\includegraphics[width = 0.4\textwidth, keepaspectratio]{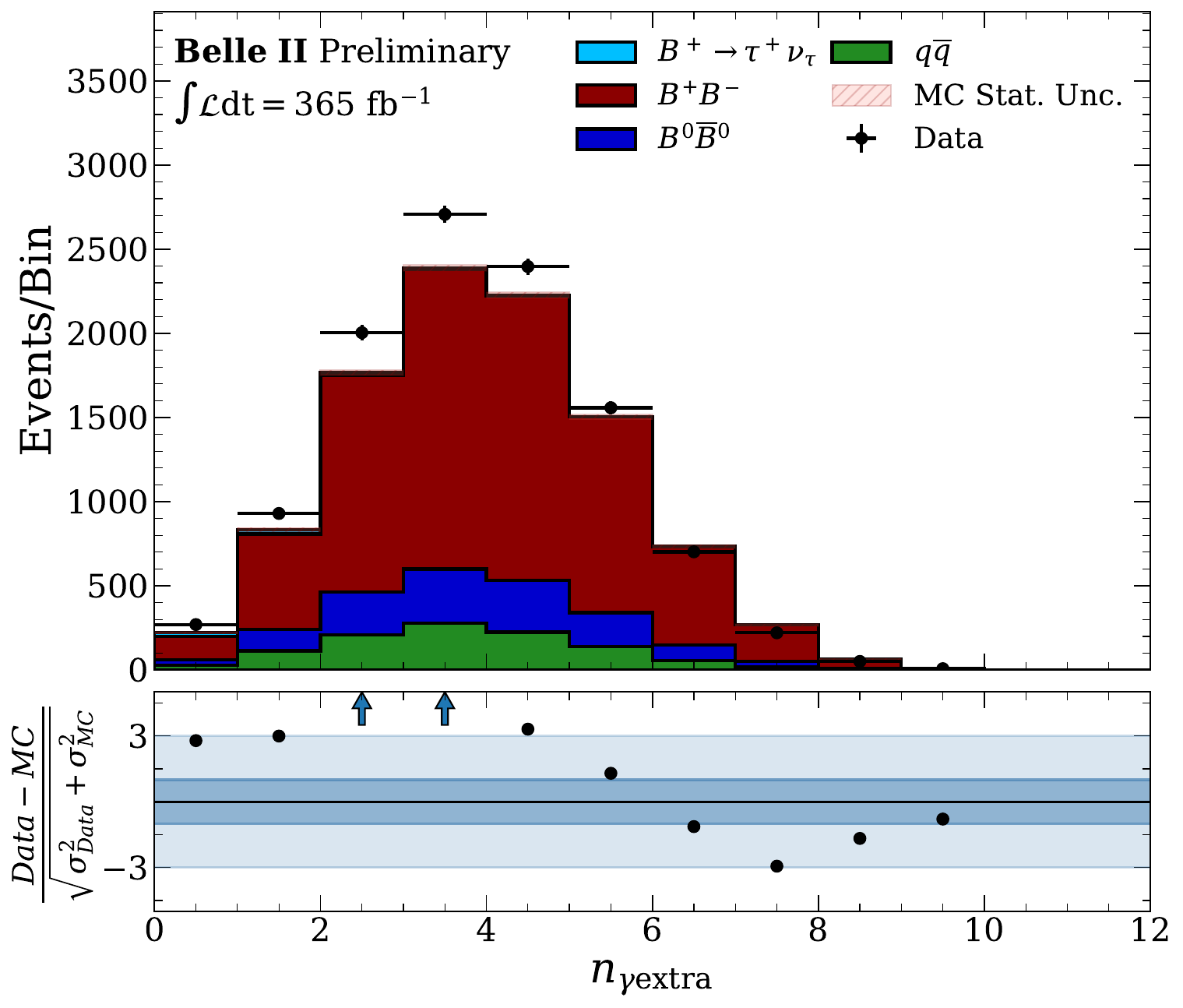}}
    \subfloat{(b)\label{fig:extra_mul_b}}{\includegraphics[width = 0.4\textwidth, keepaspectratio]{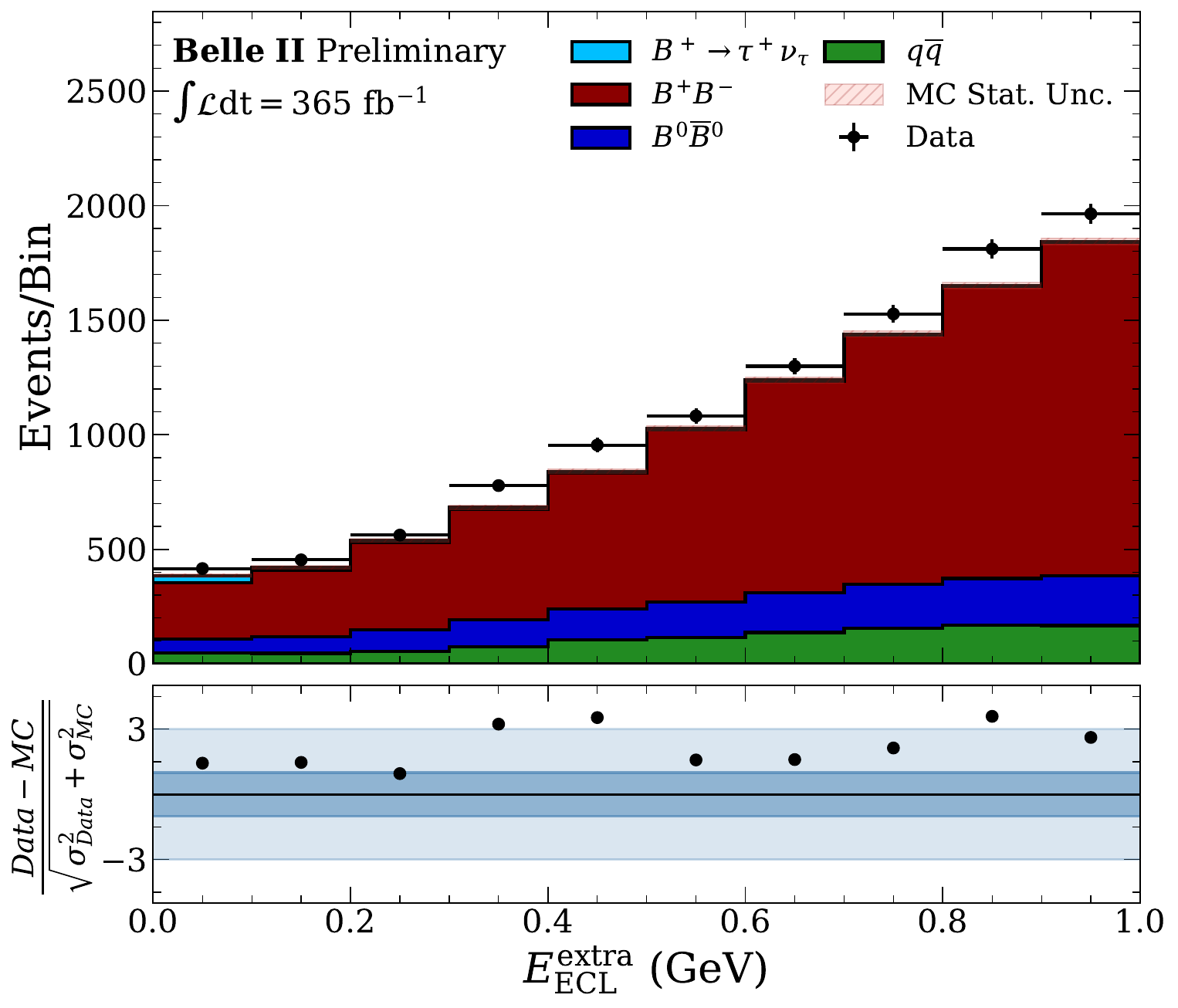}}\\
    \subfloat{(c)\label{fig:extra_mul_c}}{\includegraphics[width = 0.4\textwidth, keepaspectratio]{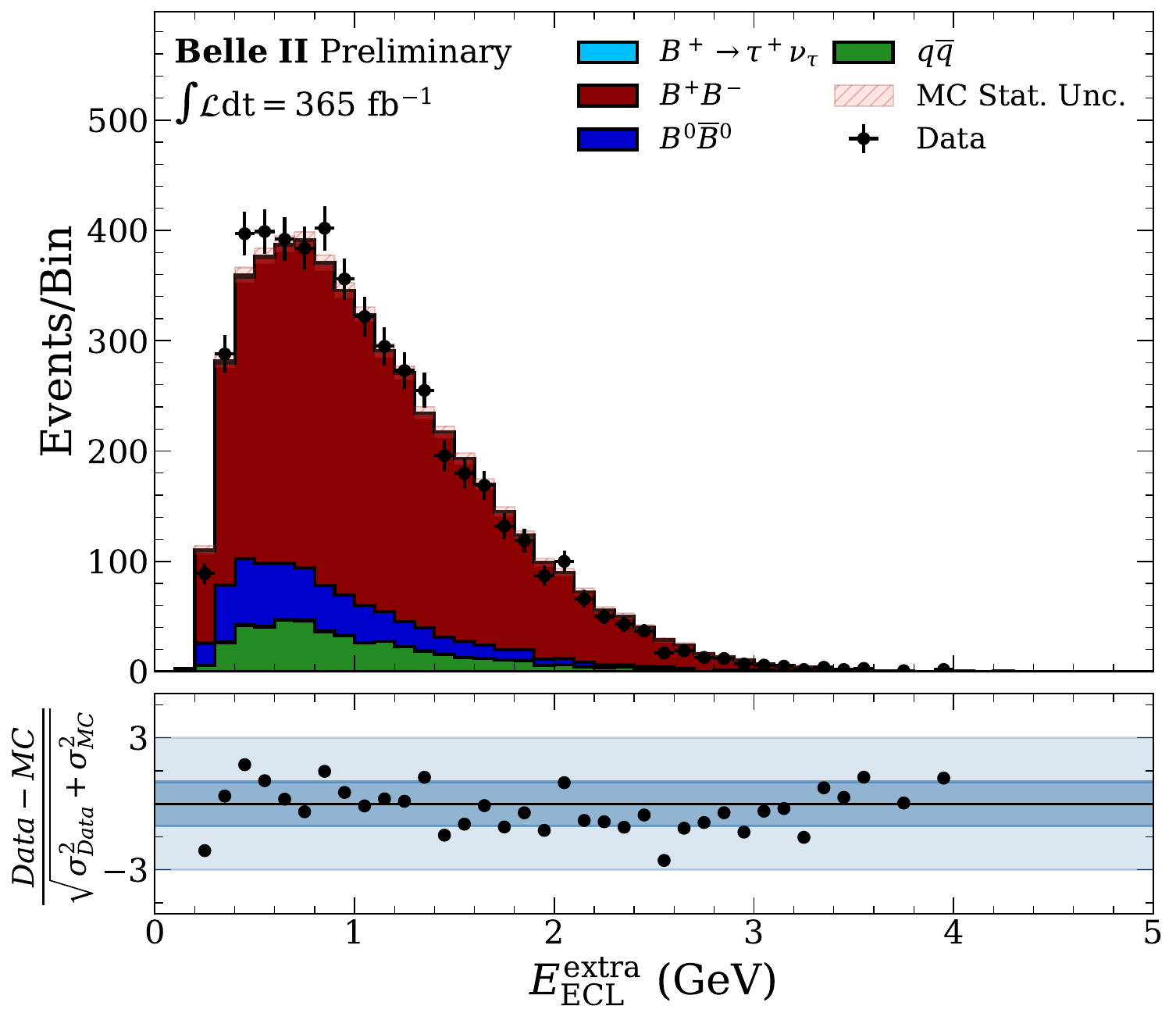}}
    \subfloat{(d)\label{fig:extra_mul_d}}{\includegraphics[width = 0.4\textwidth, keepaspectratio]{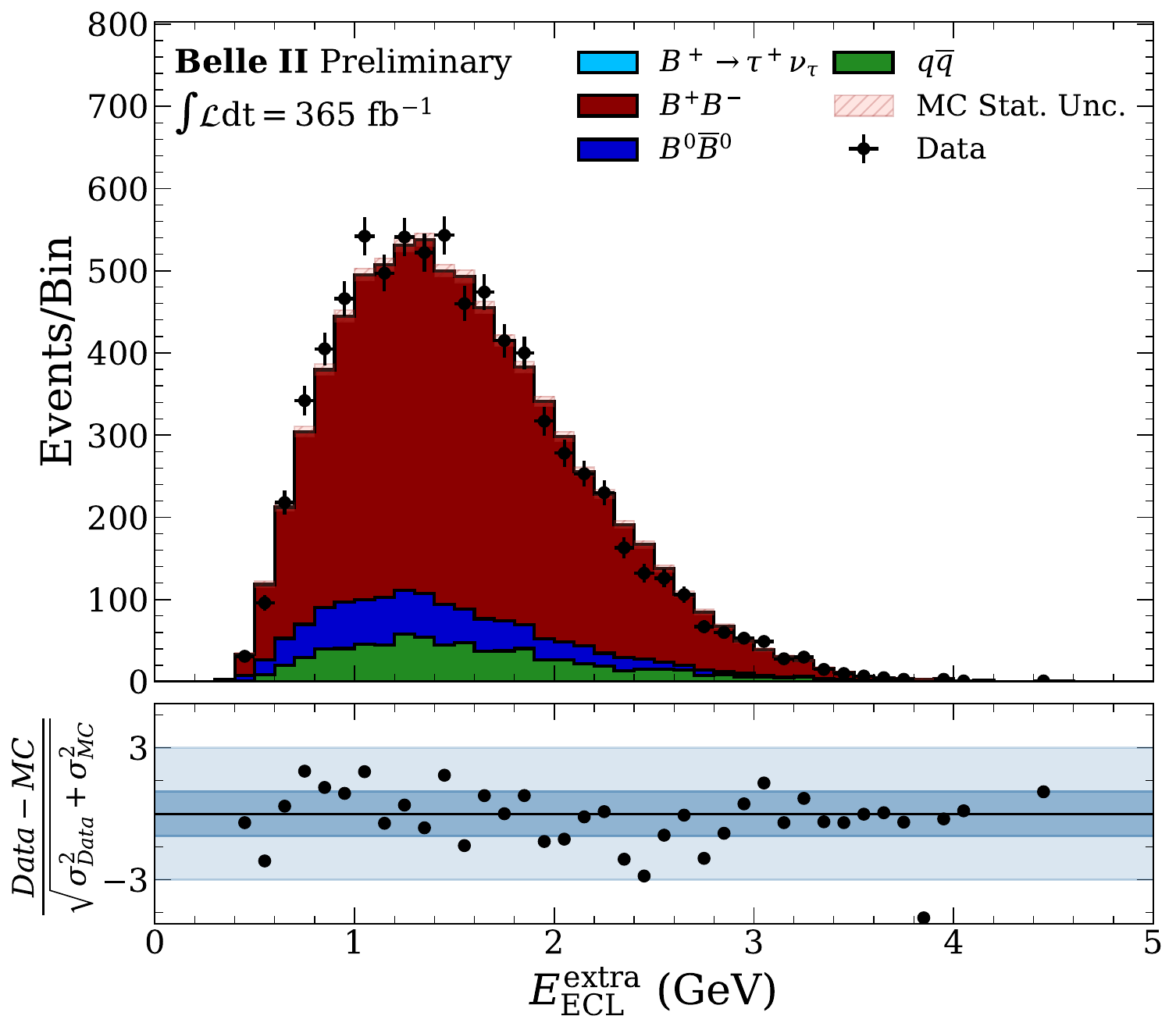}}
    \caption{First row: distributions of \ng (a) and \eextra (b) in data and simulation for \eextra$<\SI{1}{\gev}$. Second row: distributions of \eextra with $\ng = 3$~(c) and $\ng = 5$~(d). The number of events in simulation is scaled to the data for (c) and (d) to compare the shapes. The \Btaunu signal events are a small component of the full sample.}
    \label{fig:extra_mult}
\end{figure*}

\begin{figure*}[htbp]
    \centering
    \begin{minipage}{0.4\textwidth}
        \centering
        \includegraphics[width = \textwidth, keepaspectratio]{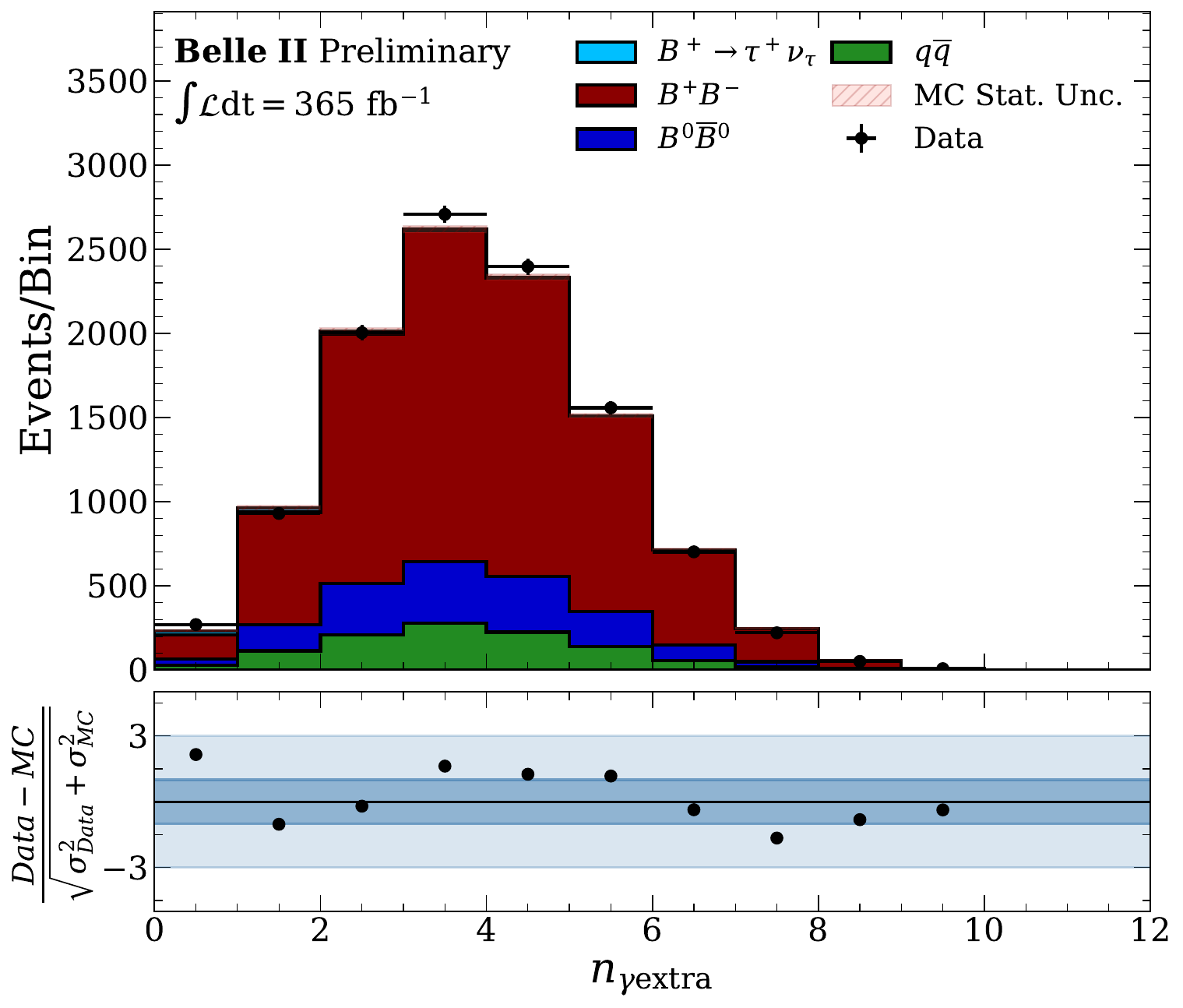}
    \end{minipage}
    \begin{minipage}{0.4\textwidth}
        \centering
        \includegraphics[width = \textwidth, keepaspectratio]{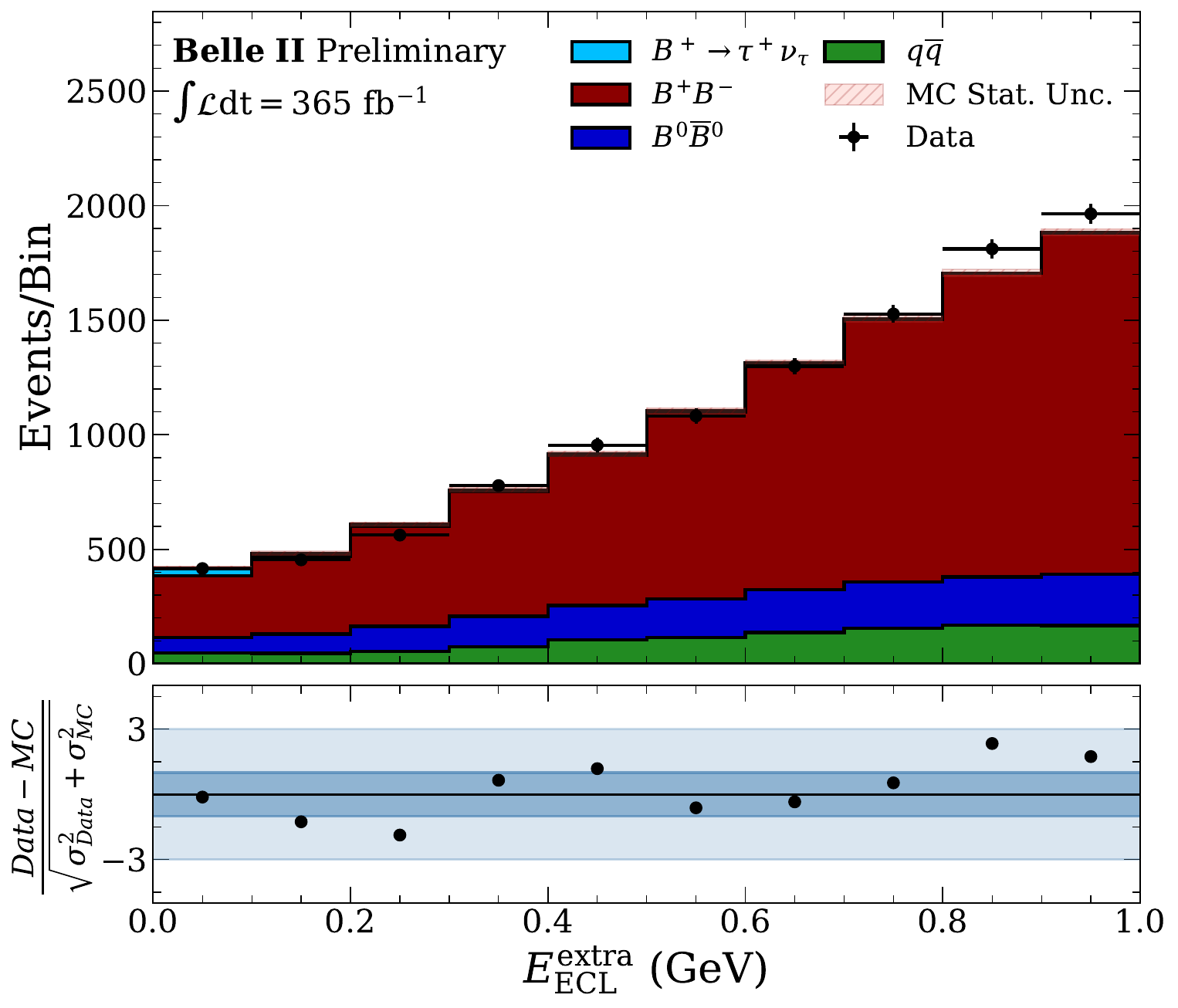}
    \end{minipage}
    \caption{Distributions of \ng (left) and \eextra (right) in data and simulation for \eextra$<\SI{1}{\gev}$ after applying the \ng calibration.}
    \label{fig:extra_mult_corr}
\end{figure*}

\subsection{Clusters multiplicity calibration}
After applying all the efficiency corrections, the shapes of the \eextra distribution in data and in simulation are slightly different. The discrepancy is found to be related to an incorrect modeling of the multiplicity of extra neutral clusters (\ng). Figures \ref{fig:extra_mul_a} and \ref{fig:extra_mul_b} show \ng and \eextra distributions.
Figures \ref{fig:extra_mul_c} and \ref{fig:extra_mul_d} show, as an example, \eextra distributions for \ng = 3 and \ng = 5, with the number of events in simulation scaled to data, in order to test that \eextra shapes agree at fixed multiplicity \ng. The agreement is good at any fixed \ng (see Appendix \ref{appx:A} for the complete set of plots). 
Therefore, we conclude that simulation approximately models the \eextra distribution in each multiplicity bin, while the expected normalization differs with respect to data by a few \% up to 20\%.
In order to correct the simulation expectation to match the data, we determine a bin-by-bin correction of \ng from control samples: the first is the Extra-Tracks sample, obtained by requiring two or more tracks in the ROE coming from the IP and with momenta less than 0.5 GeV. The corrections to \ng from the Extra-Tracks sample are used to reweight the \BBbar background simulation.
For the second sample, we reconstruct a \Bp\to\Dstarz\ellp\neul decay recoiling against the hadronic \btag, and the resulting corrections are used to reweight the signal simulation for leptonic modes.
The third sample is obtained by reconstructing two non-overlapping hadronic tag $B$ mesons with opposite charges in the event (Double Tag); this sample is used to reweight the signal simulation for hadronic modes. Typical correction factors vary between 0.8 and 1.2. Each control sample is discussed in detail in Appendix~\ref{appx:A}. Figure \ref{fig:extra_mult_corr} shows \ng and \eextra distributions after applying the corrections.

We validated the procedure by comparing the reweighted MC distributions with data in the following sidebands:  \eextra sideband, requiring \eextra$> \SI{500}{\mev}$; $M_\text{bc}$ sideband, requiring  $M_\text{bc} < \SI{5.26}{GeV}/c^2$; \missM sideband (leptons only), requiring \missM$< \SI{4}{GeV^2}/c^4$;  $p_\text{cand}$ sideband (hadrons only), requiring $p_\text{cand} < \SI{1.2}{GeV}/c$. In all cases, we find good agreement between MC simulation and data.

\section{Signal extraction}
\label{sec:sig_sel_opt}
We use the \eextra and \missM variables to discriminate between signal and background: signal events are characterized by low \eextra and large \missM. In contrast, for backgrounds \eextra has a smoother increasing distribution, and \missM tends to have smaller values. We exploit this behavior and the correlations by combining the \eextra and \missM distributions in a single two-dimensional binned probability density function (PDF). 
 
\begin{figure*}[htbp]
    \centering
    \begin{minipage}{0.38\textwidth}
        \centering
        \includegraphics[width = \textwidth, keepaspectratio]{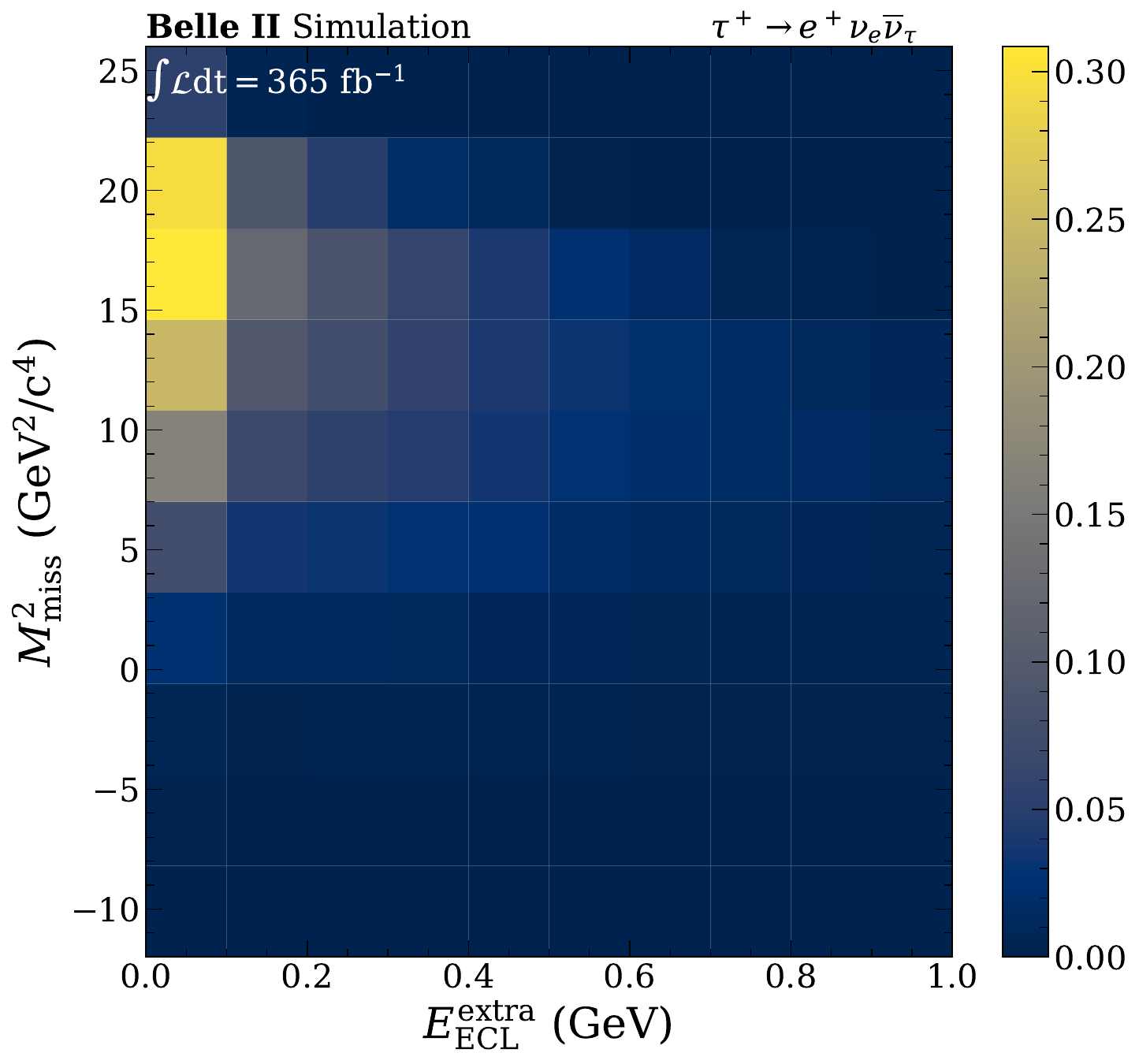}
    \end{minipage}
    \begin{minipage}{0.38\textwidth}
        \centering
        \includegraphics[width = \textwidth, keepaspectratio]{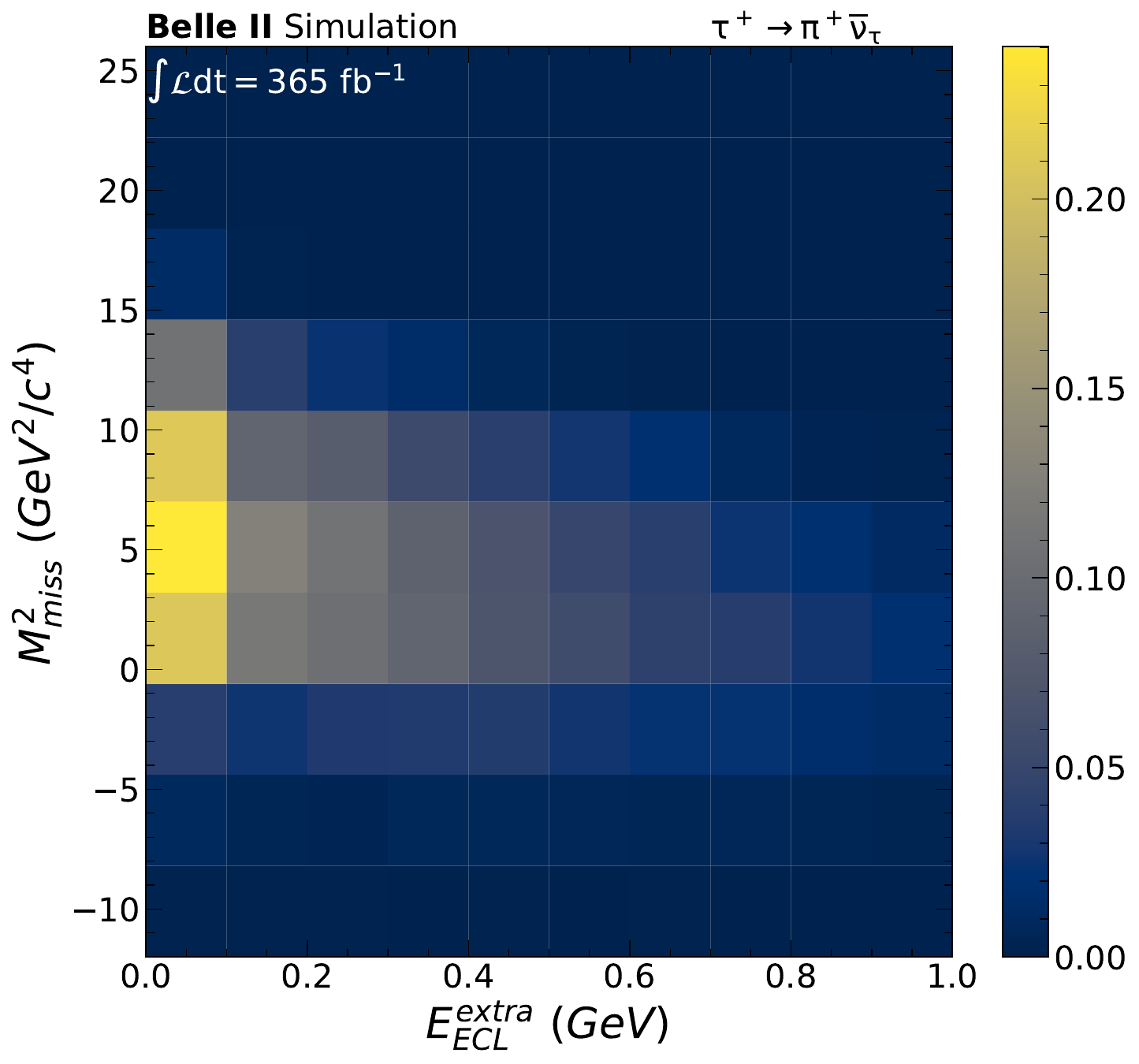}
    \end{minipage}\\
    \begin{minipage}{0.38\textwidth}
        \centering
        \includegraphics[width = \textwidth, keepaspectratio]{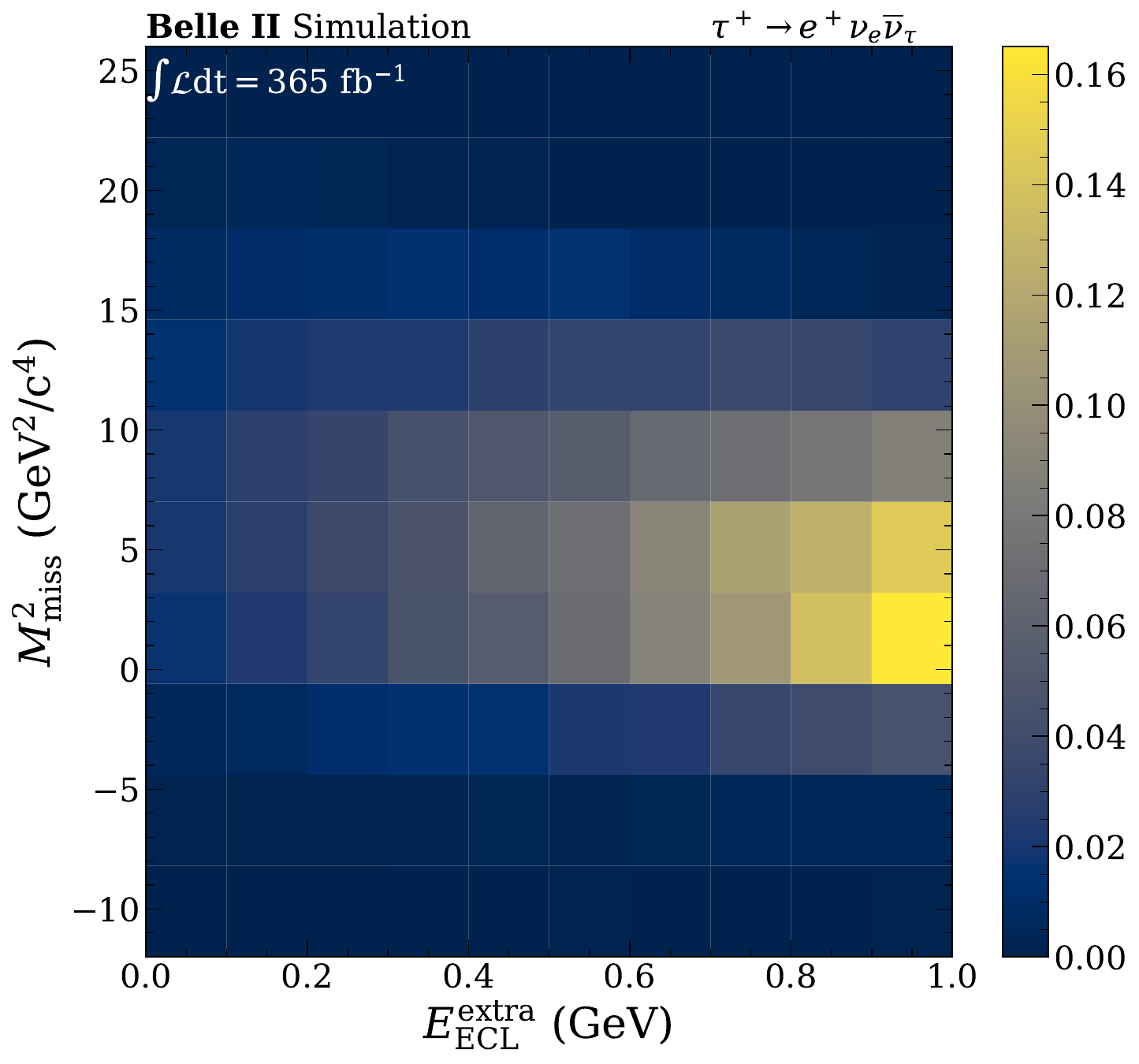}
    \end{minipage}
    \begin{minipage}{0.38\textwidth}
        \centering
        \includegraphics[width = \textwidth, keepaspectratio]{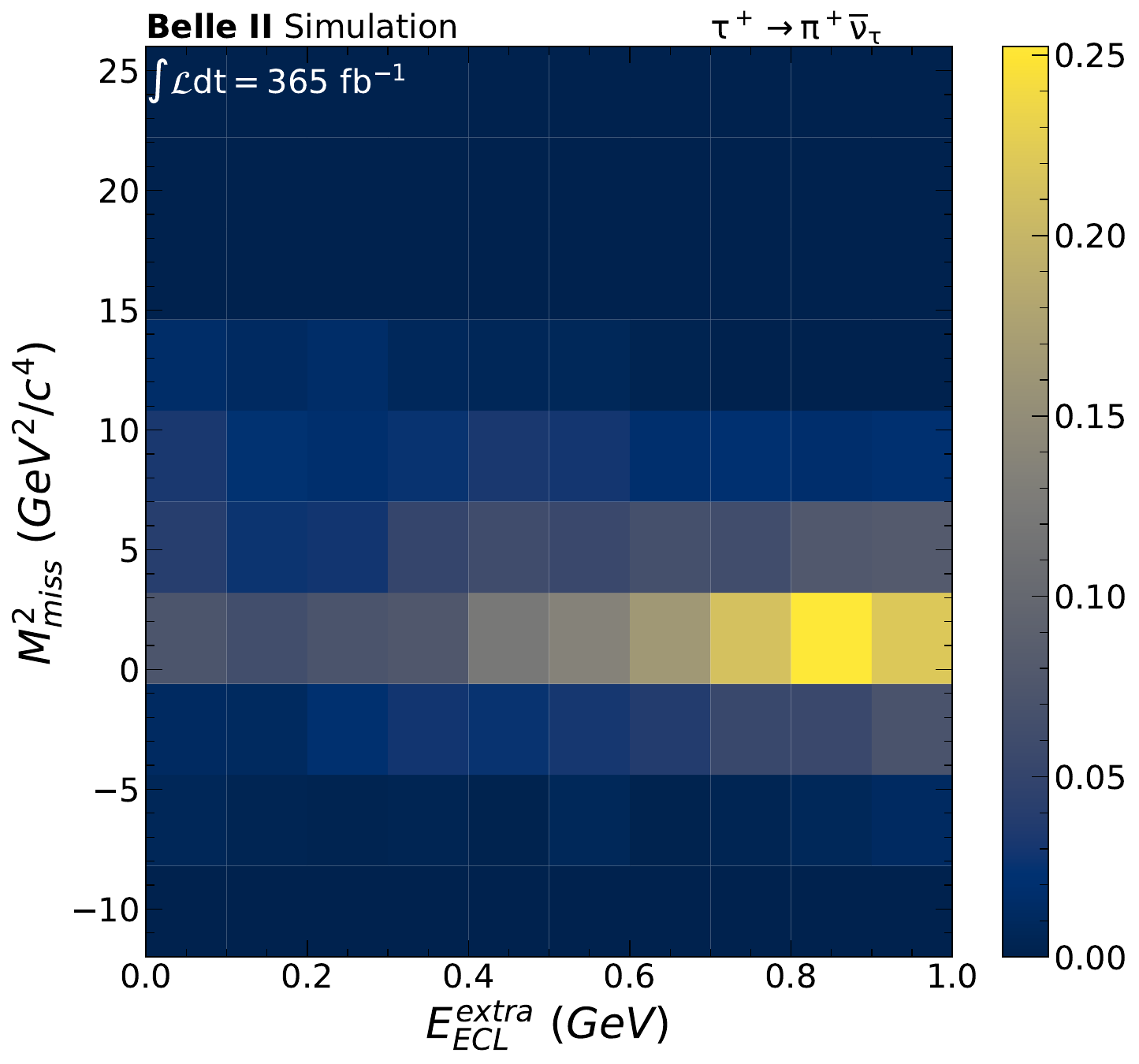}
    \end{minipage}
    \caption{Two-dimensional PDFs of \eextra and \missM from simulation for signal (top) and background (bottom) in the \taup\to\ep\neue\neutb channel (left) (similar for the \taup\to\mup\neum\neutb channel) and in the \taup\to\pip\neutb (right) (similar for the \taup\to\rhop\neutb channel). The color represents the PDF probability in each bin.}
    \label{fig:2d_fit_pdf}
\end{figure*}

We extract the branching fraction \brbtaunu from a simultaneous binned maximum likelihood fit to all the four $\tau^+$ categories. The PDFs are 2D histograms of \missM and \eextra with 10$\times$10 uniform binning, with $-10 < \missM < \SI{26}{GeV^2}/c^4$ and $0 \leqslant \eextra < \SI{1}{GeV}$. Figure \ref{fig:2d_fit_pdf} shows the 2D histogram PDFs of \eextra and \missM for signal and background in the \taup\to\ep\neue\neutb channel (left plots) (similar for the \taup\to\mup\neum\neutb channel) and in the \taup\to\pip\neutb (right plots) (similar for the \taup\to\rhop\neutb channel).

We float five parameters in the fit: the common branching fraction \brbtaunu and the total background yield for each of the four decay modes $n_{b,k}$, with $k=\ep, \mup, \pip\ \rm{or} \ \rhop$. 
The signal yields $n_{s,k}$ are not free parameters but depend on the common floating fit parameter \brbtaunu and fixed quantities as follows:
\begin{equation}
n_{s,k} = 2n_{\BpBm}\times \epsilon_k \times \brbtaunu
\end{equation}
with $n_{\BpBm} = \nBB f^{+-}$, where $f^{+-} = 0.5113 ^{+0.0073}_{-0.0108}$ is the branching fraction $\mathcal{B}(\FourS\to\BpBm)$ estimated in~\cite{banerjee2024averagesbhadronchadrontaulepton}; $\epsilon_k$ is the efficiency to reconstruct in the category $k$ a \Btaunu decay (for any kind of real $\tau$ decay). The efficiencies $\epsilon_k$, estimated in simulation and corrected for MC mis-modeling, are shown in Tab.~\ref{tab:best_cuts}. They include by construction the \taup branching fractions and cross-feed as predicted by MC. Table~\ref{tab:cross_feed} shows the composition of each reconstructed $\tau^+$ decay in terms of decay mode. The table shows the relevant sizes of the cross-feed contributions.

\begin{table}[ht]
    \renewcommand{\arraystretch}{1.2}
    \centering
    \caption{Composition of each reconstructed $\tau^+$ decay from \Btaunu in terms of decay mode. The row denotes the reconstructed final state, and the columns represent the generated decay mode. The off-diagonal entries reflect the amount of cross-feed between channels.}
    \label{tab:cross_feed}
    \begin{tabular*}{\linewidth}{@{\extracolsep{\fill}}cccccc}
        \toprule
        \toprule
        \diagbox[]{Reco}{True} & \ep (\%) & \mup(\%) & \pip(\%) & \rhop(\%) & other(\%) \\
        \midrule
        \ep &  97 & 0.1 & 0.1 & 0 & 2.8 \\
        \mup  &  0 & 87 & 0.9 & 0.1 & 12 \\
        \pip  &  0.1 & 3.3 & 55.7 & 16 & 24.9 \\
        \rhop &  0.4 & 4.5 & 27.8 & 61.2 & 6.1 \\
        \bottomrule
        \bottomrule
    \end{tabular*}
\end{table}

\begin{figure*}[htbp]
    \centering
    \begin{minipage}{0.4\textwidth}
        \centering
        \includegraphics[width = \textwidth, keepaspectratio]{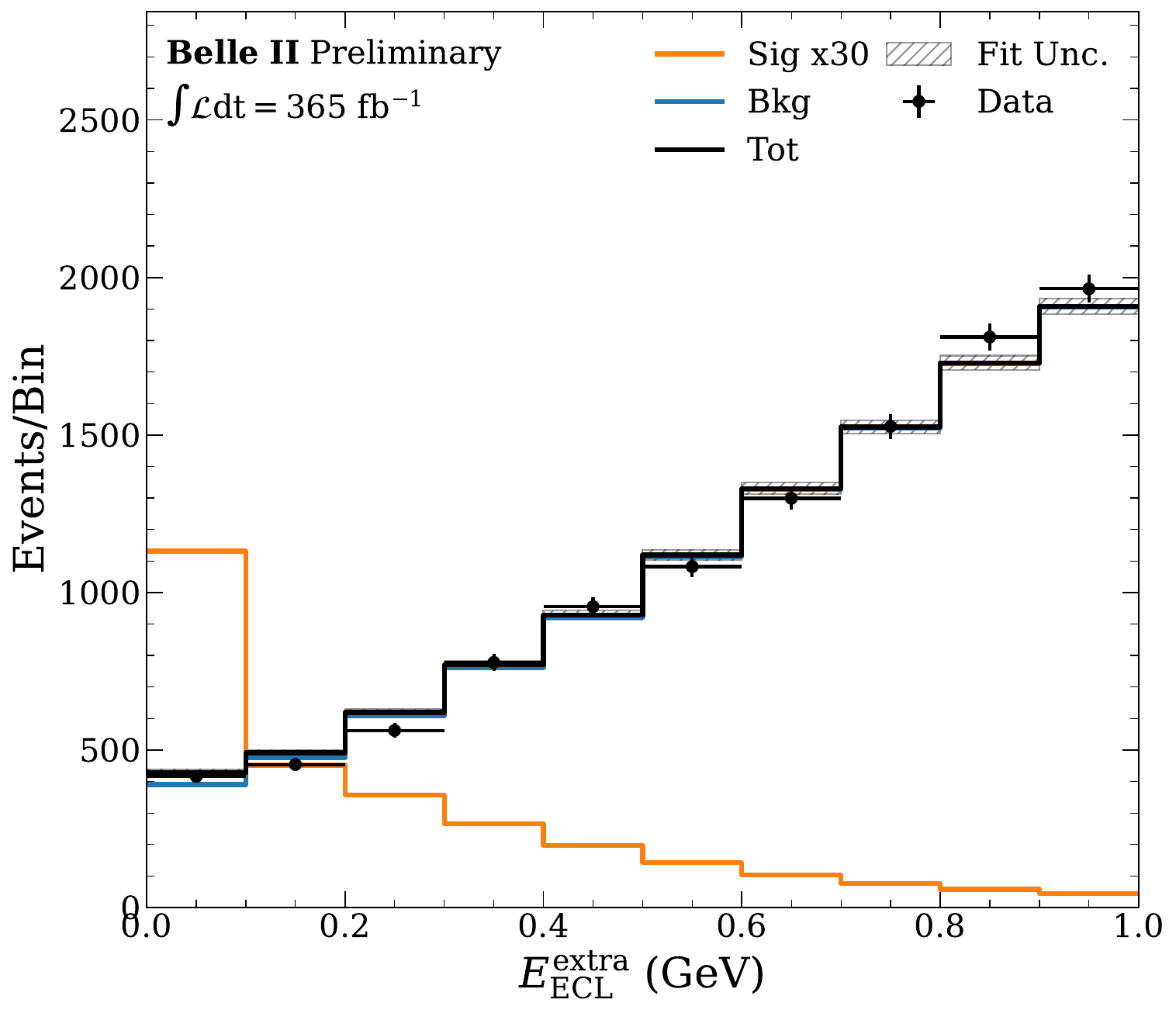}
    \end{minipage}
    \begin{minipage}{0.4\textwidth}
        \centering
        \includegraphics[width = \textwidth, keepaspectratio]{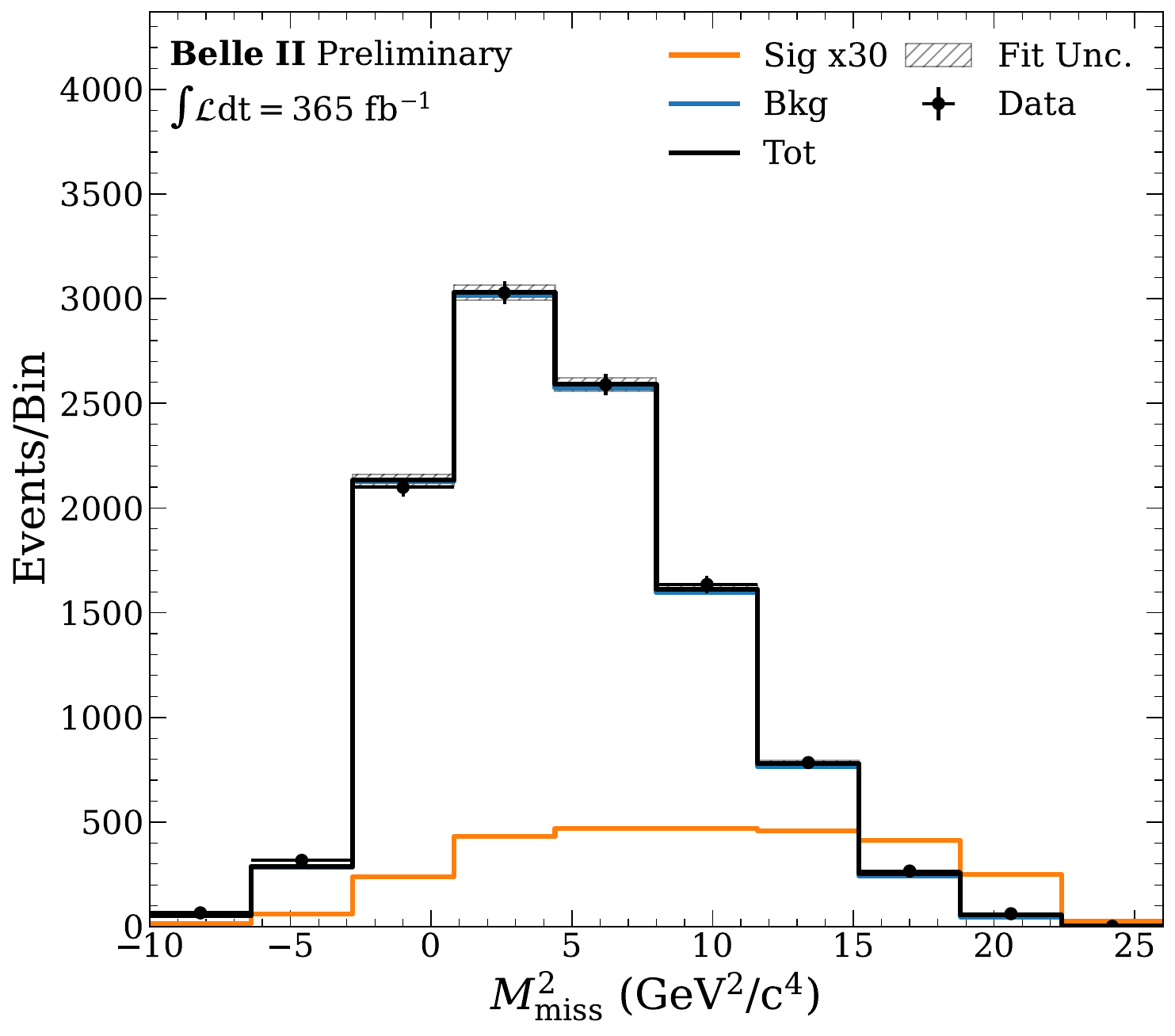}
    \end{minipage}\\
    \begin{minipage}{0.4\textwidth}
        \centering
        \includegraphics[width = \textwidth, keepaspectratio]{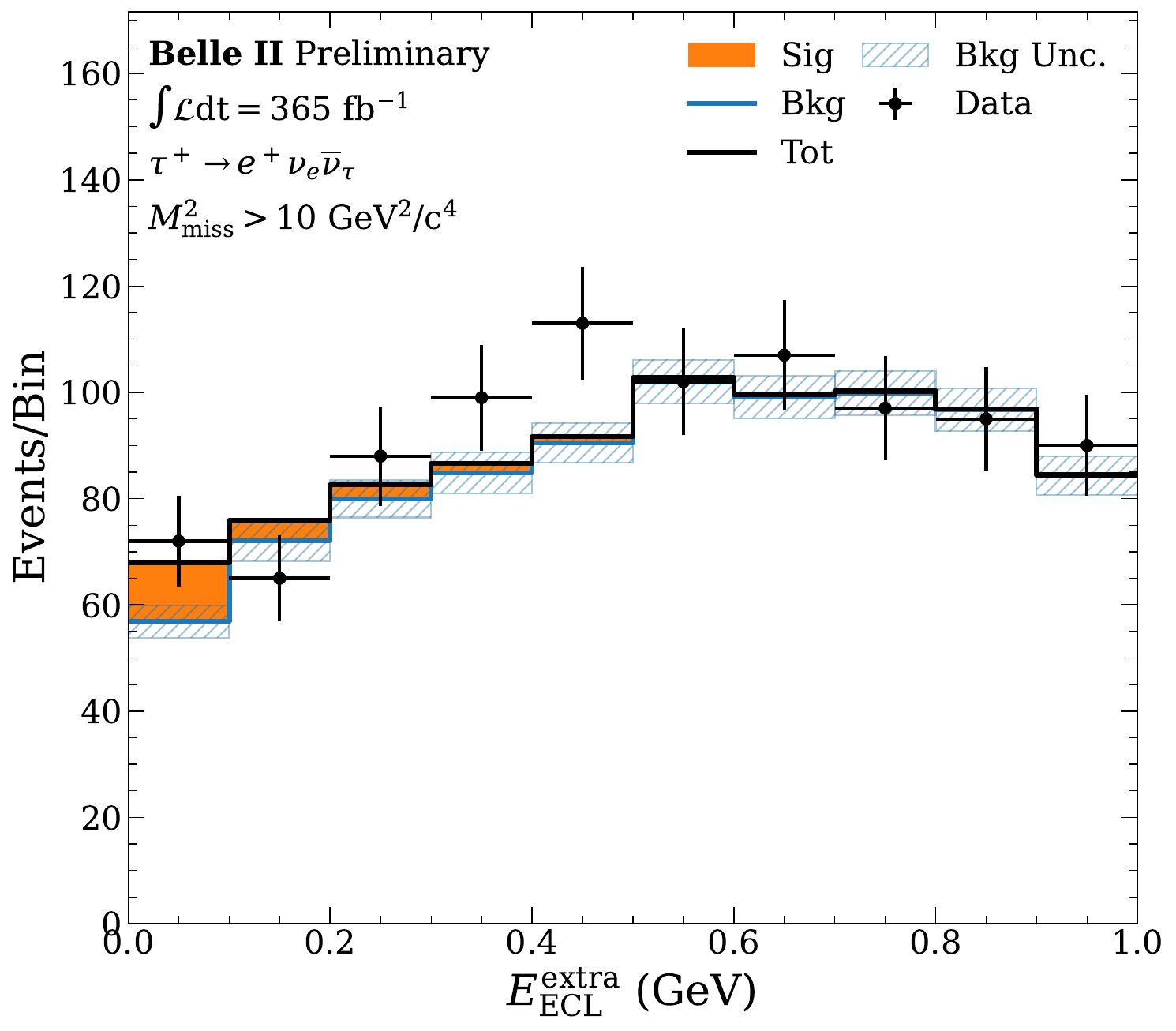}
    \end{minipage}
    \begin{minipage}{0.4\textwidth}
        \centering
        \includegraphics[width = \textwidth, keepaspectratio]{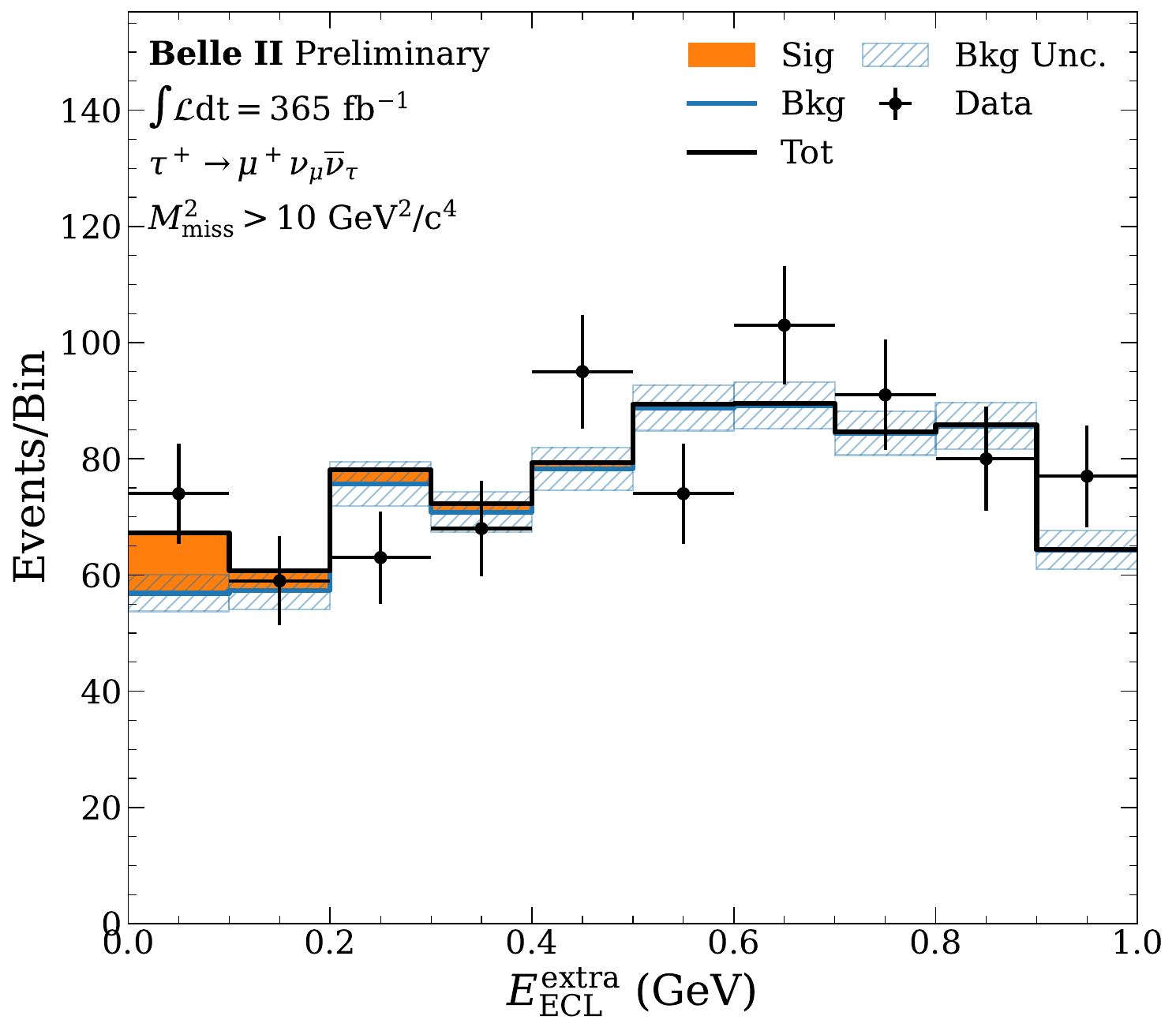}
    \end{minipage}\\
    \begin{minipage}{0.4\textwidth}
        \centering
        \includegraphics[width = \textwidth, keepaspectratio]{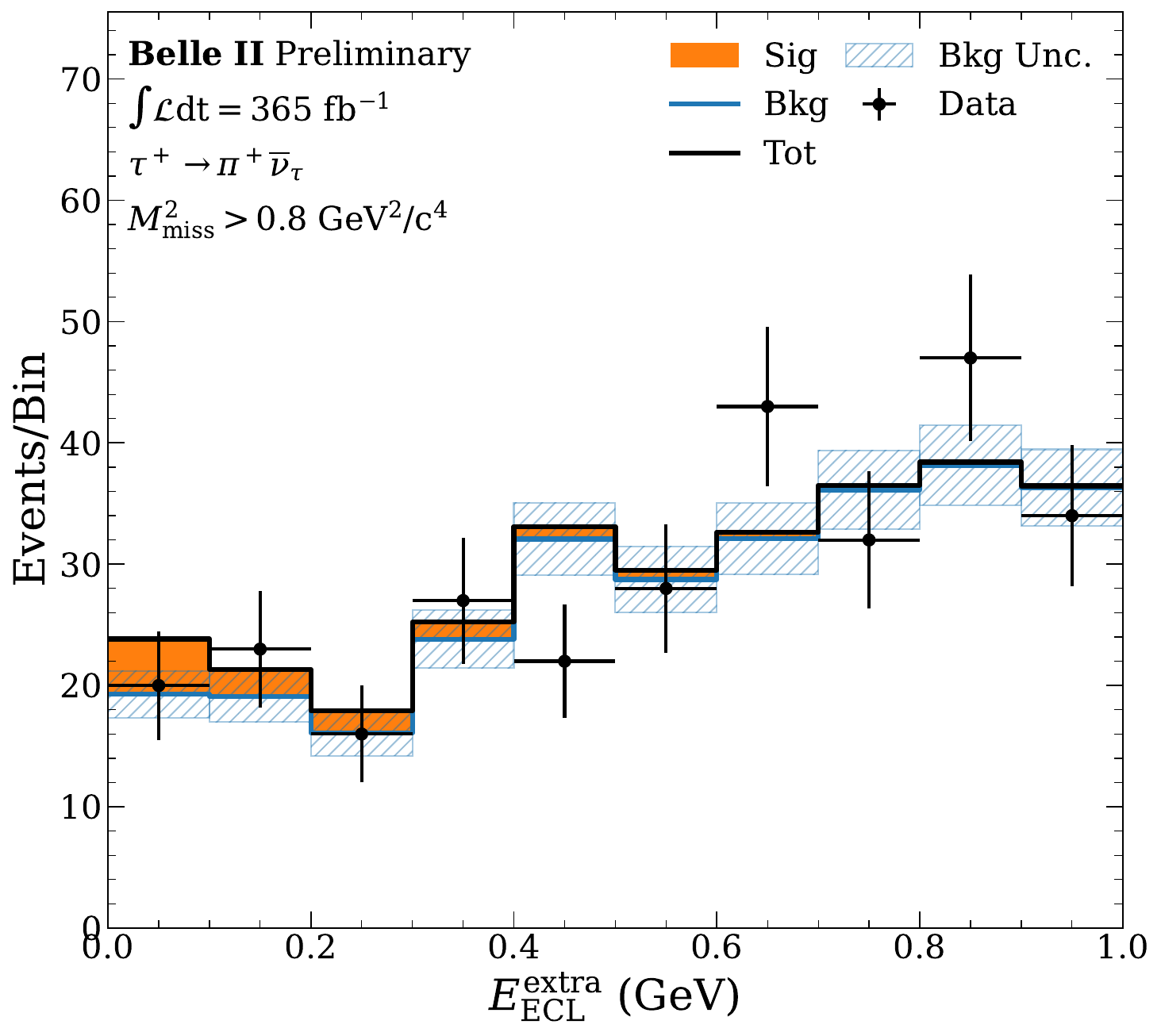}
    \end{minipage}
        \begin{minipage}{0.4\textwidth}
        \centering
        \includegraphics[width = \textwidth, keepaspectratio]{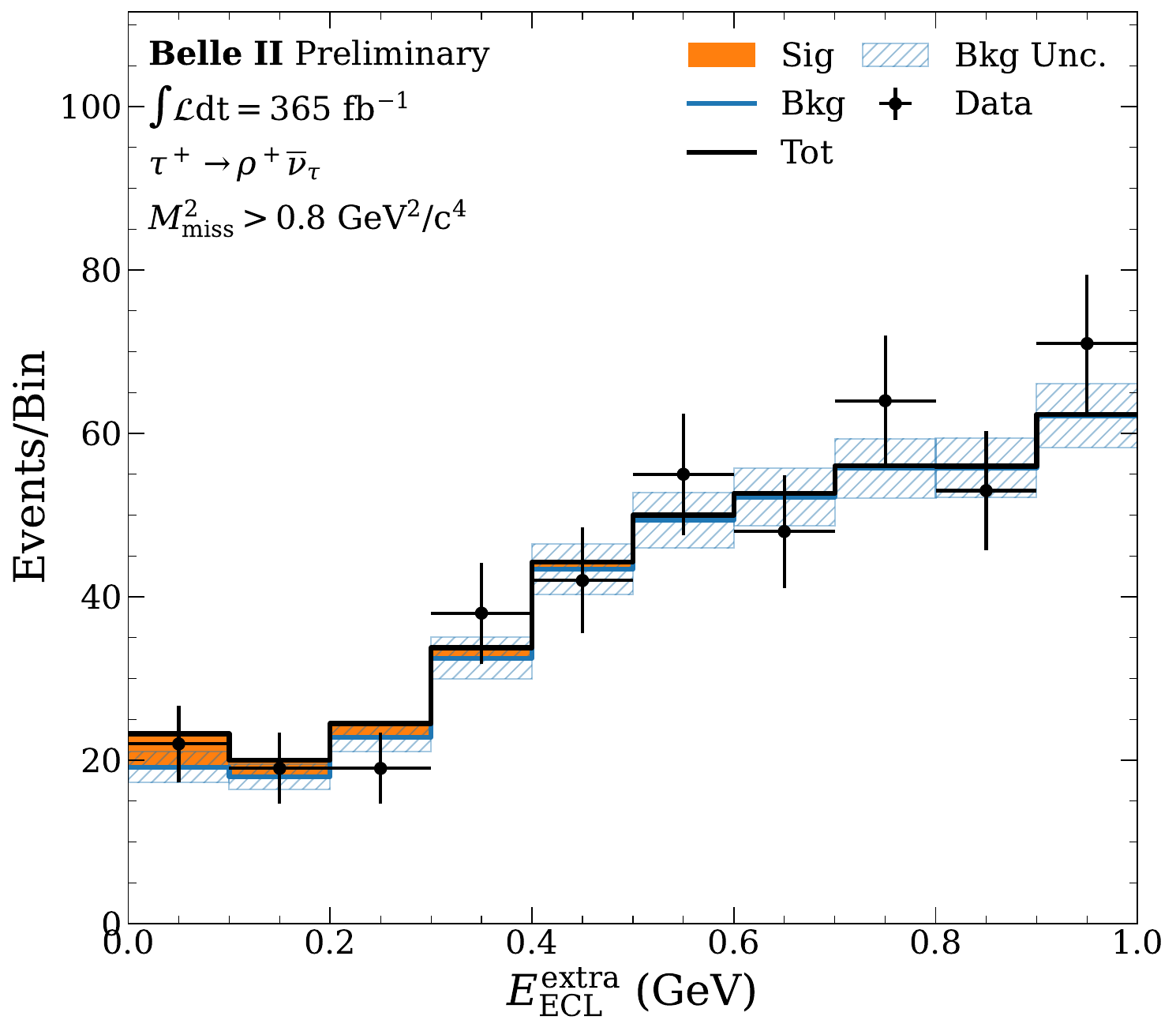}
    \end{minipage}\\
    \caption{First row: distributions of \eextra (left) and \missM (right) with the fit results superimposed. The signal MC is scaled by a factor of 30 to make it visible. Second row: distributions of \eextra with the fit results superimposed for the leptonic channels in the signal enriched region $\missM>10\ \mathrm{GeV^2/c^4}$. Third row: distributions of \eextra with the fit results superimposed for the hadronic channels in the signal enriched region $\missM>0.8\ \mathrm{GeV^2/c^4}$.}
    \label{fig:eextra_missM2_post_fit}
\end{figure*}

\section{Fit Result}
Performing the fit to the data we obtain
\begin{equation}
    \brbtaunu = (1.24\pm0.41)\times10^{-4},
\end{equation}
where the uncertainty is statistical only \stat. 

In order to check the goodness of the fit, we generate pseudo-datasets from the simulated distributions and repeat the fit on the obtained pseudo-data. We observe that the $\chi^2$ values obtained in pseudo-data are worse than the $\chi^2$ obtained in data for 11\% of the cases.
In Fig. \ref{fig:eextra_missM2_post_fit} we show the projections of the fit for \eextra and \missM distributions (in Appendix \ref{appx:B} we show the same projections for each $\tau^+$ category). The comparison of fitted background yields with respect to MC expectation is shown in  Table~\ref{tab:fit_parameters}.
Table~\ref{tab:fit_parameters_sig} shows \brbtaunu obtained fitting simultaneously the four $\tau^+$ categories and fitting each category independently from each other.   

\begin{table}[ht]
    \renewcommand{\arraystretch}{1.2}
    \centering
    \caption{Observed and expected values of the background yields in the fit. The expected values are estimated from a simulation corresponding to an integrated luminosity of $\SI{365}{\invfb}$.}
    \label{tab:fit_parameters}
    \begin{tabular*}{\linewidth}{@{\extracolsep{\fill}}ccc}
        \toprule
        \toprule
        Parameter & Observed value & Expected value \\
        \midrule
        $n_{b,e^+}$ & $4907\pm71$ & $4846\pm24$ \\
        $n_{b,\mu^+}$ & $4620\pm69$ & $4493\pm24$ \\
        $n_{b,\pi^+}$ & $454\pm22$ & $461\pm9$ \\
        $n_{b,\rho^+}$ & $772\pm28$ & $811\pm11$ \\
        
        \bottomrule
        \bottomrule
    \end{tabular*}
\end{table}  

\addtolength{\tabcolsep}{10pt}    
\begin{table}[ht]
    \renewcommand{\arraystretch}{1.2}
    \centering
    \caption{Observed values of the signal yields and branching fractions, obtained from single fits for each $\tau^+$ decay mode and the simultaneous fit.}
    \label{tab:fit_parameters_sig}
    \begin{tabular*}{\linewidth}{@{\extracolsep{\fill}}lcc}
        \toprule
        \toprule
        Decay mode & $n_s$ & $\mathcal{B}(10^{-4})$ \\
        \midrule
        Simultaneous & $94\pm31$ & $1.24\pm0.41$ \\
        \midrule
        \ep\neue\neutb & $13\pm16$ & $0.51\pm0.63$ \\
        \mup\neum\neutb & $40\pm20$ & $1.67\pm0.83$ \\
        \pip\neutb & $31\pm13$ & $2.28\pm0.93$ \\
        \rhop\neutb & $6\pm25$ & $0.42\pm1.82$ \\
        \bottomrule
        \bottomrule
    \end{tabular*}
\end{table}
\addtolength{\tabcolsep}{-10pt}

\section{Systematic Uncertainties}
\label{sec:sys}

The main systematic uncertainties affecting the measurement are listed in Table~\ref{tab:systematic_summary}.
When uncertainties do not affect the signal yields, they are propagated directly to the branching fraction, as in the case of the number of
\FourS, the fraction of \BpBm pairs (symmetrizing the uncertainty to be $f^{+-} = 0.5113\pm0.0108$ since it is not a dominant uncertainty), and the uncertainty on the tracking efficiency of the signal charged particle. Otherwise, the effect on the final result is estimated by fluctuating the assumptions and propagating the effect on the PDF shapes, generating in this way a set of alternative PDFs. The fit is repeated with all the alternative templates, and the standard deviation of the fitted \brbtaunu values is taken as the corresponding systematic uncertainty.

\begin{table}[htbp]
    \renewcommand{\arraystretch}{1.2}
    \caption{Summary of systematic uncertainties \syst on the fitted branching fraction presented as relative uncertainties. The effect of each source is evaluated in the simultaneous fit of the four signal modes. The last three sources do not affect the signal yields.}
    \label{tab:systematic_summary}
    \begin{tabular*}{\linewidth}{@{\extracolsep{\fill}}lr}
        \toprule
        \toprule
        Source                                     &  Syst.                \\
        \midrule
        Simulation statistics                      &  13.3\%               \\
        Fit variables PDF corrections              &  5.5\%                \\
        Decays branching fractions in MC           &  4.1\%                \\
        Tag \Bub reconstruction efficiency         &  2.2\%                \\
        Continuum reweighting                      &  1.9\%                \\
        \piz reconstruction efficiency             &  0.9\%                \\
        Continuum normalization                    &  0.7\%                \\
        Particle identification                    &  0.6\%                \\
        \midrule
        Number of produced \FourS                  &  1.5\%                \\
        Fraction of \BpBm pairs                    &  2.1\%                \\
        Tracking efficiency                        &  0.2\%                \\     
        \midrule
        Total                                      &  15.5\%               \\
        \bottomrule
        \bottomrule
    \end{tabular*}
\end{table}

We evaluate the systematic uncertainty related to simulation statistics by fluctuating the bin contents of the 2D histogram PDFs 200 times, varying the bin content according to MC statistical uncertainties, and assuming a Poisson distribution. We obtain an uncertainty of 13.3\%.

To evaluate the systematic corrections to the \ng multiplicity we vary the bin-by-bin correction by applying 100 Gaussian variations, taking the variance from the corrections obtained from control studies. 
The resulting PDFs are used to repeat the fit. The standard deviation of the fit results is 5.5\%, which is taken as a systematic uncertainty.

To account for possible discrepancies between data and simulation due to the branching fractions of the $B$ and $D$ decays used in the MC simulation, we apply 50 Gaussian variations to those branching fractions, with the variance set to the uncertainty of the latest PDG world average~\cite{2024pdg}. Repeating the fit with the modified MC samples, we obtain a 4.1\% systematic uncertainty.

The \btag reconstruction efficiency is calibrated with the $\Bp\to X\ellp \neul$ and $\Bp\to D^{(*)}\pip$ control samples. We generate 20 alternative sets of calibration factors from the covariance matrix of the nominal ones. Repeating the fit with the alternative corrections, we observe a 2.2\% standard deviation in the fit results, which is taken as a systematic uncertainty.

The limited size of the off-resonance sample affects the reweighting of the continuum MC. Applying a bootstrapping procedure, and resampling the training and test samples of the FBDT, we obtain 50 different sets of reweighting factors.
Repeating the fit with this change we observe a standard deviation of 1.9\% in the fit results, which is taken as a systematic uncertainty.

Events with a pion in the final state are assigned to the $\rho$ ($\pi$) category if a $\pi^0$ is (is not) found to come from a $\rhop \to \pi^+ \pi^0$ decay. Therefore, mis-modeling of the \piz reconstruction efficiency would affect only hadronic \taup decays.
We study the data and MC agreement for the \piz efficiency using $\Dstarz(\to \Dz\piz)\pip$ and $\Dz\to \Km\pip(\piz)$ decays for \piz momenta in the range [0.05,0.20] and [0.20,3.0] GeV/c, respectively, determining corrections factors to the MC for the \piz efficiency. To obtain the systematic uncertainties, we follow a $\pi^0$ removal procedure. After generating a repeatable random sequence of values between zero and one, if the value is greater than the efficiency correction, the \piz is removed, and the two $\gamma$'s are reassigned to the ROE; the event migrates from \taup\to\rhop\neutb to \taup\to\pip\neutb category. We evaluate the systematic contribution by fitting the data on 50 different modified PDFs changing the random sequence. The difference between the average of the fitted branching fractions and the nominal fit result is negligible, while the standard deviation of the fitted branching fractions is 0.9\%. Thus, we conclude that there is no bias in the result if the corrections are not applied and we set the systematic uncertainty to 0.9\%.

We change the continuum fraction of the background by the statistical uncertainty of the off-resonance sample, producing 50 alternative background PDFs, obtained assuming a Poisson distribution. Repeating the fit with the different PDFs, we observe a standard deviation of fit results of 0.7\%, which is taken as a systematic uncertainty.

The systematic uncertainty of the lepton and hadron identification efficiency and fake rates are extracted from pure samples of pions and leptons in \Dstarp\to\Dz(\to\Km\pip)\pip, \Lz\to\Pp\pim, \KS \to \pipi, \jpsi\to\ellell data and MC sample. We evaluate the impact on the branching fraction fit by changing the shapes of the PDFs and the values of selection efficiencies according to $1\,\sigma$ variations of systematic uncertainty of lepton identification, $\pi$ identification, and fake rates estimated in the control samples. We observe a standard deviation in the fit results of 0.6\%.

We check the agreement of signal selection efficiency in data and MC with a \Bp\to\Dstarz\ellp\neul control sample. After applying all the selections and calibrations, we find a Data/MC ratio equal to $0.96\pm0.04$, which implies that no further efficiency correction is needed.

\begin{figure}[htbp]
    \centering
    \begin{minipage}{0.49\textwidth}
        \centering
        \includegraphics[width = 0.9\linewidth, keepaspectratio]{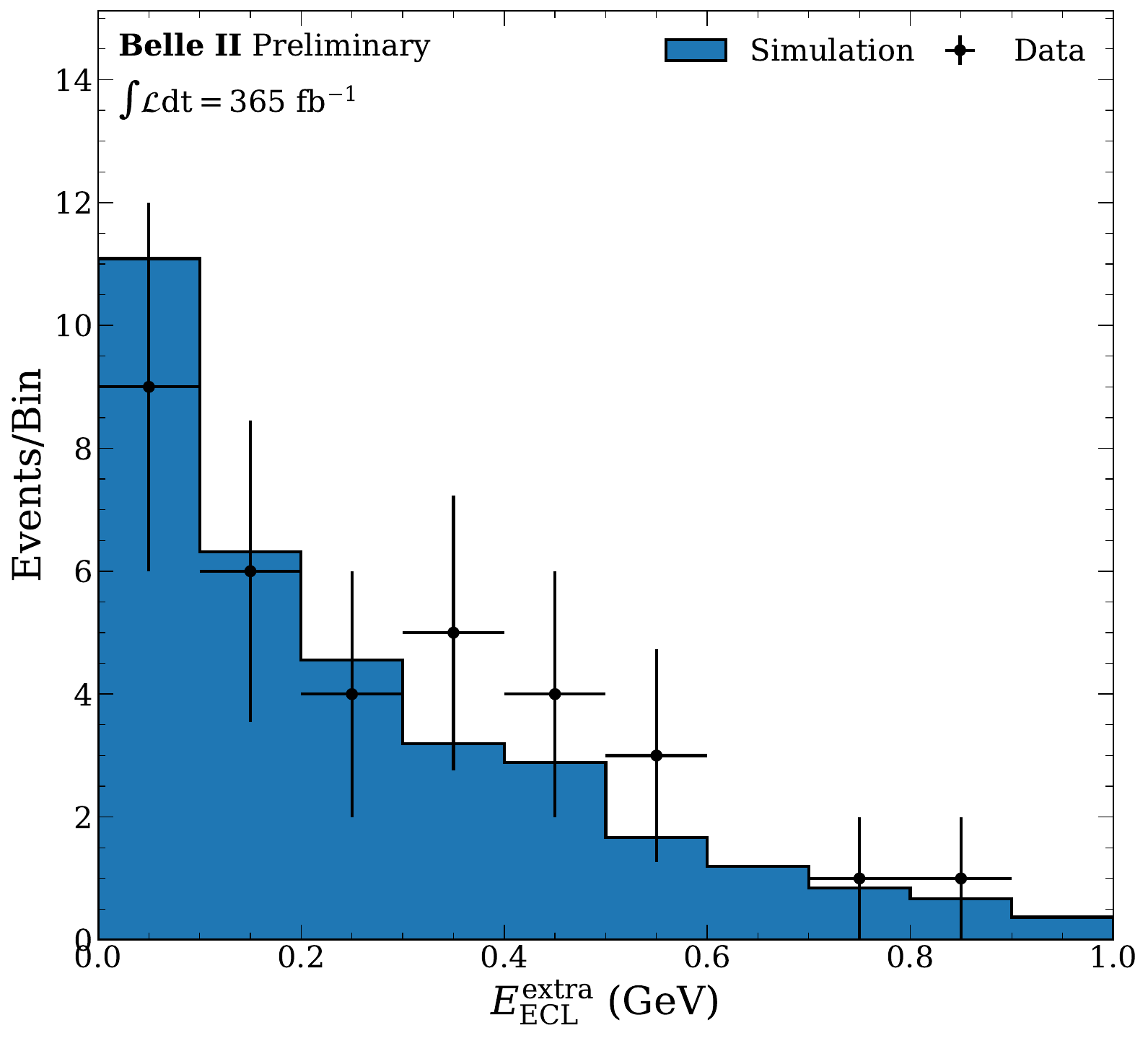}
    \end{minipage}
    \begin{minipage}{0.49\textwidth}
        \centering
        \includegraphics[width = 0.9\linewidth, keepaspectratio]{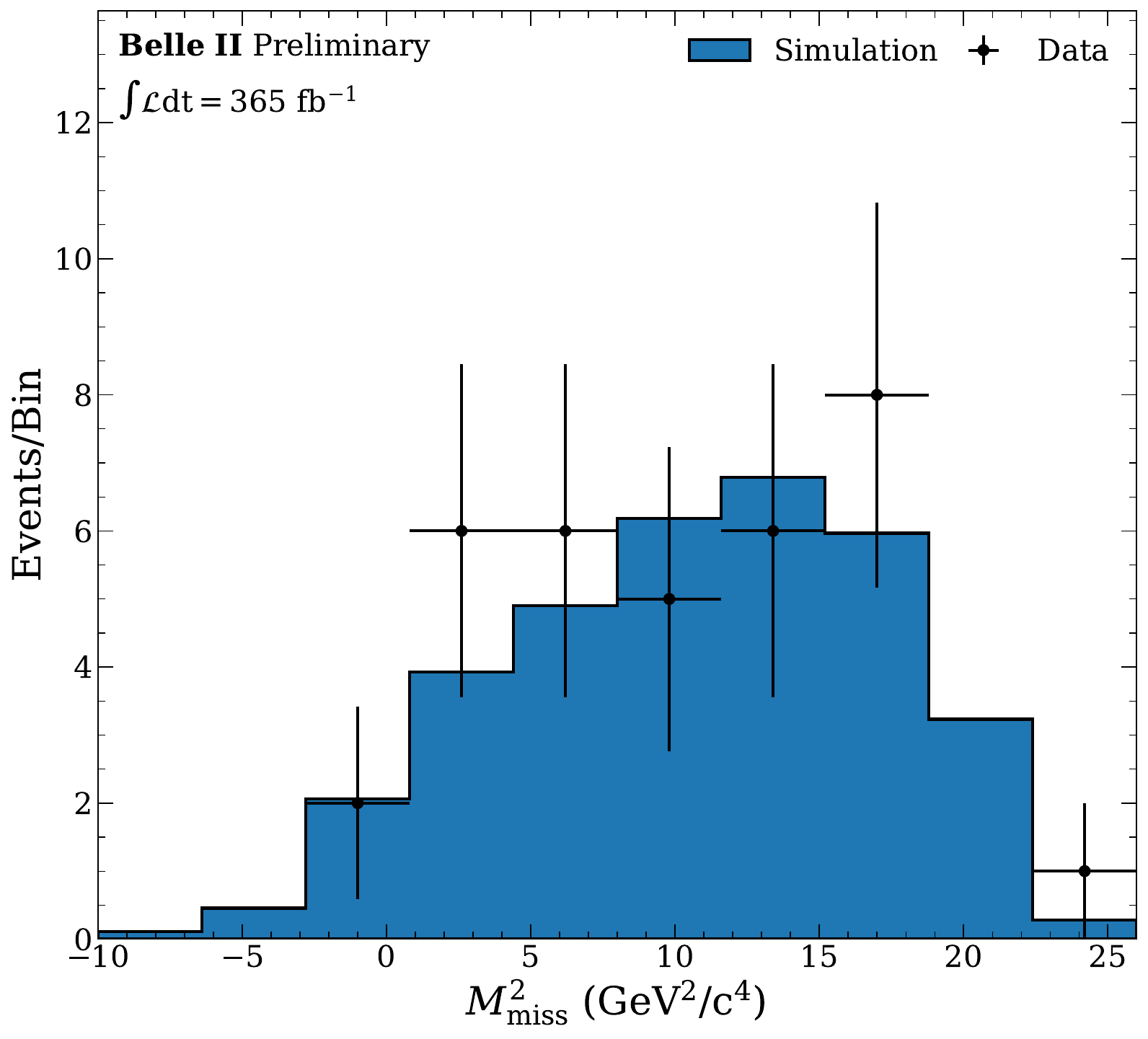}
    \end{minipage}
    \caption{Distributions of \eextra (top) and \missM (bottom) for the signal embedding control sample.}
    \label{fig:var_embed}
\end{figure}

Moreover, we implement a signal embedding procedure on a sample of \Bp\to\Kp \jpsi(\to\ellell) $(\ell=e,\mu)$, exploiting its clean experimental signature. In each event, \Bp\to\Kp\jpsi is removed, and replaced by a simulated \Btaunu. 
This procedure is performed both on data and simulation, applying the standard \Btaunu reconstruction. The ratio of signal selection efficiencies estimated between data and MC is $1.02\pm0.18$, which confirms the agreement obtained from the \Bp\to\Dstar\ellp\neul control sample. The distributions of \eextra and \missM are also in good agreement between data and MC  for this embedding sample, as shown in Fig.~\ref{fig:var_embed}.

We find evidence of signal with a significance of 3.0$\,\sigma$ from a hypothesis test after convolving the likelihood profile with a Gaussian, whose width is set to the total systematic uncertainty. The test statistic is $-2\log(\mathcal{L}/\mathcal{L}_0)$, where $\mathcal{L}$ ($\mathcal{L}_0$) is the value of the likelihood function when the signal yield is allowed to vary (is fixed to 0). We generate $10^6$ pseudo-datasets from the background-only PDF assuming no signal and repeat the fits. We obtain the significance from the p-value calculated as the fraction of fit results having a value of the test statistic smaller than the one observed in data.

\section{Conclusions}
\begin{figure}[htbp]
    \centering
    \includegraphics[width=0.9\linewidth]{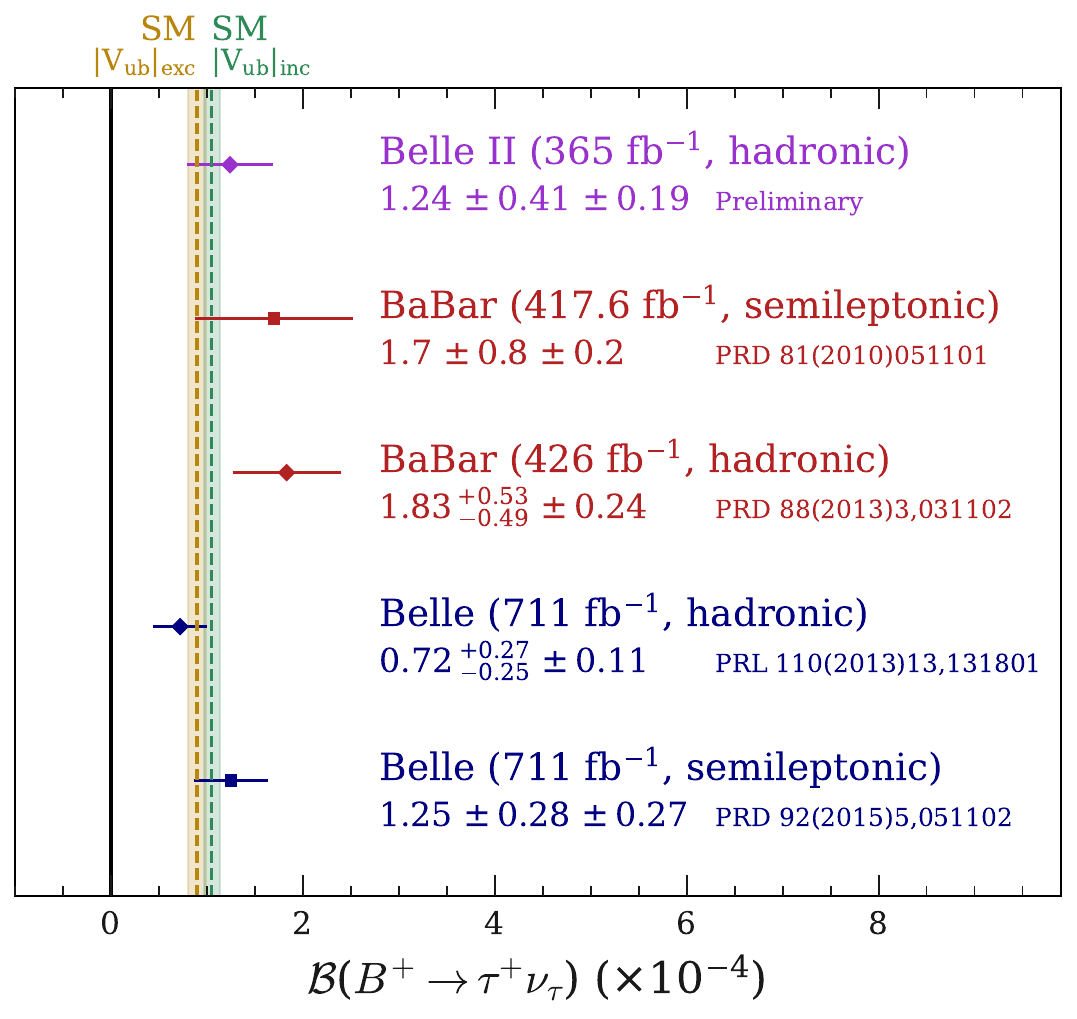}
    \caption{Branching fraction \brbtaunu measured by Belle~II compared with the past measurements and the two SM expectation values, the yellow band calculated using the exclusive value $\Vub =(3.75\pm0.06\pm0.19)\times10^{-3}$ and the green band with the inclusive value $\Vub =(4.06\pm0.12\pm0.11)\times10^{-3}$.}
    \label{fig:br_comparison}
\end{figure}

We present a measurement of the branching fraction of the \Btaunu decay using $\SI{365}{\invfb}$ of electron-positron collision data recorded at the \FourS resonance by the \belletwo detector, using hadronic $B$ tagging. For this measurement, we consider one-prong decays of the \taup lepton.
We measure \brbtaunu to be
\begin{equation}
   \brbtaunu = [1.24 \pm 0.41\stat \pm 0.19\syst]\times10^{-4}
\end{equation}
with a significance of 3.0$\,\sigma$. The measured $\cal B$ is consistent with the current world average and with the SM prediction. 
Figure~\ref{fig:br_comparison} shows a comparison of our \brbtaunu measurement with past measurements from \babar and \belle, and SM predictions based on exclusive and inclusive determinations of \Vub~\cite{banerjee2024averagesbhadronchadrontaulepton}.

Assuming the SM, and using $f_B = \SI{190.0+-1.3}{\mev}$~\cite{aoki2024flagreview2024}, we extract from the \brbtaunu a measurement of the CKM matrix element 
\begin{equation}
    \Vub_{\Btaunu}=[4.41^{+0.74}_{-0.89}]\times10^{-3}.
\end{equation}

Even though we use a smaller data sample, the statistical uncertainty of this measurement is comparable to the previous hadronic tag analysis from \babar ($\SI{426}{\invfb}$)~\cite{babar_had} and \belle ($\SI{711}{\invfb}$)~\cite{belle_had}. This improved sensitivity is due to the use of a new $B$ tagging algorithm and an optimized selection.

\begin{acknowledgments}
This work, based on data collected using the Belle II detector, which was built and commissioned prior to March 2019,
%Belle1 and data collected using the Belle detector, which was operated until June 2010,
was supported by
%Armenia
Higher Education and Science Committee of the Republic of Armenia Grant No.~23LCG-1C011;
%Australia
Australian Research Council and Research Grants
No.~DP200101792, % Jackson
No.~DP210101900, % Urquijo
No.~DP210102831, % Sevior
No.~DE220100462, % Hsu
No.~LE210100098, % Infrastructure
and
No.~LE230100085; % Infrastructure
%Austria
Austrian Federal Ministry of Education, Science and Research,
Austrian Science Fund (FWF) Grants
DOI:~10.55776/P34529,
DOI:~10.55776/J4731,
DOI:~10.55776/J4625,
DOI:~10.55776/M3153,
and
DOI:~10.55776/PAT1836324,
and
Horizon 2020 ERC Starting Grant No.~947006 ``InterLeptons'';
%Canada
Natural Sciences and Engineering Research Council of Canada, Compute Canada and CANARIE;
%China
National Key R\&D Program of China under Contract No.~2022YFA1601903,
National Natural Science Foundation of China and Research Grants
No.~11575017,
No.~11761141009,
No.~11705209,
No.~11975076,
No.~12135005,
No.~12150004,
No.~12161141008,
No.~12475093,
and
No.~12175041,
and Shandong Provincial Natural Science Foundation Project~ZR2022JQ02;
%Czech Republic
the Czech Science Foundation Grant No.~22-18469S 
and
Charles University Grant Agency project No.~246122;
%EU
European Research Council, Seventh Framework PIEF-GA-2013-622527,
Horizon 2020 ERC-Advanced Grants No.~267104 and No.~884719,
Horizon 2020 ERC-Consolidator Grant No.~819127,
Horizon 2020 Marie Sklodowska-Curie Grant Agreement No.~700525 ``NIOBE''
and
No.~101026516,
and
Horizon 2020 Marie Sklodowska-Curie RISE project JENNIFER2 Grant Agreement No.~822070 (European grants);
%France
L'Institut National de Physique Nucl\'{e}aire et de Physique des Particules (IN2P3) du CNRS
and
L'Agence Nationale de la Recherche (ANR) under Grant No.~ANR-21-CE31-0009 (France);
%Germany
BMBF, DFG, HGF, MPG, and AvH Foundation (Germany);
%India
Department of Atomic Energy under Project Identification No.~RTI 4002,
Department of Science and Technology,
and
UPES SEED funding programs
No.~UPES/R\&D-SEED-INFRA/17052023/01 and
No.~UPES/R\&D-SOE/20062022/06 (India);
%Israel
Israel Science Foundation Grant No.~2476/17,
U.S.-Israel Binational Science Foundation Grant No.~2016113, and
Israel Ministry of Science Grant No.~3-16543;
%Italy
Istituto Nazionale di Fisica Nucleare and the Research Grants BELLE2,
and
the ICSC – Centro Nazionale di Ricerca in High Performance Computing, Big Data and Quantum Computing, funded by European Union – NextGenerationEU;
%Japan
Japan Society for the Promotion of Science, Grant-in-Aid for Scientific Research Grants
No.~16H03968,
No.~16H03993,
No.~16H06492,
No.~16K05323,
No.~17H01133,
No.~17H05405,
No.~18K03621,
No.~18H03710,
No.~18H05226,
No.~19H00682, % Niigata
No.~20H05850,
No.~20H05858,
No.~22H00144,
No.~22K14056,
No.~22K21347,
No.~23H05433,
No.~26220706,
and
No.~26400255,
%the National Institute of Informatics, and Science Information NETwork 5 (SINET5), 
and
the Ministry of Education, Culture, Sports, Science, and Technology (MEXT) of Japan;  
%Korea
National Research Foundation (NRF) of Korea Grants
No.~2016R1-D1A1B-02012900,
No.~2018R1-A6A1A-06024970,
No.~2021R1-A6A1A-03043957,
No.~2021R1-F1A-1060423,
No.~2021R1-F1A-1064008,
No.~2022R1-A2C-1003993,
No.~2022R1-A2C-1092335,
No.~RS-2023-00208693,
No.~RS-2024-00354342
and
No.~RS-2022-00197659,
Radiation Science Research Institute,
Foreign Large-Size Research Facility Application Supporting project,
the Global Science Experimental Data Hub Center, the Korea Institute of
Science and Technology Information (K24L2M1C4)
and
KREONET/GLORIAD;
%Malaysia
Universiti Malaya RU grant, Akademi Sains Malaysia, and Ministry of Education Malaysia;
%Mexico
% CINVESTAV-IPN, UNAM, UAS, BUAP and CONACYT are funded under
Frontiers of Science Program Contracts
No.~FOINS-296,
No.~CB-221329,
No.~CB-236394,
No.~CB-254409,
and
No.~CB-180023, and SEP-CINVESTAV Research Grant No.~237 (Mexico);
%Poland
the Polish Ministry of Science and Higher Education and the National Science Center;
%Russia
the Ministry of Science and Higher Education of the Russian Federation
and
the HSE University Basic Research Program, Moscow;
%Saudi Arabia
University of Tabuk Research Grants
No.~S-0256-1438 and No.~S-0280-1439 (Saudi Arabia), and
Researchers Supporting Project number (RSPD2025R873), King Saud University, Riyadh,
Saudi Arabia;
%Slovenia
Slovenian Research Agency and Research Grants
No.~J1-9124
and
No.~P1-0135;
%Spain
%Belle1 Ikerbasque, Basque Foundation for Science,
%Belle1 the State Agency for Research of the Spanish Ministry of Science and Innovation through Grant No. PID2022-136510NB-C33,
Agencia Estatal de Investigacion, Spain
Grant No.~RYC2020-029875-I
and
Generalitat Valenciana, Spain
Grant No.~CIDEGENT/2018/020;
%Swiss (Belle 1)
%Belle1 the Swiss National Science Foundation;
%Sweden
The Knut and Alice Wallenberg Foundation (Sweden), Contracts No.~2021.0174 and No.~2021.0299;
%Taiwan
National Science and Technology Council,
and
Ministry of Education (Taiwan);
%Thailand
Thailand Center of Excellence in Physics;
%Turkey
TUBITAK ULAKBIM (Turkey);
%Ukraine
National Research Foundation of Ukraine, Project No.~2020.02/0257,
and
Ministry of Education and Science of Ukraine;
%USA
the U.S. National Science Foundation and Research Grants
No.~PHY-1913789 % Indiana CEEM
and
No.~PHY-2111604, % Luther
and the U.S. Department of Energy and Research Awards
No.~DE-AC06-76RLO1830, % PNNL
No.~DE-SC0007983, % Wayne State
No.~DE-SC0009824, % Florida
No.~DE-SC0009973, % VPI
No.~DE-SC0010007, % Duke
No.~DE-SC0010073, % South Carolina
No.~DE-SC0010118, % Carnegie Mellon
No.~DE-SC0010504, % Hawaii
No.~DE-SC0011784, % Cincinnati
No.~DE-SC0012704, % BNL
No.~DE-SC0019230, % Duke
No.~DE-SC0021274, % Mississippi
No.~DE-SC0021616, % Mississippi
No.~DE-SC0022350, % Louisville
No.~DE-SC0023470; % South Alabama
%last group
and
%Vietnam
the Vietnam Academy of Science and Technology (VAST) under Grants
No.~NVCC.05.12/22-23
and
No.~DL0000.02/24-25.

% Policy from October 20, 2022
These acknowledgements are not to be interpreted as an endorsement of any statement made
by any of our institutes, funding agencies, governments, or their representatives.

We thank the SuperKEKB team for delivering high-luminosity collisions;
the KEK cryogenics group for the efficient operation of the detector solenoid magnet and IBBelle on site;
the KEK Computer Research Center for on-site computing support; the NII for SINET6 network support;
and the raw-data centers hosted by BNL, DESY, GridKa, IN2P3, INFN, 
%Belle1 PNNL/EMSL, 
and the University of Victoria.

\end{acknowledgments}

\bibliography{references}

\appendix
\section{Calibration of \texorpdfstring{\ng}{n_gammaextra}}
\label{appx:A}
\begin{figure*}[htbp]
    \centering
    \begin{minipage}{0.31\textwidth}
        \centering
        \includegraphics[width = \textwidth, keepaspectratio]{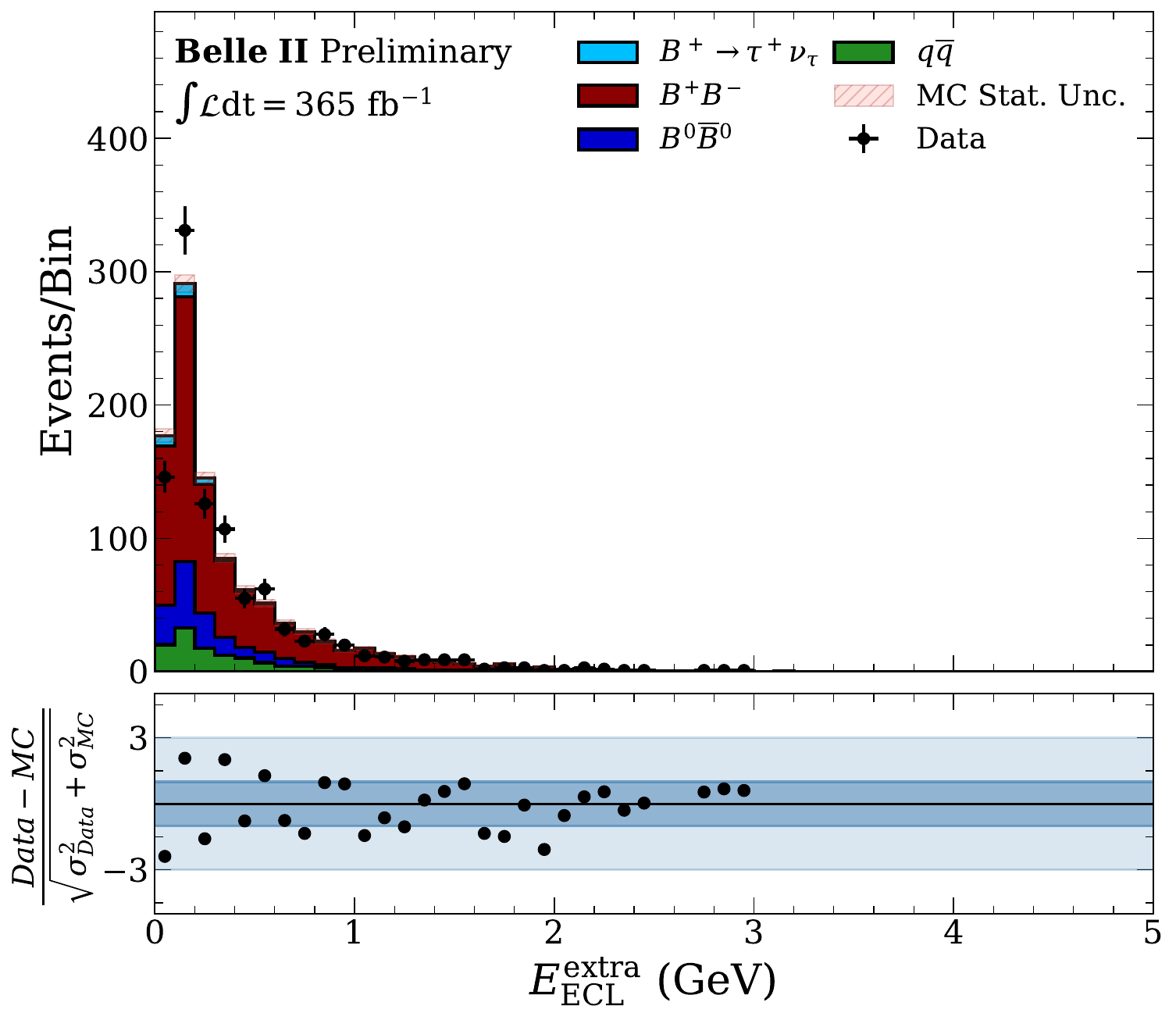}
    \end{minipage}
    \begin{minipage}{0.31\textwidth}
        \centering
        \includegraphics[width = \textwidth, keepaspectratio]{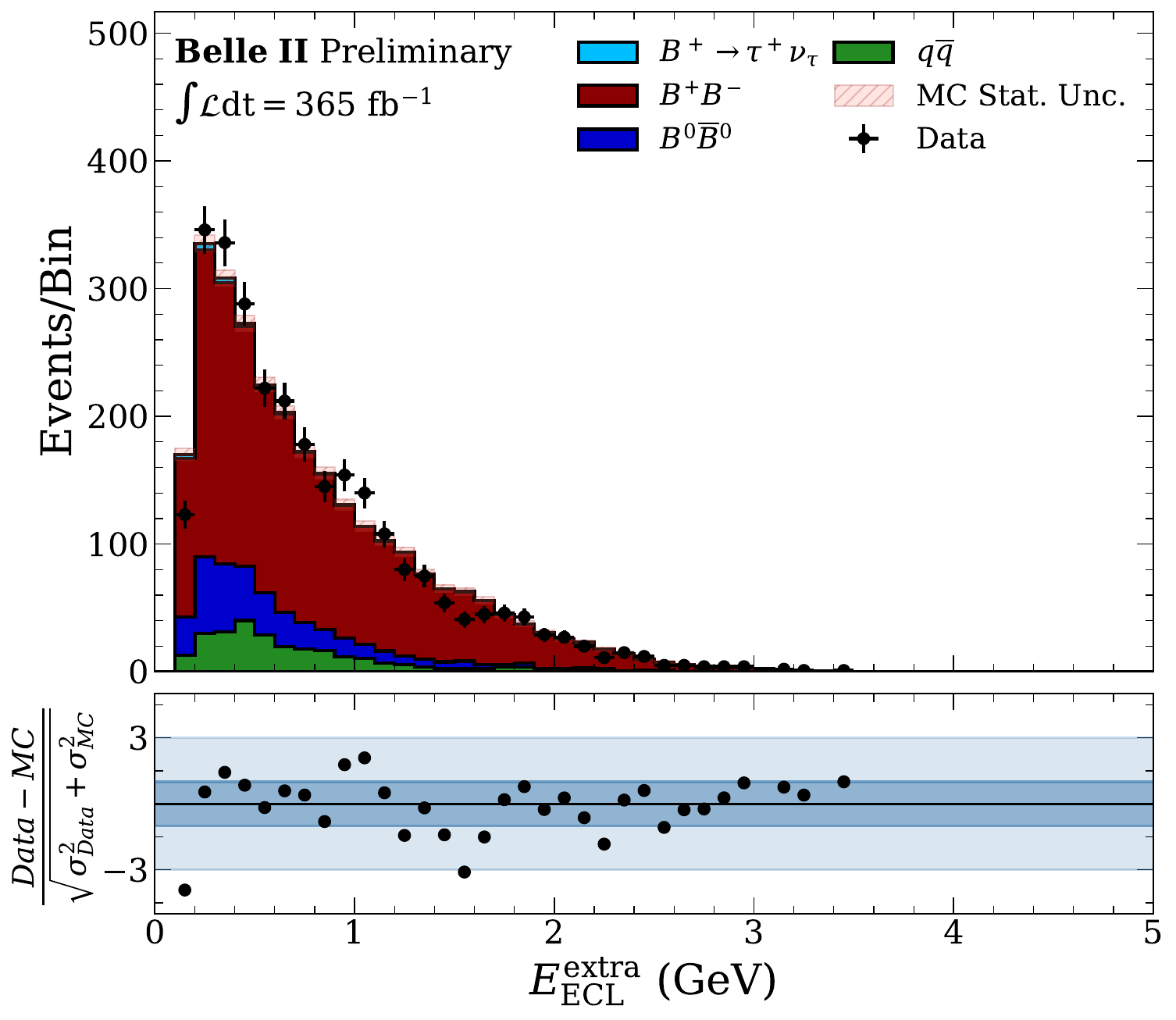}
    \end{minipage}
    \begin{minipage}{0.31\textwidth}
        \centering
        \includegraphics[width = \textwidth, keepaspectratio]{Extra_Photons/Extra_Energy_3.pdf}
    \end{minipage}\\
    \begin{minipage}{0.31\textwidth}
        \centering
        \includegraphics[width = \textwidth, keepaspectratio]{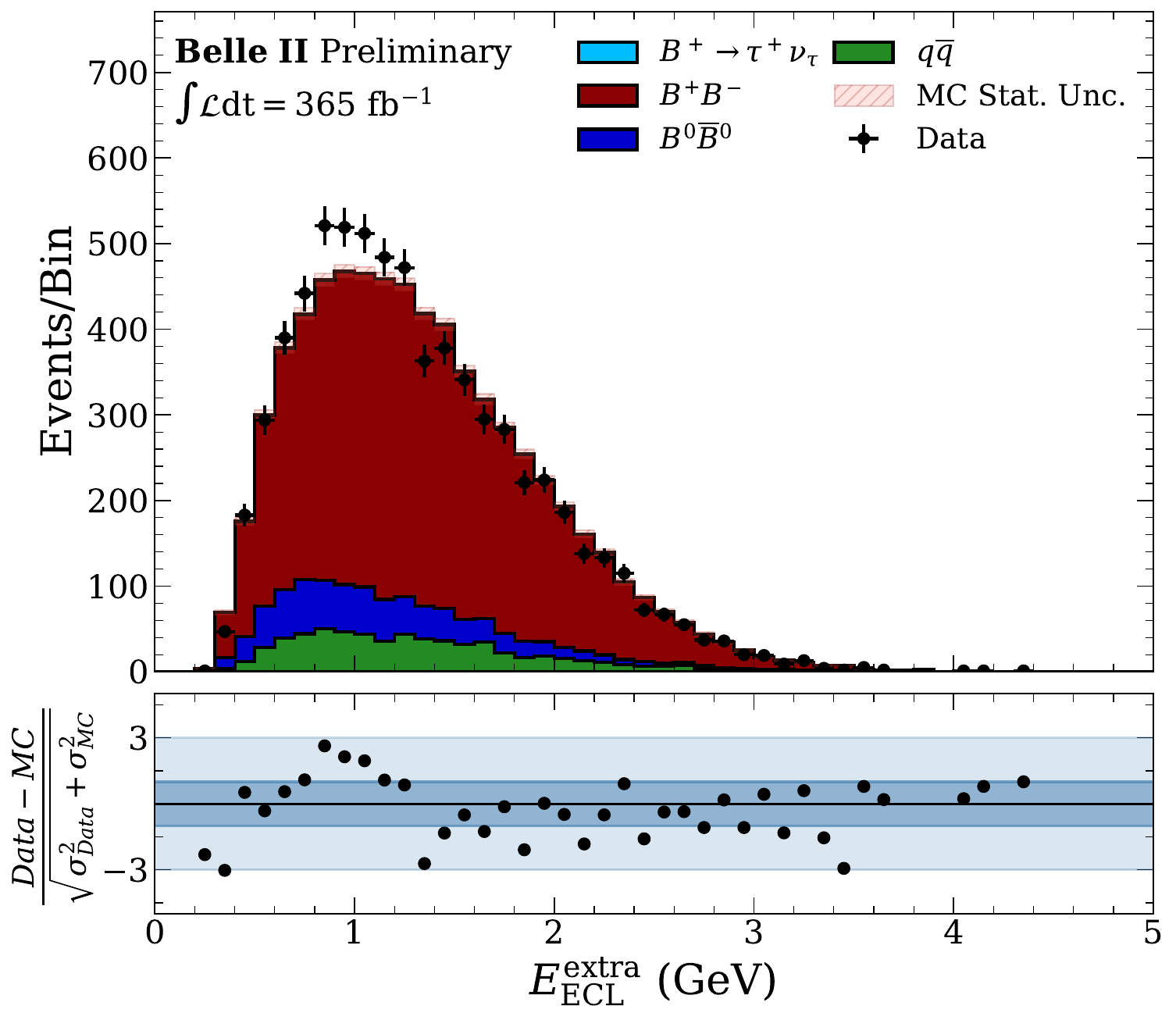}
    \end{minipage}
    \begin{minipage}{0.31\textwidth}
        \centering
        \includegraphics[width = \textwidth, keepaspectratio]{Extra_Photons/Extra_Energy_5.pdf}
    \end{minipage}
    \begin{minipage}{0.31\textwidth}
        \centering
        \includegraphics[width = \textwidth, keepaspectratio]{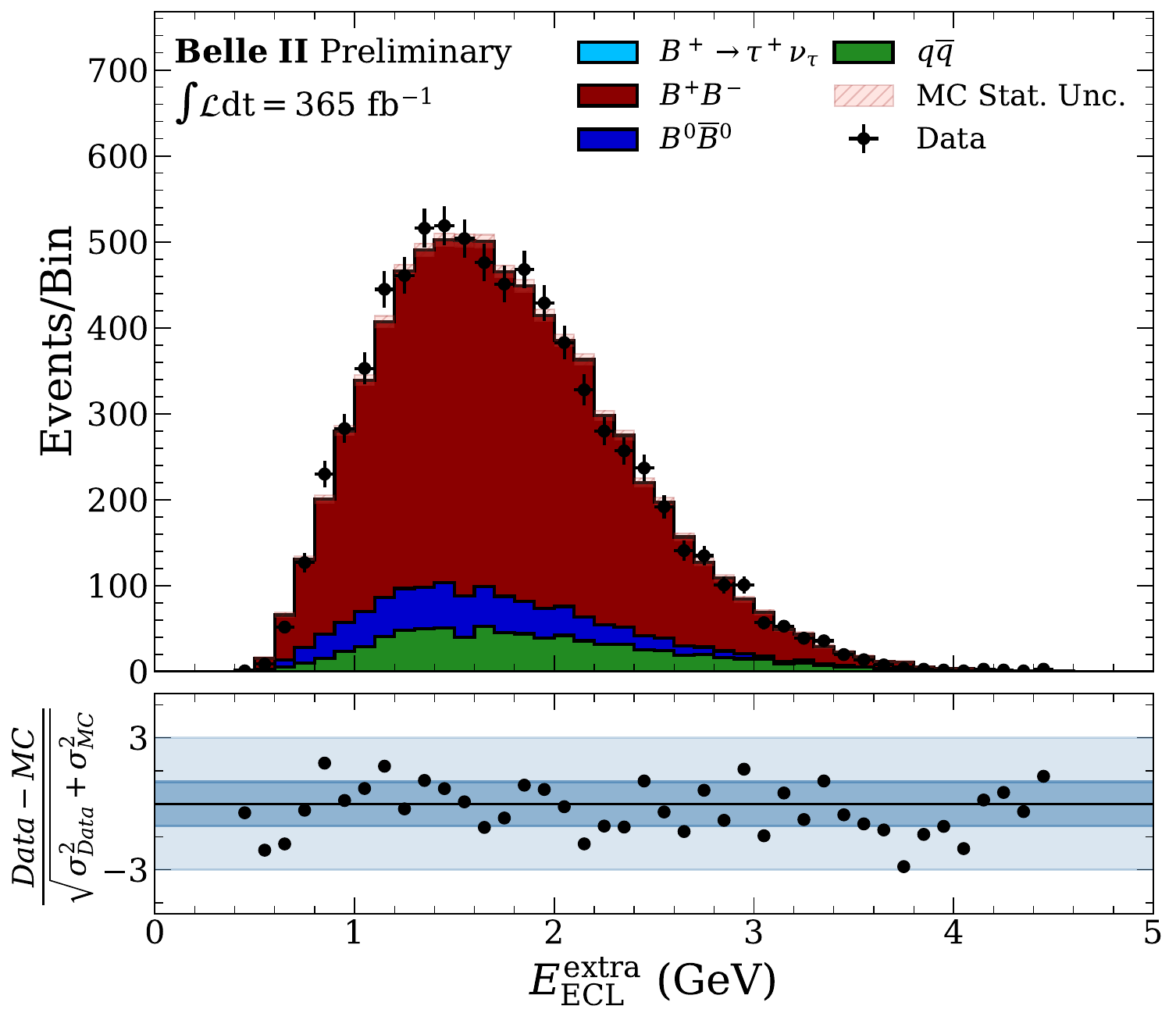}
    \end{minipage}\\
    \begin{minipage}{0.31\textwidth}
        \centering
        \includegraphics[width = \textwidth, keepaspectratio]{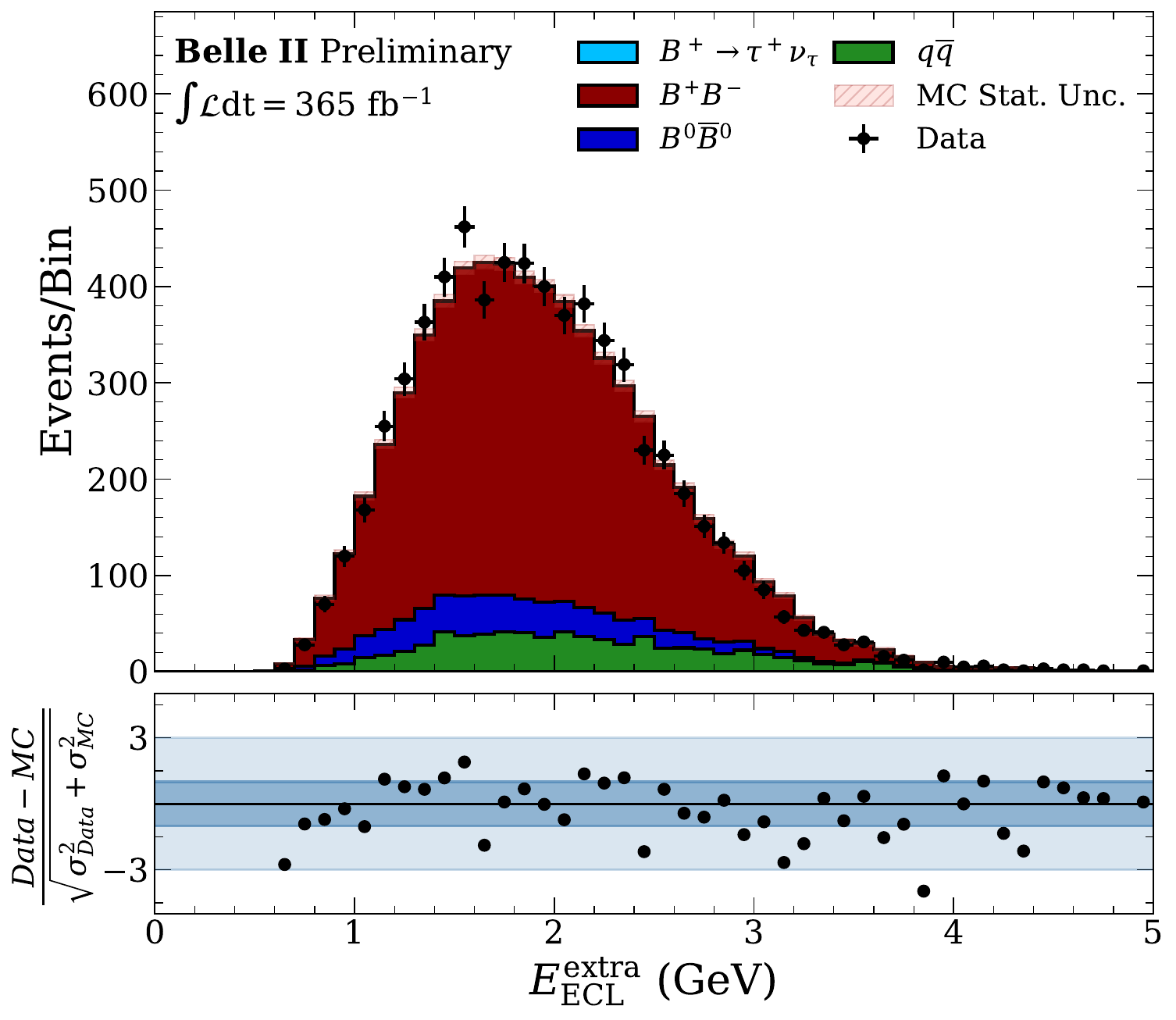}
    \end{minipage}
    \begin{minipage}{0.31\textwidth}
        \centering
        \includegraphics[width = \textwidth, keepaspectratio]{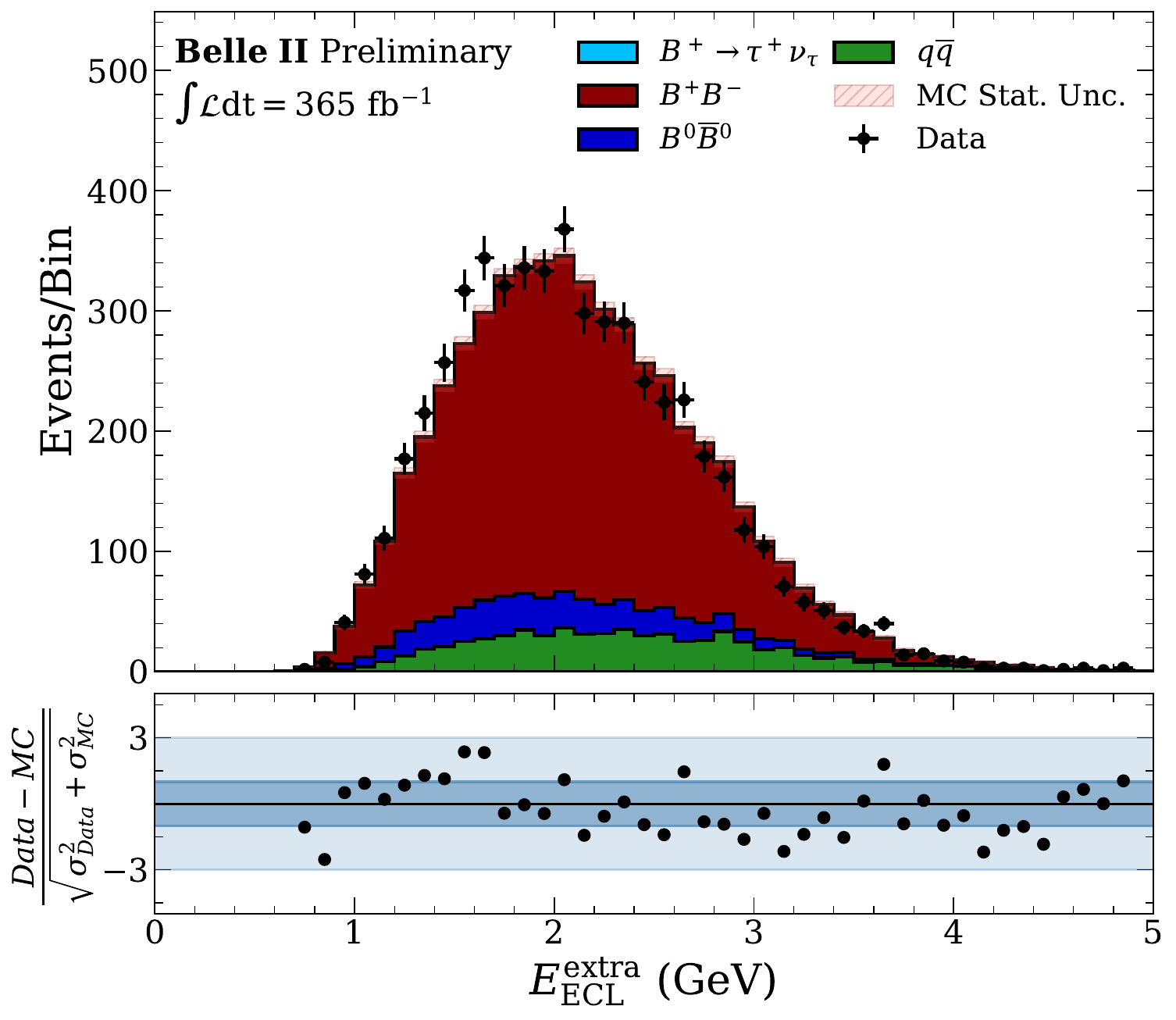}
    \end{minipage}
    \begin{minipage}{0.31\textwidth}
        \centering
        \includegraphics[width = \textwidth, keepaspectratio]{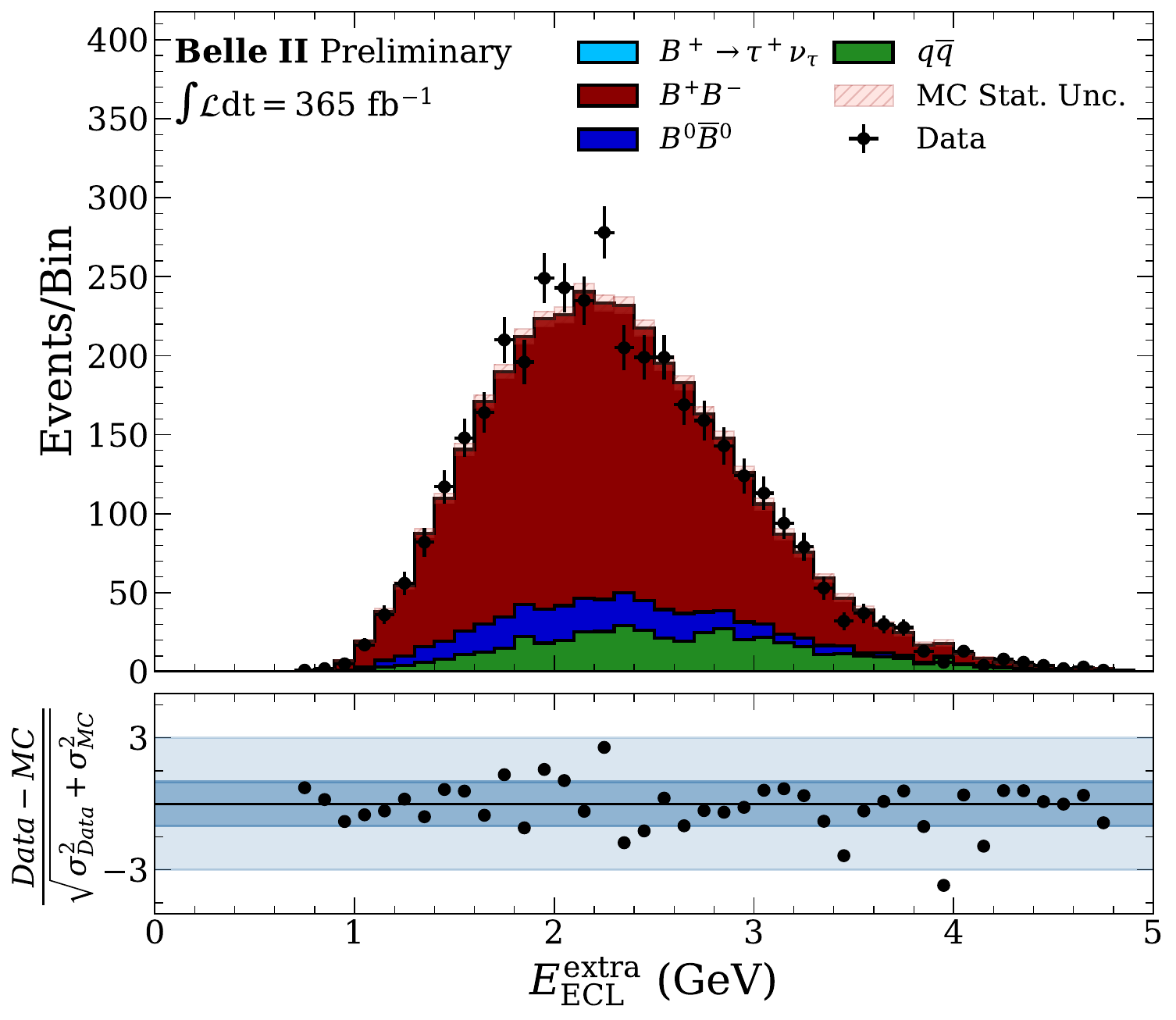}
    \end{minipage}\\
    \caption{Distribution of \eextra as a function of \ng. Simulations are normalized to data. The first row shows \eextra with $\ng = 1, 2, \mathrm{and}\ 3$. The second row shows \eextra with $\ng = 4, 5, \mathrm{and}\ 6$. The third row shows \eextra $\ng = 7, 8, \mathrm{and}\ 9$.}
    \label{fig:extra_mult_all}
\end{figure*}

The main source of the discrepancy in the \eextra distribution between data and simulation is related to incorrect modeling of \ng. Figure \ref{fig:extra_mult_all} shows the \eextra shape for $\ng=1,2,...9$, normalizing simulation to data. The \Btaunu signal affects only \ng$\leqslant 2$.
The simulation agrees well with the data for this variable. Therefore, we correct the normalization through bin-by-bin corrections of the \ng distribution using different control samples.

\subsection{Extra-Tracks control sample}
\label{sec:extratrack}
\begin{figure*}[htbp]
    \centering
    \begin{minipage}{0.41\textwidth}
        \centering
        \includegraphics[width = \textwidth, keepaspectratio]{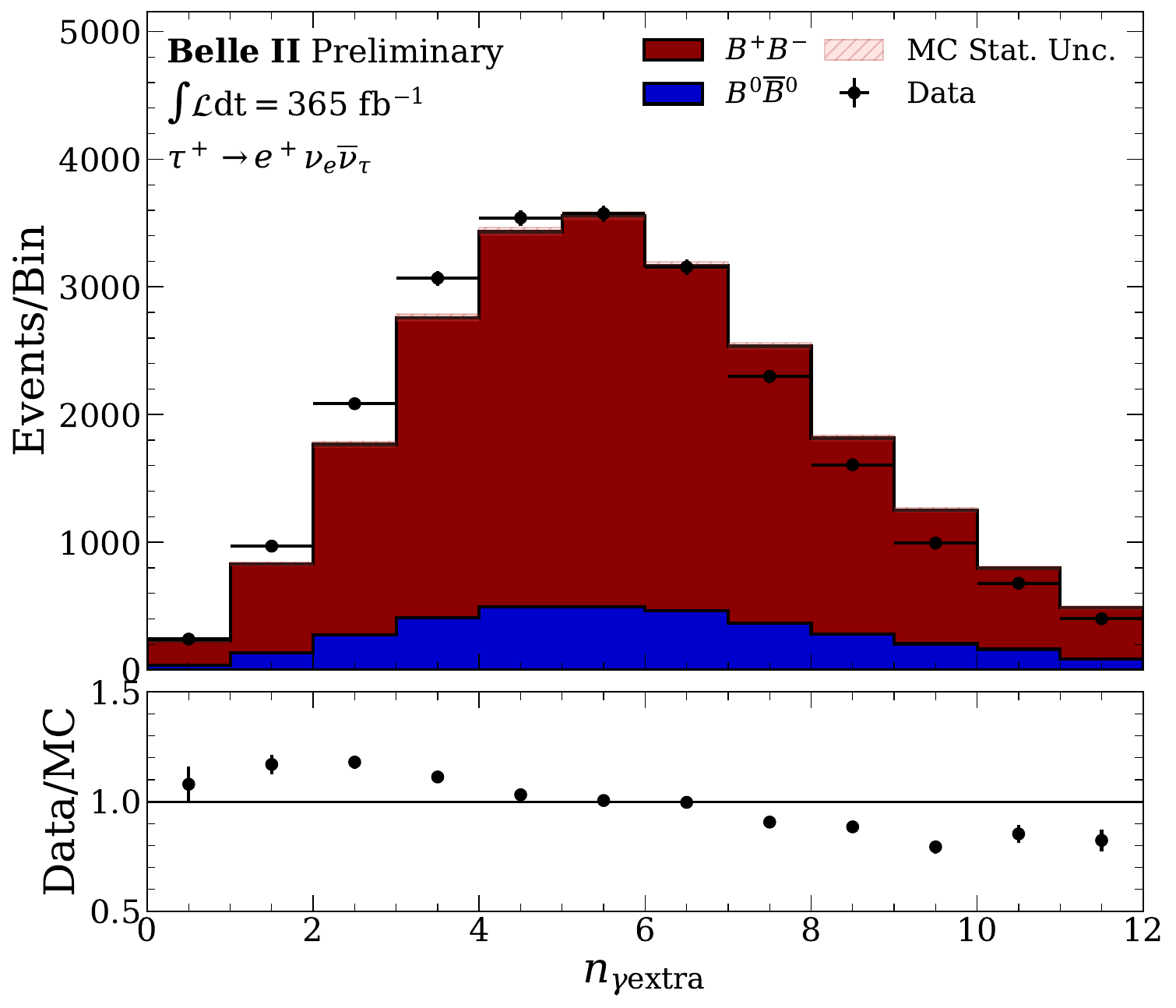}
    \end{minipage}
    \begin{minipage}{0.41\textwidth}
        \centering
        \includegraphics[width = \textwidth, keepaspectratio]{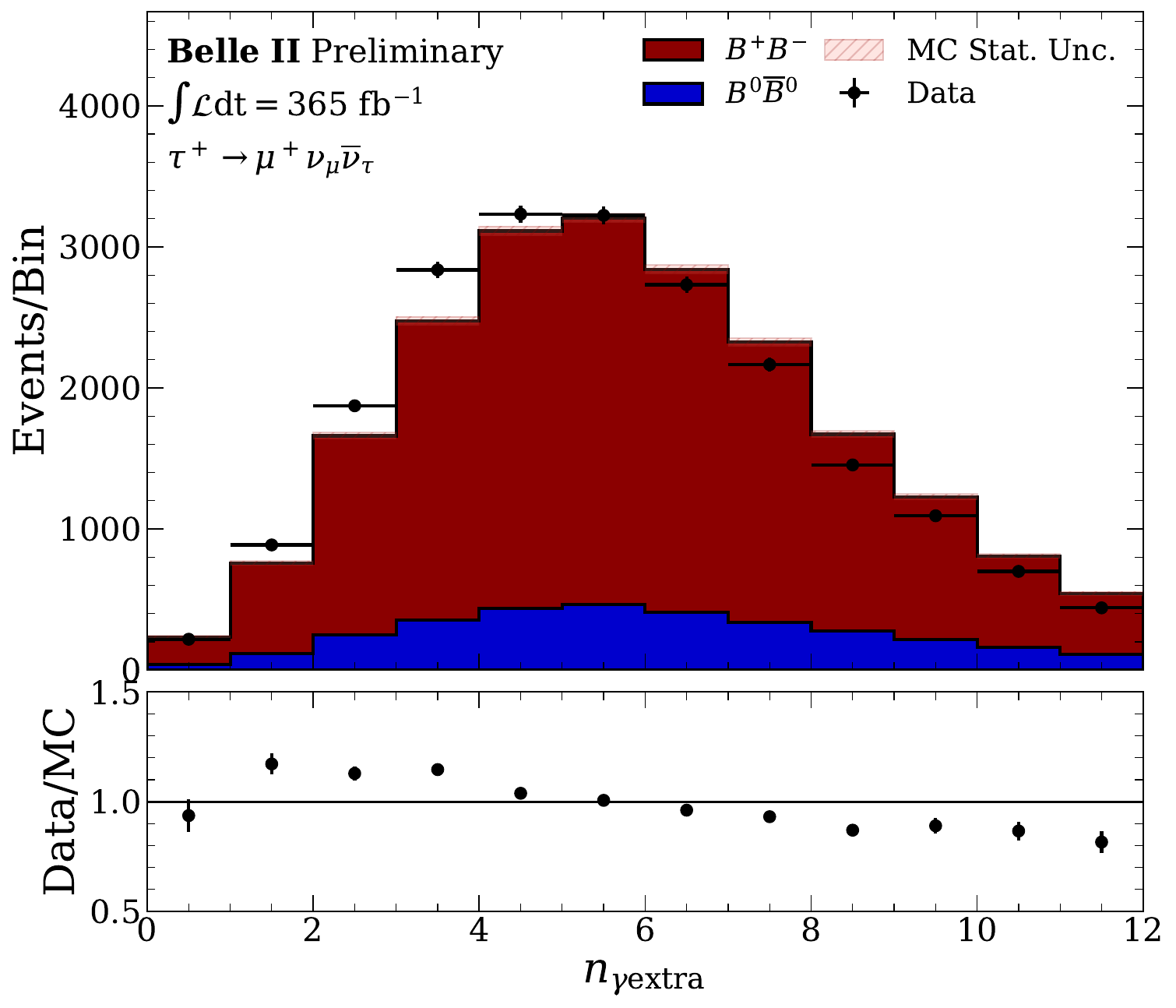}
    \end{minipage}\\
    \begin{minipage}{0.41\textwidth}
        \centering
        \includegraphics[width = \textwidth, keepaspectratio]{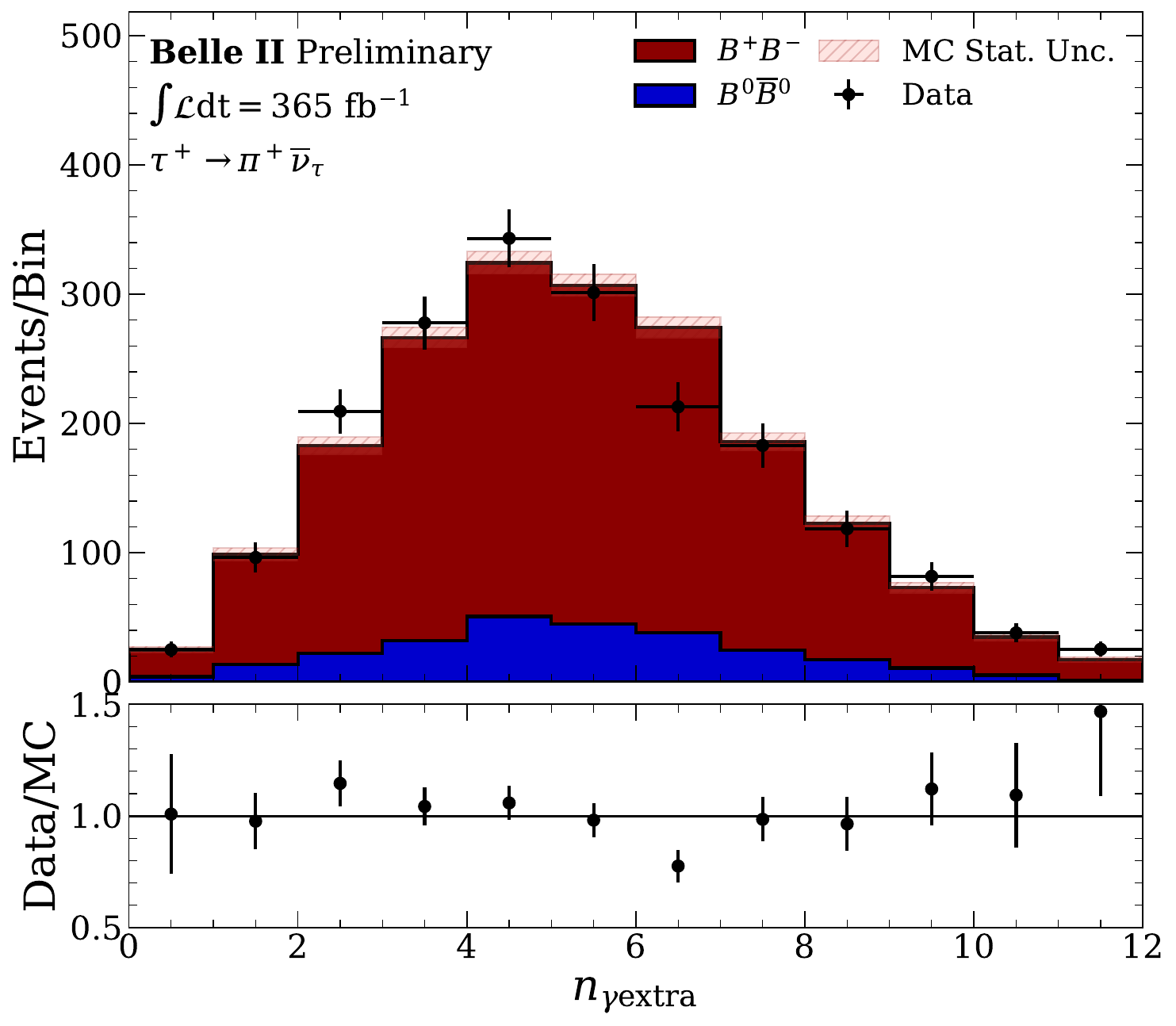}
    \end{minipage}
        \begin{minipage}{0.41\textwidth}
        \centering
        \includegraphics[width = \textwidth, keepaspectratio]{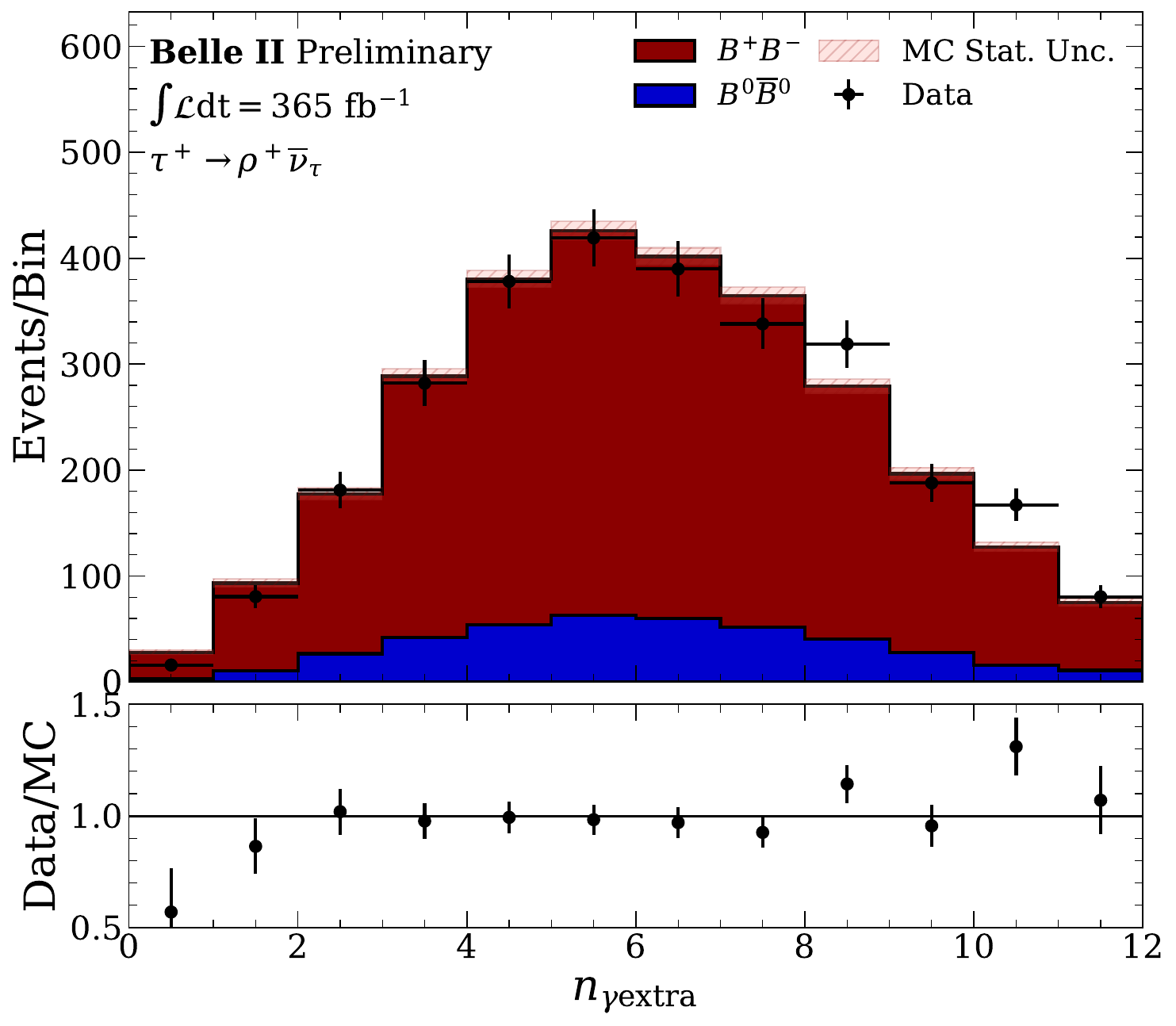}
    \end{minipage}\\
    \caption{Distributions of \ng in the Extra-Tracks control sample for the four signal channels. The bottom panel of each distribution shows the Data/MC ratio values and their uncertainties to be used as corrections for MC background for the analysis. The continuum component is subtracted since the correction factors are used only to correct the \BBbar background component.}
    \label{fig:extra_control_sample_multiplicity}
\end{figure*}

We select a control sample of \Btaunu events that have more than 2 tracks in the ROE to evaluate the corrections to the \eextra distribution for the \BBbar background simulation. The control sample has the same \BBbar background composition but no signal events.
The control sample is defined using 2 different requirements on the charged tracks: \texttt{IP tracks} and \texttt{Signal-like Tracks}. The \texttt{IP tracks} include all the tracks in the ROE with an impact parameter with respect to the IP being less than 2\cm along the $z$ axis, and 0.5\cm in the transverse plane, without any momentum requirement. The \texttt{Signal-like Tracks} criterion has the same definition but also requires a momentum greater than 0.5\gev. We then require $N_\texttt{IP Tracks} > 1$ and $N_\texttt{Signal-like Tracks} = 0$. 
The average number of $N_\texttt{IP Tracks}$ in each event for this control sample is $3.2$.
Figure \ref{fig:extra_control_sample_multiplicity} shows the \ng distribution for data and MC in this control sample for each signal channel, from which we extract correction factors to reweight \BBbar MC background.

\subsection{\texorpdfstring{\Bp\to\Dstarz\ellp\neul}{BDstarellnu} control sample}
\label{sec:dstarellnu}
We use the \Bp\to\Dstarz\ellp\neul (\ellp=\xspace\ep,\xspace\mup) control sample to extract the correction to reweight signal MC for leptonic \taup decay categories. As for the \Btaunu case, \eextra (calculated from ROE) is expected to peak at zero. 
We reconstruct a hadronic \btag candidate in each event with the hadronic FEI algorithm, and a signal \bsig decaying in \Dstarz\ellp\neul and no extra tracks in the ROE. The requirements on the hadronic \btag are the same as in the main analysis. For the signal side, we reconstruct the \Dstarz in its decays to \Dz\piz and \Dz\Pgamma, with \Dz\to\Km\pip,  \KS\pip\pim, \Km\pip\pip\pim. After the reconstruction, we apply all the corrections used in the main analysis to this sample. This includes continuum-reweighting and normalization, PID corrections, ROE clusters clean-up, and corrections. To increase the purity of the sample, we require the momentum of the lepton to be greater than $1.5\gev/c$ and the energy of the \g from the \Dstarz\to\Dz\g decay greater than 170\mev.
The main component is correctly reconstructed \Bp\to\Dstarz\ellp\neul, which is 84\% of the full sample. The remaining 16\% is composed of misreconstructed \Bp\to\Dstarstar\ellp\neul, \Bp\to\Dz\ellp\neul and other channels.
Figure \ref{fig:dstar_control_sample_multiplicity} shows the \ng distribution for data and MC in this control sample, from which we extract correction factors to reweight MC signal for leptonic $\tau^+$ decay categories.
\\
\subsection{Double Tag control sample}
\label{sec:doubletag}
\begin{figure*}[htbp]
    \centering
    \subfloat{(a)\label{fig:dstar_control_sample_multiplicity}}{\includegraphics[width = 0.44\textwidth, keepaspectratio]{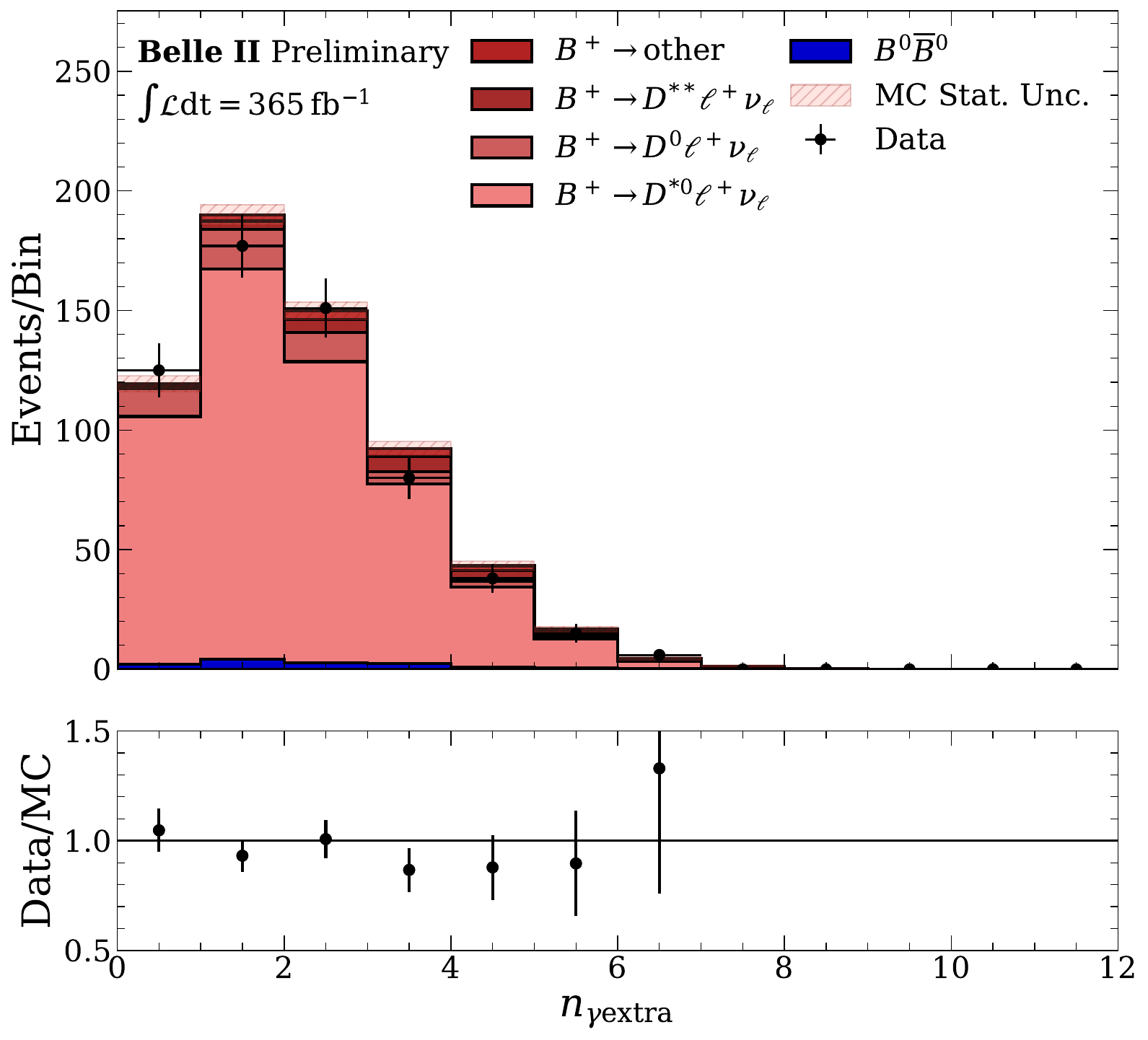}}
    \subfloat{(b)\label{fig:doubletag_control_sample_multiplicity}}{\includegraphics[width = 0.44\textwidth, keepaspectratio]{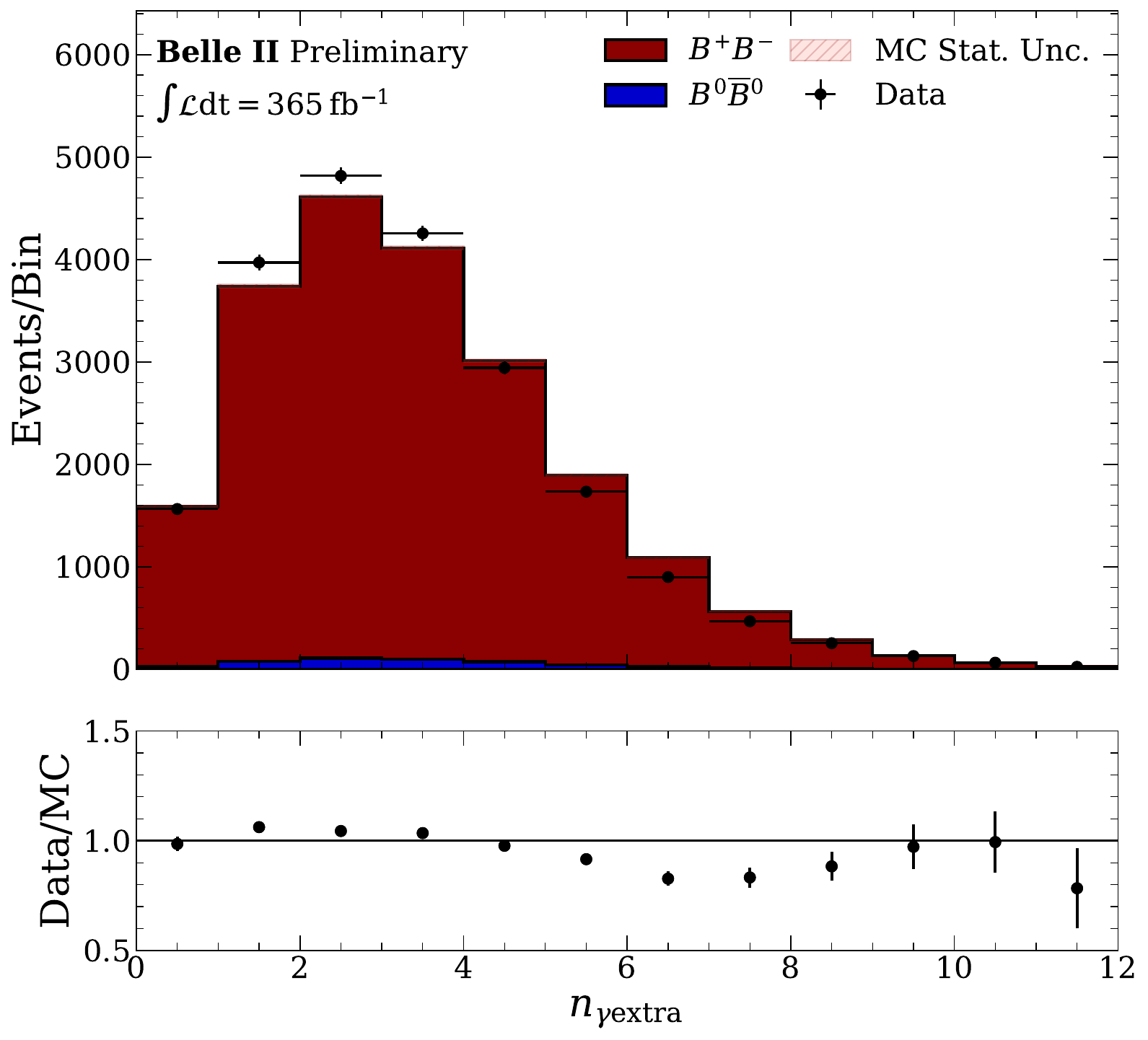}}
    \caption{Distributions of \ng in the \Bp\to\Dstarz\ellp\neul (a) and Double Tag (b) control samples. The bottom panel of each distribution shows the Data/MC ratio values and their uncertainties to be used as corrections for leptonic and hadronic signal MC, respectively. The continuum component is subtracted.}
    \label{fig:signal_control_sample_multiplicity}
\end{figure*}

We use the Double Tag control sample to evaluate the corrections for the hadronic \taup decays in the signal simulation. We reconstruct events with two non-overlapping $B$ candidates with opposite charges with the hadronic FEI algorithm. We require no extra tracks in the ROE. The selection requirements for the hadronic \btag are the same as the main analysis. The signal side is a second $B$ reconstructed by the hadronic FEI.
In the Double Tag control sample, \eextra is expected to peak at $\SI{0}{GeV}$ as for \Btaunu.
We apply the same continuum-reweighting and normalization, PID corrections, ROE clusters clean-up, and corrections as for the main analysis.
Figure \ref{fig:doubletag_control_sample_multiplicity} shows the data and MC agreement in this control sample, from which we extract correction factors to reweight signal MC for hadronic $\tau^+$ decay categories.\\

\section{Distributions of fit variables} \label{appx:B}

\begin{figure*}[htbp]
    \centering
    \begin{minipage}{0.24\textwidth}
        \centering
        \includegraphics[width = \textwidth, keepaspectratio]{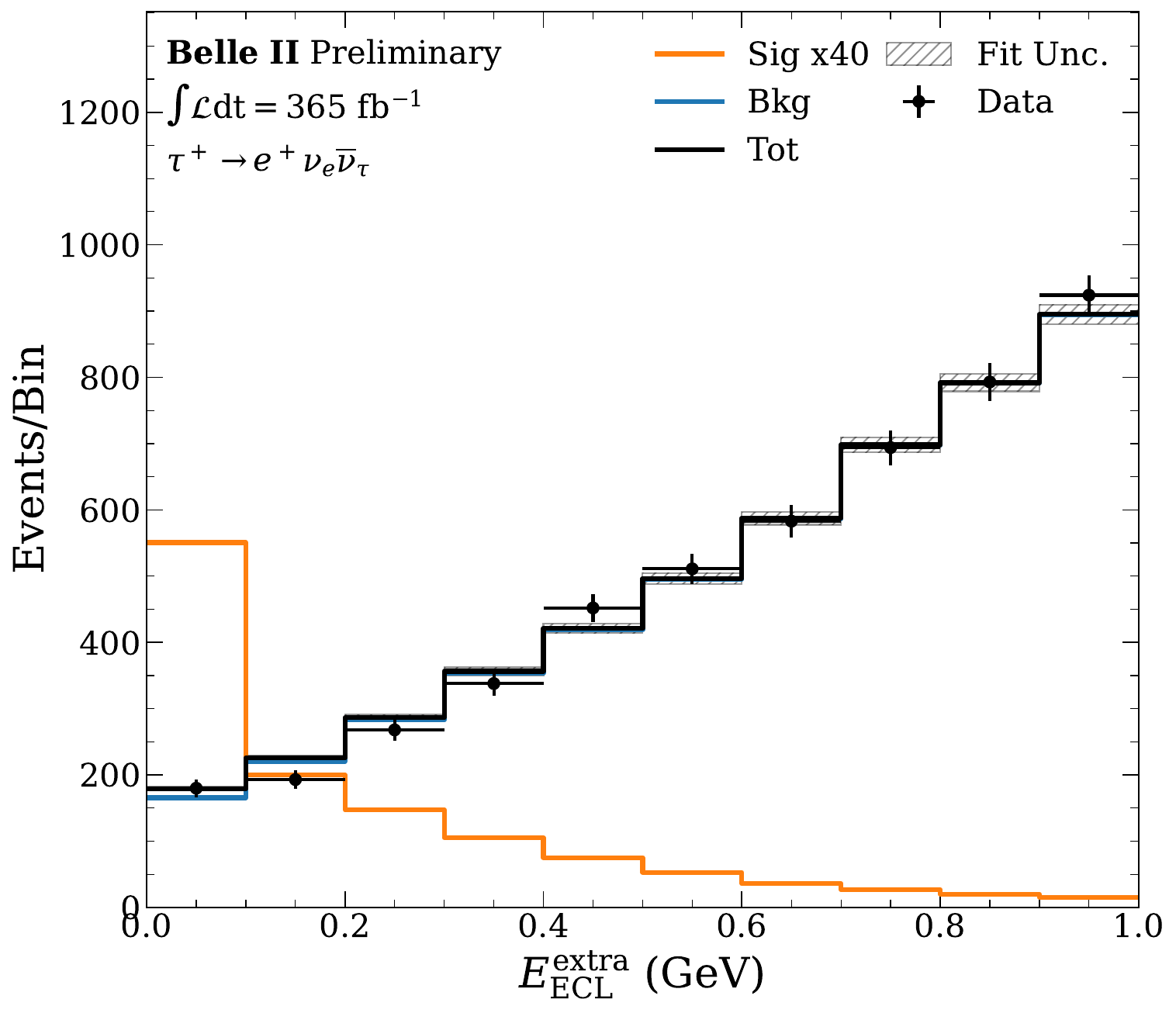}
    \end{minipage}
    \begin{minipage}{0.24\textwidth}
        \centering
        \includegraphics[width = \textwidth, keepaspectratio]{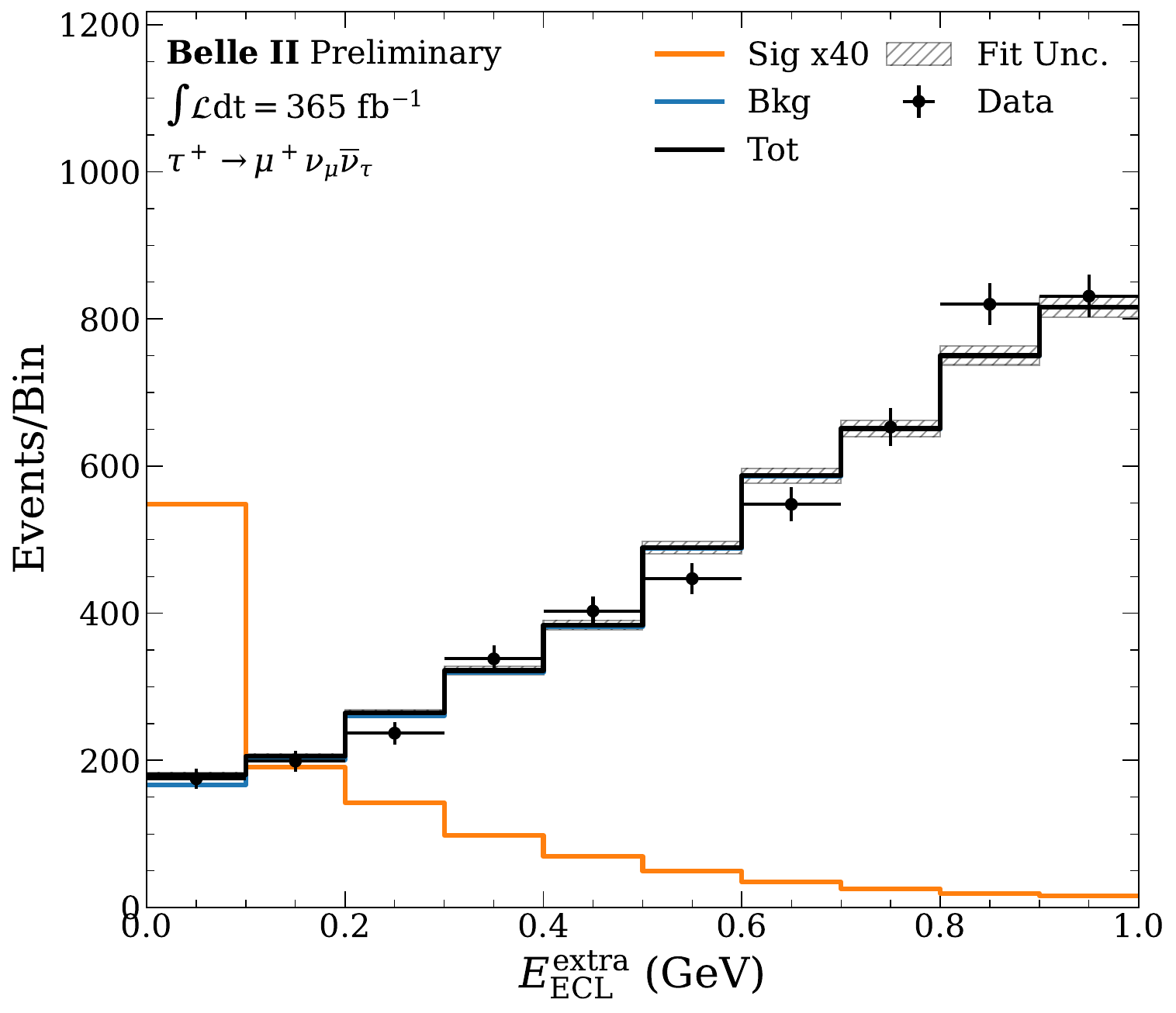}
    \end{minipage}
    \begin{minipage}{0.24\textwidth}
        \centering
        \includegraphics[width = \textwidth, keepaspectratio]{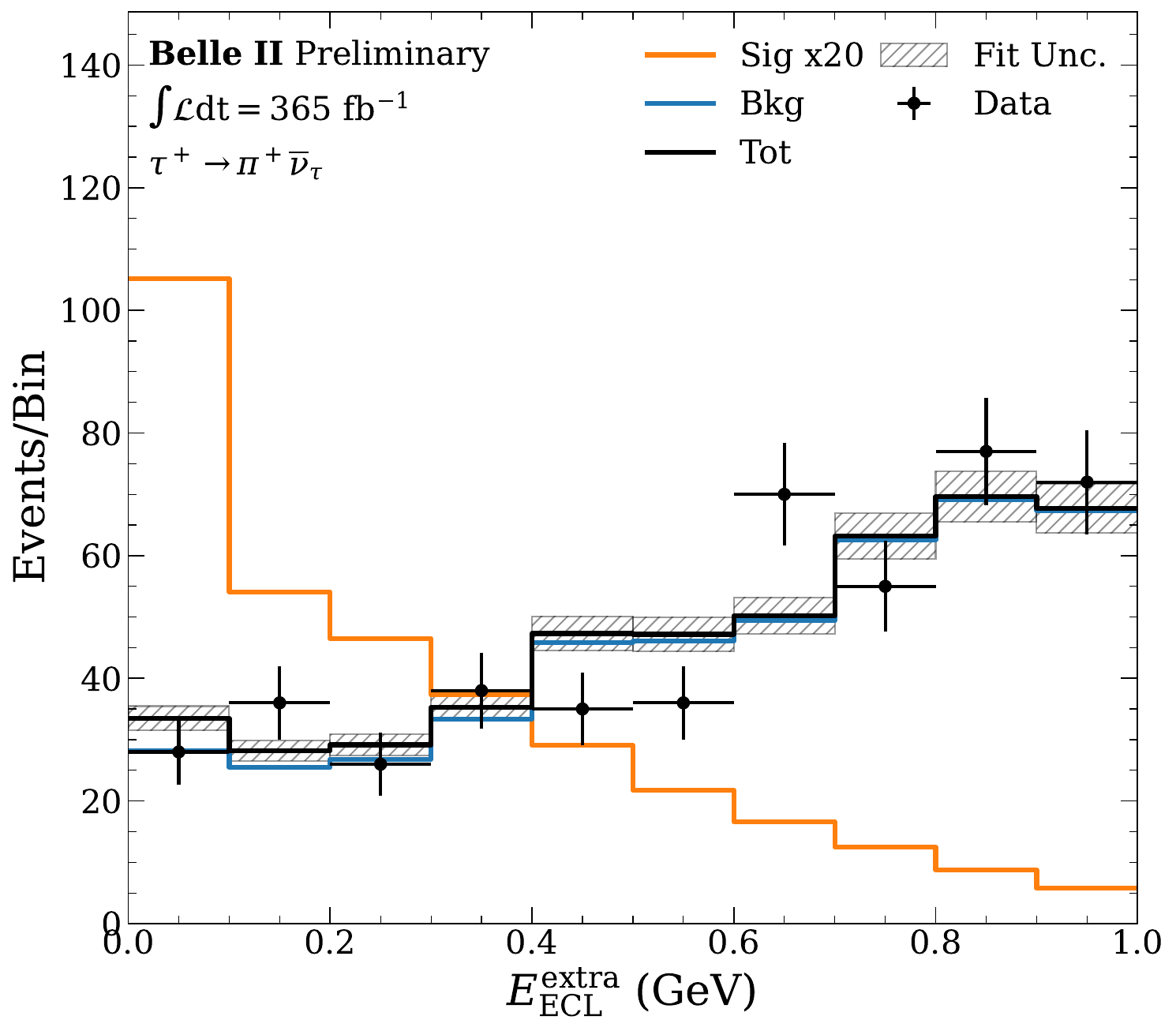}
    \end{minipage}
        \begin{minipage}{0.24\textwidth}
        \centering
        \includegraphics[width = \textwidth, keepaspectratio]{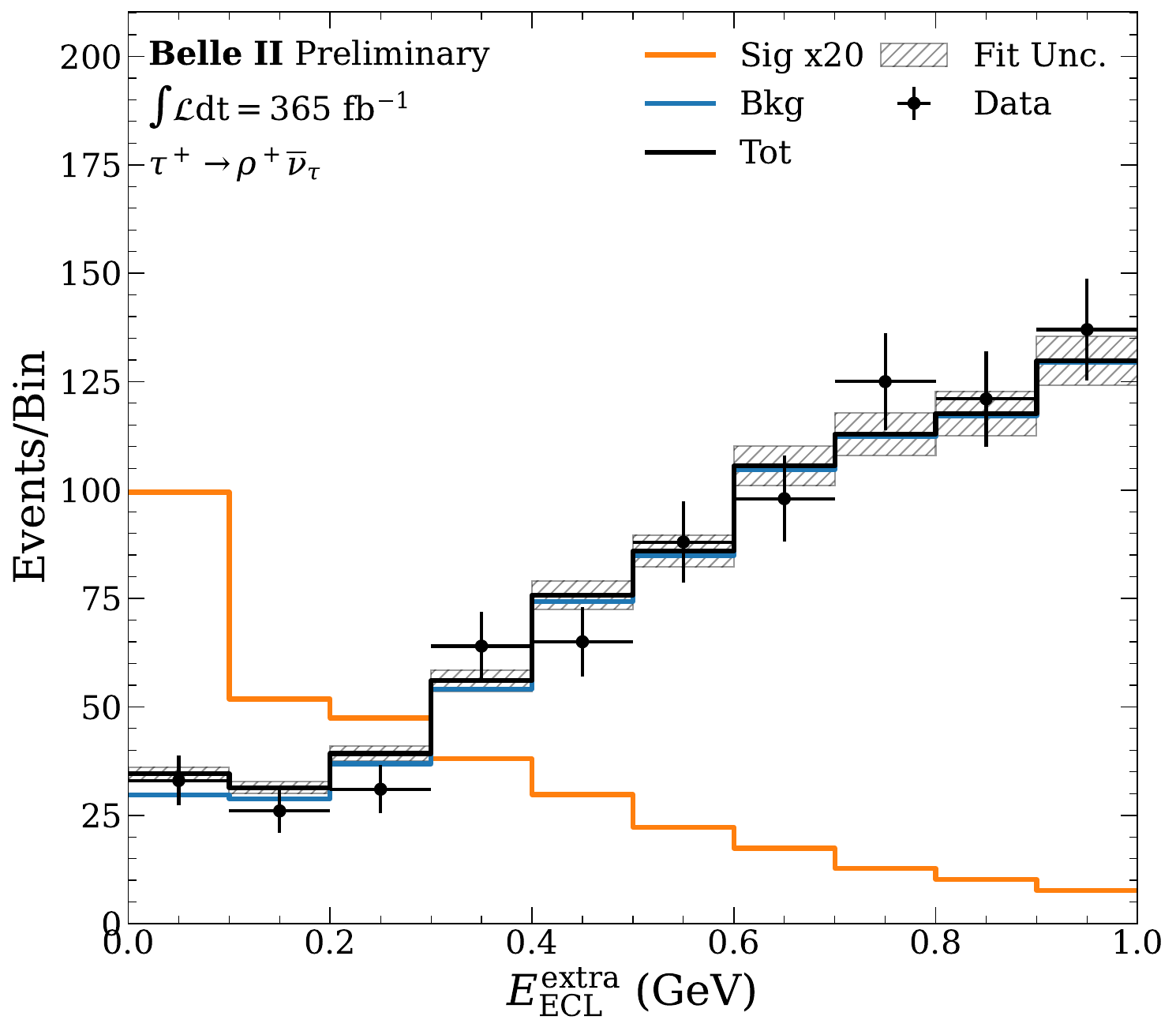}
    \end{minipage}\\
    \begin{minipage}{0.24\textwidth}
        \centering
        \includegraphics[width = \textwidth, keepaspectratio]{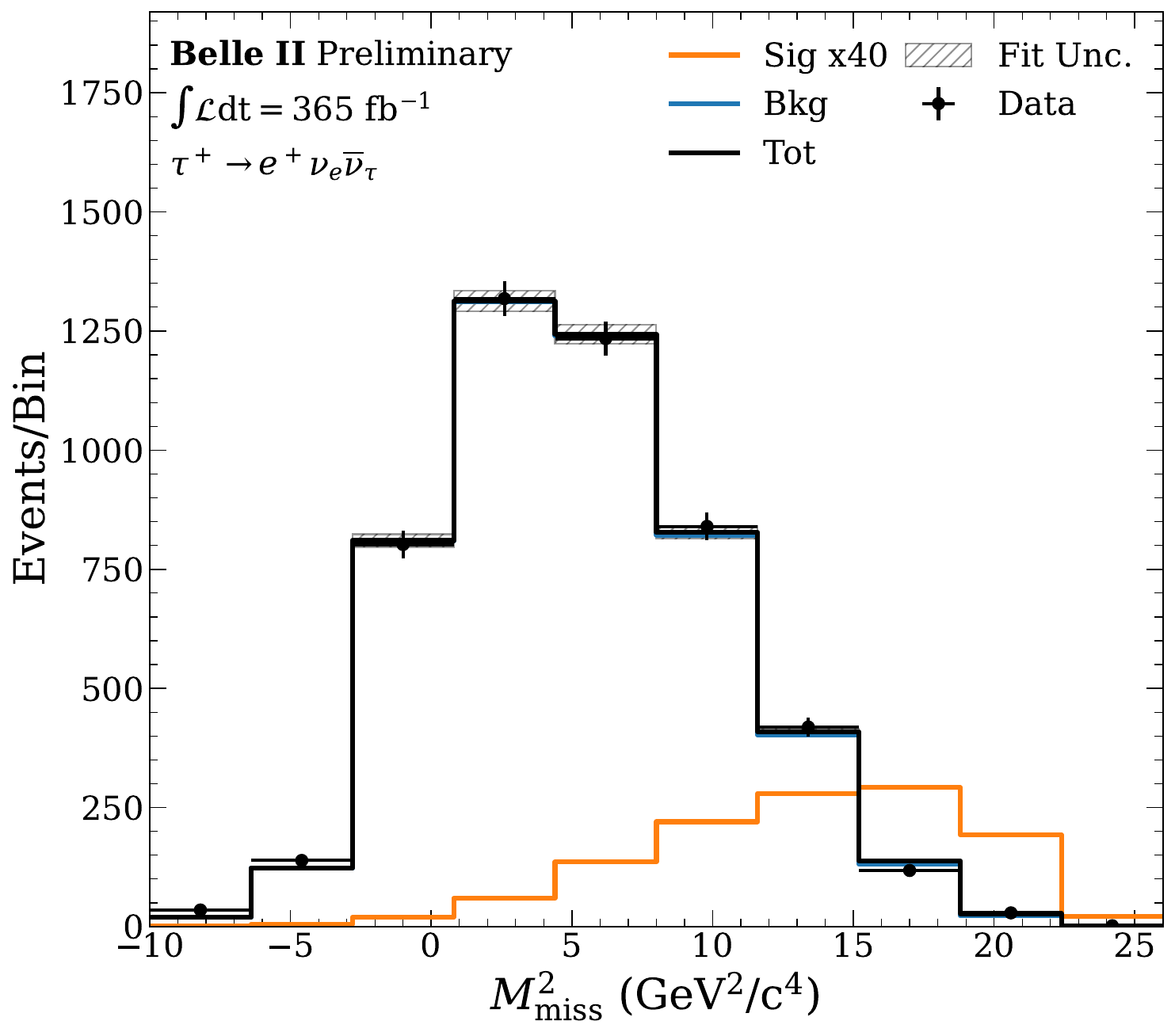}
    \end{minipage}
    \begin{minipage}{0.24\textwidth}
        \centering
        \includegraphics[width = \textwidth, keepaspectratio]{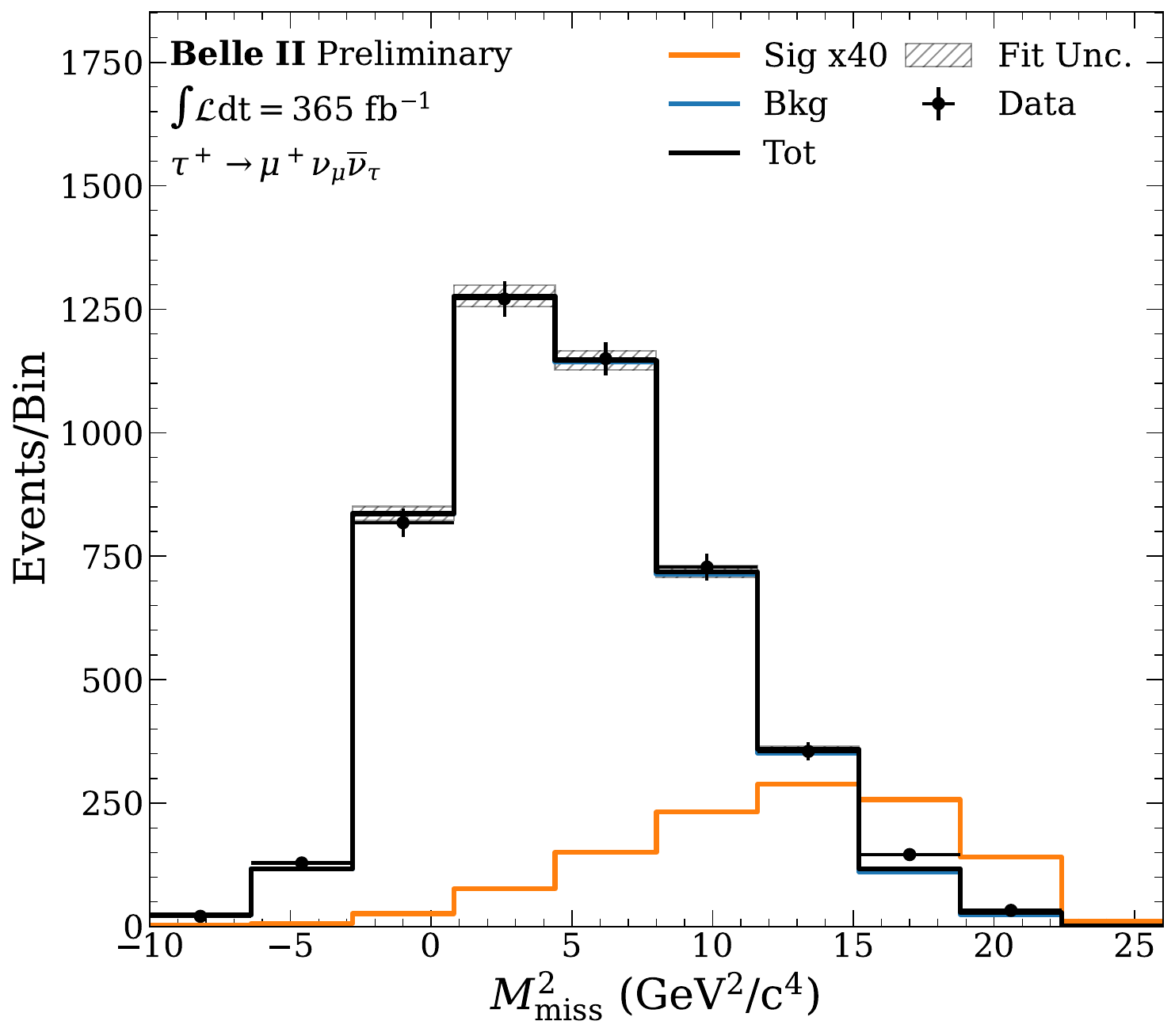}
    \end{minipage}
    \begin{minipage}{0.24\textwidth}
        \centering
        \includegraphics[width = \textwidth, keepaspectratio]{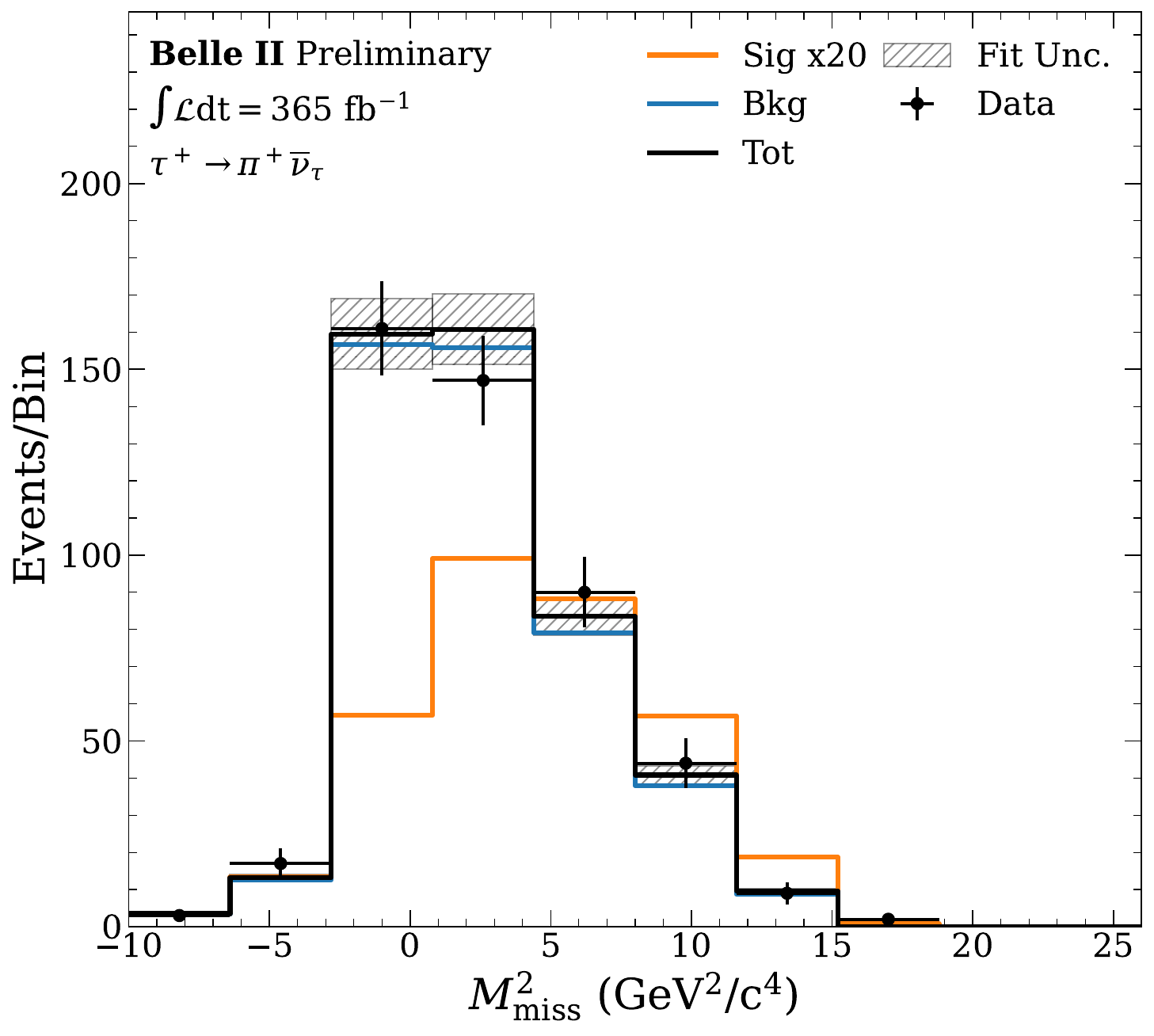}
    \end{minipage}
        \begin{minipage}{0.24\textwidth}
        \centering
        \includegraphics[width = \textwidth, keepaspectratio]{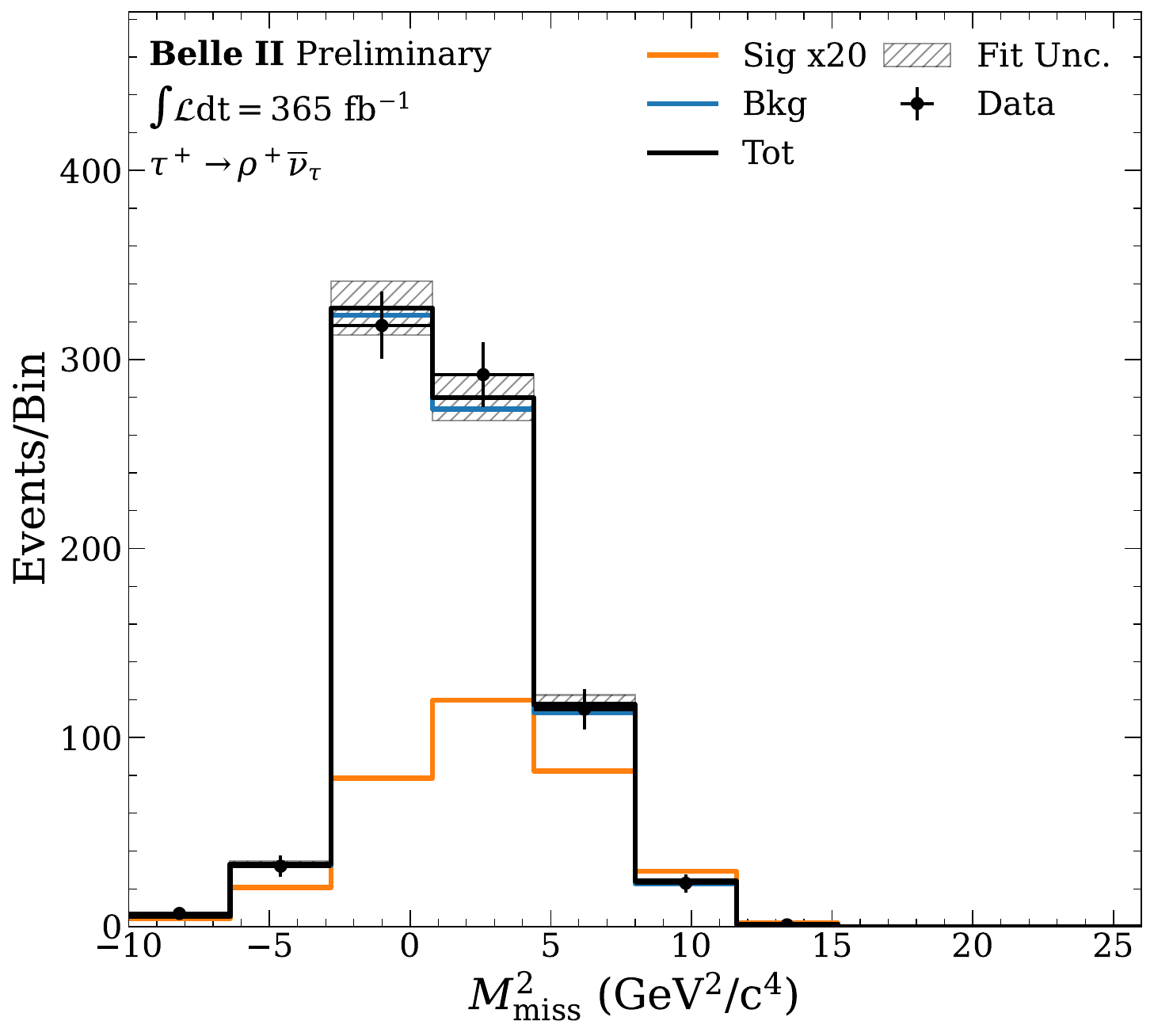}
    \end{minipage}
    \caption{Distributions of \eextra (first row) and \missM (second row) with the fit results superimposed for each $\tau^+$ category. The signal component is scaled by a factor of 40 for leptonic channels and 20 for hadronic channels in order to make it visible.}
    \label{fig:eextra_missM2_post_fit_proj}
\end{figure*}

In Fig. \ref{fig:eextra_missM2_post_fit_proj} we show the projections of the fit for \eextra and \missM distributions for each $\tau^+$ category.

\end{document}